\documentclass[12pt,onepage,oneside]{article}
\usepackage{fullpage}
\usepackage[utf8]{inputenc}
\usepackage[margin=1in]{geometry}
\usepackage{footmisc}             

\usepackage{natbib} 

\usepackage{dsfont}
\usepackage{amsmath}
\usepackage{amssymb}
\usepackage[hidelinks]{hyperref}
\usepackage{booktabs}
\usepackage{chngcntr}

\usepackage{xcolor}
\usepackage{longtable}
\usepackage{appendix}
\usepackage{graphicx}
\usepackage{graphics}
\usepackage{float}
\usepackage{comment}
\usepackage{pgfplots}
\usepackage{subcaption}
\usepackage{import}
\usepackage{setspace,caption}
\usepackage[font=footnotesize,labelfont=bf,justification=raggedright,singlelinecheck=false]{caption}
\usepackage{algpseudocode}
\usepackage{algorithm} 
\usepackage{mathrsfs}
\usepackage{adjustbox}
\usepackage{lscape}

\usepackage{tikz}
\usetikzlibrary{shapes,arrows,positioning}
\usetikzlibrary{calc} 
\usetikzlibrary{fit}
\usetikzlibrary{intersections}
\usetikzlibrary{positioning,shapes,arrows}
\tikzset{tips=proper}

\def\layersep{2.5cm}

\usepackage[colorinlistoftodos]{todonotes}

\def\cdf(#1)(#2)(#3){0.5*(1+(erf((#1-#2)/(#3*sqrt(2)))))}%

\definecolor{color2}{HTML}{1D91C0}
\definecolor{color1}{HTML}{C7E9B4}
\definecolor{color3}{HTML}{225EA8}

\title{Forecasting stock return distributions around the globe with quantile neural networks\thanks{We are grateful to the editor, Dick van Dijk, associate editor, and two anonymous reviewers for their useful comments and suggestions, which greatly improved the paper. We appreciate helpful comments from participants at various seminars, workshops and conferences. The support from the Czech Science Foundation under the 24-11555S project is gratefully acknowledged.}}

\author{%
Jozef {\sc Barun\'{i}k}$^{\rm a,b,}$\thanks{Corresponding author, Tel. +420 (776) 259 273, Email address: barunik@fsv.cuni.cz} and
Martin {\sc Hronec}$^{\rm a,b}$ and 
Ond\v{r}ej {\sc Tobek}$^{\rm c,}$\thanks{The views expressed in this work are those of the author and are in no way related to UBS AG.}
\vspace{5mm} \\
 \small $^{\rm a}$ Charles University, Institute of Economic Studies\vspace{-0.5mm}\\ 
 \small $^{\rm b}$ The Czech Academy of Sciences, Institute of Information Theory and Automation \\
\small $^{\rm c}$ UBS, Quant Hub, Zurich. }

\date{\today}

\begin{document}

\maketitle
 
\begin{abstract}
\noindent We propose a novel machine learning approach for forecasting the distribution of stock returns using a rich set of firm-level and market predictors. Our method combines a two-stage quantile neural network with spline interpolation to construct smooth, flexible cumulative distribution functions without relying on restrictive parametric assumptions. This allows accurate modelling of non-Gaussian features such as fat tails and asymmetries.  Furthermore, we show how to derive other statistics from the forecasted return distribution such as mean, variance, skewness, and kurtosis. The derived mean and variance forecasts offer significantly improved out-of-sample performance compared to standard models. We demonstrate the robustness of the method in US and international markets.\\

\end{abstract}
\maketitle

\newpage

\section{Introduction} \label{sec:Introduction}

Predicting the distribution of stock returns has become increasingly important because modern financial decision-making requires more than just point estimates. Traditional approaches that focus on predicting the mean or variance of returns impose restrictive assumptions, such as multivariate normality or quadratic utility, that rarely hold in practice. These methods fail to capture the complexity of investor preferences, which depend on higher order moments and distributional properties such as asymmetry, tail behaviour and discontinuities \citep{gul1991theory,kimball1993standard,manski1988ordinal,kai1979prospect}, as well as features of the data such as non-Gaussianity and heavy tails. To address these limitations, we develop a machine learning-based framework for forecasting the distribution of stock returns using a rich set of firm-level and market variables.
 
Specifically, this paper presents a novel approach that uses a two-stage quantile neural network combined with spline interpolation to generate predictive distributions. Our method flexibly models non-Gaussian, heavy-tailed return distributions without relying on parametric assumptions. It captures non-linear interactions among a large number of predictors, allows estimation of key distributional moments, and provides accurate forecasts in high-dimensional settings with low signal-to-noise ratios. Our approach performs well in both US and international markets and delivers significant improvements in out-of-sample loss metrics compared to existing parametric and machine learning benchmarks in the prediction of next month returns.

The ability to forecast next month returns distributions rather than isolated moments has a number of key benefits. It supports decision frameworks that require knowledge of tail risk or asymmetry - such as those used in portfolio optimisation, risk management and asset pricing. It improves the estimation of standard moments (mean, variance, skewness, kurtosis) through numerical integration, which we show improves predictive performance over direct moment forecasting. Finally, it also enables new empirical investigations into the pricing of distributional characteristics across assets.

We contribute to the literature in two main ways. First, we propose a new method for distribution forecasting that combines quantile neural networks with cubic B-spline interpolation to construct smooth and flexible cumulative distribution functions. Unlike existing approaches, our method is non-parametric, data-driven and generalises well across the globe.

Second, we show how distributional forecasts can be used to improve asset pricing applications. Specifically, we examine the pricing of predicted quantiles and moments in the cross-section of stock returns. We show that the mean return derived from the interpolated distribution significantly outperforms direct mean forecasts in out-of-sample portfolio returns. Moreover, we find that only the central quantiles are systematically priced, with the profitability of long-short portfolios declining as one moves towards the tails. Contrary to some previous findings, we find no consistent pricing of volatility, skewness or kurtosis using our distribution-derived estimates. These results provide new insights into the role of higher-order moments in asset pricing and help reconcile previous mixed evidence in the literature \citep{bollerslev2020good,ang2006cross,bali2016empirical}.

Our work builds on and extends a growing body of research applying machine learning to asset pricing and return forecasting. Recent studies have demonstrated the superiority of machine learning methods in capturing complex non-linear patterns and improving out-of-sample prediction accuracy (e.g. \cite{gu2020empirical,gu2021autoencoder,bryzgalova2020forest,tobek2021does,dong2022anomalies}). However, most of these studies focus on predicting mean returns or linear factor models. In contrast, our approach targets the entire distribution and goes beyond previous quantile prediction methods by interpolating across a dense grid of quantiles to construct smooth distributions.

We also complement recent contributions to distributional forecasting in asset pricing. For example, \cite{liu2023quantile} demonstrates the usefulness of quantile regression for forecasting returns and constructing quantile-based risk premia. Our approach differs in that we interpolate a richer set of quantile forecasts onto a denser representation of the distribution and derive distributional moments via numerical integration rather than relying on pre-specified interquantile ranges. We also show how to adjust the predicted higher moments to account for the lack of extreme observations in the tails of the distribution. Furthermore, our method generalises to international markets and is validated using models trained only on US data.

Another related strand is the work of \cite{yang2024asset}, who use a conditional quantile variational autoencoder (CQVAE) to approximate future return distributions. While both approaches use machine learning to generate distribution forecasts, our method uses observable firm-level predictors and explicit quantile targets, in contrast to the latent factor representations used in CQVAE. In addition, we provide a detailed decomposition of distributional moments and their pricing implications for both US and global equity markets. In contrast, \cite{yang2024asset} constructs quantile forecasts with a primary focus on mean estimation.

Our contribution is also relevant to the literature on distributional properties in asset pricing \citep{hou2020replicating,bessembinder2018stocks,kim2004more,an2014joint,huang2019option,alexiou2022option}. Existing evidence on the pricing of volatility, skewness and kurtosis is mixed and often relies on ex-post sample estimates or option-implied measures, which have limited availability. Our model provides an alternative by using out-of-sample distribution forecasts based on observable predictors to study the pricing of distributional characteristics. We show that only the mean and certain central quantiles carry significant pricing information, while higher moments do not appear to be systematically priced.

In summary, this paper presents a novel machine learning-based method for forecasting stock return distributions and provides empirical evidence on its predictive accuracy and relevance for asset pricing. Our results underscore the value of distribution forecasting in financial applications and suggest that incorporating richer distributional information can improve return forecasting and portfolio performance.

The rest of the paper is organised as follows: \autoref{subsubsec:TwoStage} presents the distributional forecasting methodology, including quantile neural networks and spline interpolation. \autoref{sec:Results} describes the empirical setup and data, reports results on out-of-sample forecasting performance and asset pricing applications. Section 4 concludes.


\section{Two-stage quantile neural networks} \label{subsubsec:TwoStage}



Let the returns $r_{i,t}$ collected over $t = 1, \dots, T$ months and $i = 1, \dots, N_t$ individual stocks be random variables with a cumulative distribution function (CDF)\footnote{In the case of the forecast distribution of stock returns, $Q_{r_{i,t}}(0) = -1$ and $Q_{r_{i,t}}(1) = \infty$.} $F(r_{i,t})$. Let $\tau_1, \dots, \tau_K \in (0,1)$ denote the quantile levels of interest, and define the set $\mathcal{T} = \{\tau_1, \ldots, \tau_K\}$. We are interested in forecasting the corresponding conditional quantiles 
$Q_{r_{i,t+1}}^{\mathcal{T}} = \big\{Q_{r_{i,t+1}}(\tau_1|\mathcal{F}_t), \ldots, \\ Q_{r_{i,t+1}}(\tau_K|\mathcal{F}_t)\big\}$, where each quantile is defined as 
\[
Q_{r_{i,t}}(\tau \mid \mathcal{F}_t) = \inf\left\{r \in \mathbb{R} : F(r) \geq \tau \mid \mathcal{F}_t \right\},
\]
that is, the $\tau$-quantile of the return distribution at time $t+1$ conditional on the information set $\mathcal{F}_t$ available at time $t$. Forecasting multiple $\tau$-quantiles, where $\tau \in (0,1)$ is possible with a multi-output feed-forward neural network model based on aggregating individual $\tau$-quantile losses into a single loss function (as described later in this section). In addition to being computationally more efficient than training individual models for each $\tau$, joint training for all $\tau$ quantiles allows the model to capture the relationship across the entire distribution. As a side effect, even without introducing penalisation for quantile crossing\footnote{Quantile crossing refers to the situation where $Q(\tau_l) > Q(\tau_h)$ when $\tau_l < \tau_h$, violating the quantile monotonicity condition.}, the model naturally captures the monotonicity requirement of quantile predictions since there are no quantiles that are crossed in our quantile forecasts.

\subsection{Network architecture} \label{subsubsubsec:NetworkArchitecture}

While neural networks have been successfully used in stock return forecasting \citep{gu2020empirical}, quantile forecasting is challenging because we need to capture both cross-sectional variation across stocks and market-wide variation over time. 
One way to address this challenge is to use cross-sectional standardisation of returns to remove the time series noise. This is possible from an asset pricing perspective, where the goal is to rank individual stocks cross-sectionally to create long-short portfolios. However, since our goal is to forecast quantiles of future raw (non-standardised) stock returns, we propose a two-stage neural network architecture that exploits the predictability of standardised returns in the first stage and generates quantile forecasts of true (non-standardised) stock returns in the second stage. In this way, we allow the neural networks to exploit the information in standardised returns and to adjust for expected market volatility.

More specifically, a set of quantiles of stock returns are modelled using a large set of variables, including stock-level characteristics (e.g., size, value, or momentum) stored in a vector $x_{i,t}$, as well as market variables (e.g., market volatility) stored in a vector $z_t$, both of which proxy the information available at period $t$. We use both raw and standardised returns to address the discussion above, and we standardise the return of a stock by dividing the raw stock returns by the cross-sectional average of the stock-level volatilities at period $t$, such that 
\begin{equation} \label{eq:standardisation}
    \widetilde{r}_{i,t+1} = \frac{r_{i,t+1}}{\overline{\sigma}_{t}},
\end{equation}
where $\overline{\sigma}_{t} =\frac{1}{N_{t}}\sum_{i=1}^{N_{t}}\sigma_{i,{t}}$ is a cross-sectional average of stock-level volatilities, $\sigma_{i,t}$ is a stock-level volatility estimate of stock $i$ at period $t$, and $N_{t}$ is the number of stocks in the panel at period $t$. Ideally, we would use $\overline{\sigma}_{t+1}$ instead of $\overline{\sigma}_{t}$ to rescale the standardised quantile forecasts back to the raw quantile forecasts of stock returns, but this is not known at period $t$ when the forecast is made. Therefore, we use the cross-sectional average of stock-level volatilities at period $t$ instead of the future value of $\overline{\sigma}_{t+1}$ to avoid look-ahead bias. $\sigma_{i,t}$ is estimated using an exponentially weighted moving average of the squared returns of stock $i$ with a decay factor of $0.94$ and is one of the input variables to the model.\footnote{The 0.94 parameter is widely used in both academia and the financial industry, an example being the RiskMetrics GARCH specification.} See Appendix~\ref{sec:AppendixTwoStageMotivation} for a visualisation of $\overline{\sigma}$ over time in the US and more detailed arguments for the two-stage setup.

The idea behind standardising individual stock returns using the cross-sectional average of stock-level volatilities rather than the stock volatility itself is to reduce the noise from individual stock volatility estimates. For the same noise-reducing reason, we standardise stock returns only by dividing by the average volatility
and not by subtracting the average return, either stock or market, because average returns are not persistent.

Let $\mathcal{T} = \{\tau_1, \ldots, \tau_K\}$ be a set of quantile levels with $\tau_k \in (0,1)$. A stock return quantile is then modelled by the two-stage quantile neural network with two sub-networks as
\begin{eqnarray}
\nonumber Q_{r_{i,t+1}}^{\mathcal{T}} &=& \Big\{Q_{r_{i,t+1}}(\tau_1|x_{i,t},z_t,\overline{\sigma}_t),\ldots,Q_{r_{i,t+1}}(\tau_K|x_{i,t},z_t,\overline{\sigma}_t)\Big\} \\
\nonumber &=& \Bigg[  \Big( \underbrace{g^{(L)}_{W_{S_1}^{(L)},b_{S_1}^{(L)}} \circ \ldots \circ g^{(1)}_{W_{S_1}^{(1)},b_{S_1}^{(1)}} \left(x_t\right) }_{\text{standardised $\tau$ quantiles sub-network}}\Big) \times \overline{\sigma}_t \Bigg] \times \Big( \underbrace{g^{(2)}_{W_{S_2}^{(2)},b_{S_2}^{(2)}} \circ g^{(1)}_{W_{S_2}^{(1)},b_{S_2}^{(1)}} \left(z_t\right)}_{\text{market-wide volatility sub-network}}\Big) \\
&=& \underbrace{\Bigg[ \Big\{ \underbrace{Q_{\widetilde{r}_{i,t+1}}(\tau_1|x_{i,t}),\ldots,Q_{\widetilde{r}_{i,t+1}}(\tau_K|x_{i,t})}_{\text{Stage I}}\Big\}\times \overline{\sigma}_t\Bigg] \times \widehat{\sigma}_t^M}_{\text{Stage II}},
\end{eqnarray}
where $W_{S}=\left(W_S^{(1)},\ldots,W_S^{(L)}\right)$ and $b_S=\left(b_S^{(1)},\ldots,b_S^{(L)}\right)$ are weight matrices and bias vectors from a stage $S$. Each weight matrix $W_S^{(\ell)} \in \mathbb{R}^{m\times n}$ contains $m$ neurons as $n$ column vectors $W_S^{(\ell)} = [w_{\cdot,1}^{(\ell)},\ldots,w_{\cdot, n}^{(\ell)}]$, and $b_S^{(\ell)}$ are thresholds or activation levels that contribute to the output of a hidden layer that allows the function to be shifted. 

It is important to note that, in contrast to the literature, we consider a multi-output (deep) neural network to characterise the collection of quantiles. Our network has two sub-networks, namely a standardised $\tau$ quantile network and a market-wide volatility network\footnote{Note that the market-wide volatility sub-network is forecasting optimal multiplicative adjustment factor to $\overline{\sigma}_{t}$ to produce quantile predictions for non-standardised stock returns. It does not forecast market volatility directly.} in which $l \in {1,...,L}$ hidden layers transform the input data into a chain using a collection of nonlinear activation functions $g_S^{(1)},\ldots,g_S^{(L)}$. An activation function, $g^{(\ell)}_{W_S^{(\ell)},b_S^{(\ell)}}$, is used as a\begin{equation*}
g^{(\ell)}_{W_S^{(\ell)},b_S^{(\ell)}}(u) := g_{\ell}\left( W_S^{(\ell)}u + b_S^{(\ell)} \right) = g_{\ell}\left( \sum_{i=1}^{m} W^{(\ell)}_{i,S} u + b_{i,S}^{(\ell)} \right)
\label{eq: ffneuron}
\end{equation*}
is rectified linear units $g_{\ell}(u) = \max\{u,0\}$, or $g_{\ell,S}(u) = \tanh(u)$.

\begin{figure}

\begin{tikzpicture}[shorten >=1pt,draw=black!50, node distance=\layersep]
    \tikzstyle{every pin edge}=[<-,shorten <=1pt]
    \tikzstyle{neuron}=[circle, draw=none,
    fill=black!15, minimum size=17pt, inner sep=0pt]
    \tikzstyle{neuronin}=[circle, draw=none,
    fill=color1, minimum size=17pt, inner sep=0pt]
    \tikzstyle{neuronout}=[circle, draw=none,
    fill=color2, minimum size=17pt, inner sep=0pt]
    \tikzstyle{neuron1}=[circle, draw,minimum size=17pt, inner sep=0pt]
    \tikzstyle{annot} = [text width=4em, text centered]

    \path[yshift=0.5cm]
        node[neuronin] (S-1) at (7.5cm, -1cm) {};

    \node[annot,above of=S-1, node distance=0.5cm] (A-S-1) {\footnotesize{$\overline{\sigma}_{t}$}};


    \path[yshift=0.5cm]
        node[neuronin] (N1-L1-1) at (0cm, -1cm) {};
    \path[yshift=0.5cm]
        node (N1-L1-2) at (0cm,-2cm) {\vdots};
    \path[yshift=0.5cm]
        node[neuronin] (N1-L1-3)at (0cm, -3cm) {};
    \path[yshift=0.5cm]
        node (N1-L1-4) at (0cm,-4cm) {\vdots};
    \path[yshift=0.5cm]
        node[neuronin] (N1-L1-5) at (0cm, -5cm) {};

    \path[yshift=0.5cm]
        node[neuron] (N1-L2-1) at (1cm, -1cm) {};
    \path[yshift=0.5cm]
        node (N1-L2-2) at (1cm,-2cm) {\vdots};
    \path[yshift=0.5cm]
        node[neuron] (N1-L2-3)at (1cm, -3cm) {};
    \path[yshift=0.5cm]
        node (N1-L2-4) at (1cm,-4cm) {\vdots};
    \path[yshift=0.5cm]
        node[neuron] (N1-L2-5) at (1cm, -5cm) {};

    \path[yshift=0.5cm]
        node[neuron] (N1-L3-1) at (2cm, -1cm) {};
    \path[yshift=0.5cm]
        node (N1-L3-2) at (2cm,-2cm) {\vdots};
    \path[yshift=0.5cm]
        node[neuron] (N1-L3-3)at (2cm, -3cm) {};
    \path[yshift=0.5cm]
        node (N1-L3-4) at (2cm,-4cm) {\vdots};
    \path[yshift=0.5cm]
        node[neuron] (N1-L3-5) at (2cm, -5cm) {};
    
    \path[yshift=0.5cm]
        node[neuron] (N1-L4-1) at (3cm, -2cm) {};
    \path[yshift=0.5cm]
        node (N1-L4-2) at (3cm, -3cm) {\vdots};
    \path[yshift=0.5cm]
        node[neuron] (N1-L4-3) at (3cm,-4cm) {};

    \path[yshift=0.5cm]
        node[neuron] (N1-L5-1) at (4cm, -1cm) {};
    \path[yshift=0.5cm]
        node (N1-L5-2) at (4cm,-2cm) {\vdots};
    \path[yshift=0.5cm]
        node[neuron] (N1-L5-3)at (4cm, -3cm) {};
    \path[yshift=0.5cm]
        node (N1-L5-4) at (4cm,-4cm) {\vdots};
    \path[yshift=0.5cm]
        node[neuron] (N1-L5-5) at (4cm, -5cm) {};

    \path[yshift=0.5cm]
        node[neuron] (N1-L6-1) at (5cm, -1cm) {};
    \path[yshift=0.5cm]
        node (N1-L6-2) at (5cm,-2cm) {\vdots};
    \path[yshift=0.5cm]
        node[neuron] (N1-L6-3)at (5cm, -3cm) {};
    \path[yshift=0.5cm]
        node (N1-L6-4) at (5cm,-4cm) {\vdots};
    \path[yshift=0.5cm]
        node[neuron] (N1-L6-5) at (5cm, -5cm) {};

    \path[yshift=0.5cm]
        node[neuronout] (NQ-1) at (6cm, -1cm) {};
    \path[yshift=0.5cm]
        node (NQ-2) at (6cm,-2cm) {\vdots};
    \path[yshift=0.5cm]
        node[neuronout] (NQ-3)at (6cm, -3cm) {};
    \path[yshift=0.5cm]
        node (NQ-4) at (6cm,-4cm) {\vdots};
    \path[yshift=0.5cm]
        node[neuronout] (NQ-5) at (6cm, -5cm) {};

    \node[annot,above of=NQ-1, node distance=1cm] (A-NQ-1) {\footnotesize{$Q^{\mathcal{T}}_{\widetilde{r}_{i,t}}$ \\ (37$\times$1)}};

    \path[yshift=0.5cm]
        node[neuron1] (NSQ-1) at (9cm, -2cm) {$\times$};
    \path[yshift=0.5cm]
        node (NSQ-2) at (9cm,-3cm) {\vdots};
    \path[yshift=0.5cm]
        node[neuron1] (NSQ-3)at (9cm, -4cm) {$\times$};
    \path[yshift=0.5cm]
        node (NSQ-4) at (9cm,-5cm) {\vdots};
    \path[yshift=0.5cm]
        node[neuron1] (NSQ-5) at (9cm, -6cm) {$\times$};

    \path[yshift=0.5cm]
        node (A-N1-H1) at (1.5cm,0cm) {\footnotesize{(128$\times$2)}};
    \path[yshift=0.5cm]
        node (A-N1-H2) at (4.5cm,0cm) {\footnotesize{(128$\times$2)}};
    \node[annot] (A-N1-L4) at (3cm,0.5cm) {\footnotesize{Bottleneck \\ (4$\times$1) }};
    \node[annot,above of=N1-L1-1, node distance=1cm] (A-N1-L1-1) {\footnotesize{$x_t$} \\ \footnotesize{(176$\times$1)}};

    \path[yshift=0.5cm]
        node[neuronin] (N2-L1-1) at (0cm, -7cm) {};
    \path[yshift=0.5cm]
        node (N2-L1-2) at (0cm,-8cm) {\vdots};
    \path[yshift=0.5cm]
        node[neuronin] (N2-L1-3)at (0cm, -9cm) {};
    \path[yshift=0.5cm]
        node (N2-L1-4) at (0cm,-10cm) {\vdots};
    \path[yshift=0.5cm]
        node[neuronin] (N2-L1-5) at (0cm, -11cm) {};

    \path[yshift=0.5cm]
        node[neuron] (N2-L2-1) at (1cm, -7cm) {};
    \path[yshift=0.5cm]
        node (N2-L2-2) at (1cm,-8cm) {\vdots};
    \path[yshift=0.5cm]
        node[neuron] (N2-L2-3)at (1cm, -9cm) {};
    \path[yshift=0.5cm]
        node (N2-L2-4) at (1cm,-10cm) {\vdots};
    \path[yshift=0.5cm]
        node[neuron] (N2-L2-5) at (1cm, -11cm) {};

    \path[yshift=0.5cm]
        node[neuron] (N2-L3-1) at (2cm, -9cm) {};

    \path[yshift=0.5cm]
        node[neuron1] (MULT-1) at (11cm, -7cm) {$\times$};
    \path[yshift=0.5cm]
        node (MULT-2) at (11cm,-8cm) {\vdots};
    \path[yshift=0.5cm]
        node[neuron1] (MULT-3)at (11cm, -9cm) {$\times$};
    \path[yshift=0.5cm]
        node (MULT-4) at (11cm,-10cm) {\vdots};
    \path[yshift=0.5cm]
        node[neuron1] (MULT-5) at (11cm, -11cm) {$\times$};

    \path[yshift=0.5cm]
        node[neuronout] (MULT1-1) at (13cm, -7cm) {};
    \path[yshift=0.5cm]
        node (MULT-2) at (13cm,-8cm) {\vdots};
    \path[yshift=0.5cm]
        node[neuronout] (MULT1-3)at (13cm, -9cm) {};
    \path[yshift=0.5cm]
        node (MULT-4) at (13cm,-10cm) {\vdots};
    \path[yshift=0.5cm]
        node[neuronout] (MULT1-5) at (13cm, -11cm) {};

    \node[annot,below of=N2-L1-5, node distance=1cm] (A-N2-L1-5) {\footnotesize{$z_t$} \\ \footnotesize{(18$\times$1)}};
    \node[annot,below of=N2-L2-5, node distance=1cm] (A-N2-L2-5) {\footnotesize{(8$\times$1)}};
    \node[annot,below of=N2-L3-1, node distance=3cm] (A-N2-L3-1) {\footnotesize{\footnotesize{$\widehat{\sigma}_t^M$} \\ (1$\times$1)}};
    \node[annot,below of=MULT1-5, node distance=1cm] () {\footnotesize{$Q^{\mathcal{T}}_{r_{i,t}}$} \\ \footnotesize{(37$\times$1)}};

    \foreach \i in {1,3,5}
        \foreach \j in {1,3,5}
            \path (N1-L1-\i) edge (N1-L2-\j);

    \foreach \i in {1,3,5}
        \foreach \j in {1,3,5}
            \path (N1-L2-\i) edge (N1-L3-\j);

    \foreach \i in {1,3,5}
        \foreach \j in {1,3}
            \path (N1-L3-\i) edge (N1-L4-\j);

    \foreach \i in {1,3}
        \foreach \j in {1,3,5}
            \path (N1-L4-\i) edge (N1-L5-\j);

    \foreach \i in {1,3,5}
        \foreach \j in {1,3,5}
            \path (N1-L5-\i) edge (N1-L6-\j);

    \foreach \i in {1,3,5}
        \foreach \j in {1,3,5}
            \path (N1-L6-\i) edge (NQ-\j);

    \foreach \i in {1,3,5}
        \path (S-1) edge (NSQ-\i);

    \foreach \i in {1,3,5}
        \path (NQ-\i) edge (NSQ-\i);

    \foreach \i in {1,3,5}
        \foreach \j in {1,3,5}
            \path (N2-L1-\i) edge (N2-L2-\j);
            
    \foreach \i in {1,3,5}
        \foreach \j in {1,3,5}
            \path (N2-L2-\i) edge (N2-L3-1);

    \foreach \i in {1,3,5}
        \foreach \j in {1,3,5}
            \path (NSQ-\i) edge (MULT-\i);

    \foreach \i in {1,3,5}
        \foreach \j in {1,3,5}
            \path (MULT-\i) edge (MULT1-\i);

    \foreach \i in {1,3,5}
        \path (MULT-\i) edge (N2-L3-1);

    \draw [decorate,decoration={brace,amplitude=10pt,raise=2ex, mirror}]
    (N1-L1-5.south) -- (NQ-5.south) node[midway,below=4ex]{Stage I};

    \draw [decorate,decoration={brace,amplitude=10pt,mirror,raise=8ex}]
    (N2-L1-5.south) -- (MULT1-5.south) node[midway,below=10.5ex]{Stage II};

    \node[draw, dashed, fit=(N1-L1-1) (NQ-5), inner sep=1ex, label={[align=center]left:{\footnotesize{Standardized} \\ \footnotesize{$\tau$-quantiles} \\ \footnotesize{sub-network}}}] {};
    \node[draw, dash pattern=on 4pt off 2pt on 1pt off 2pt, rectangle, fit=(N2-L1-1) (N2-L2-5) (N2-L3-1), inner sep=1ex, label={[align=center]left:{\footnotesize{Market-wide} \\ \footnotesize{volatility} \\ \footnotesize{sub-network}}}] {};

\end{tikzpicture}
    \caption{\textbf{Two-stage neural network architecture.}
    Stage I generates cross-sectionally standardised $\tau$-quantile forecasts, $Q^{\mathcal{T}}_{\widetilde{r}}$, as output.
    Cross-sectional standardisation is performed by dividing individual future stock returns $r_{i,t+1}$ (labels) by the 
    cross-sectional average of the individual stock standard deviation estimates, $\overline{\sigma}_{t}$.
    The standard deviation estimates for the individual stocks are an exponential weighted moving average of their daily squared returns with a smoothing factor of $0.94$.
    Step II uses $\overline{\sigma}_{t}$, i.e. based on data from the previous period,
    to rescale the standardised $\tau$-quantile forecasts generated in Stage I back to the original scale.
    The rescaling takes place at the multiplication nodes marked with $\times$, representing the multiplication operation between the inputs of the node.
    The market-wide volatility sub-network then generates a forward-looking scaling factor to $\overline{\sigma}_{t}$ in order to further refine the $\tau$-quantile forecasts produced as the output of stage II.
    The forward pass generates a tensor of form $(B,37,2)$, where $B$ is the number of observations in the batch,
    37 is the number of predicted quantiles, and 2 corresponds to the standardised and raw versions.
    } \label{fig:TwoStage}
\end{figure}

Figure \ref{fig:TwoStage} details the architecture of the two-stage quantile neural network model defined above. This is our main model used for quantile forecasting, and we refer to it as the \textit{two-stage model} throughout the paper. The first stage of the model is used to forecast quantiles of cross-sectionally standardised next month stock returns $Q^{\mathcal{T}}_{\widetilde{r}_{i,t}}$.
In our empirical application, we use 37 $\tau$s ranging from the extreme left tail of the distribution (0.00005) to the extreme right tail (0.99995).

Once the set of quantile levels $\mathcal{T}$ is specified, the two-stage architecture proceeds to generate forecasts for each quantile through a sequence of rescaling operations. The second stage of the model generates quantile forecasts of the next month's raw stock returns $Q_{r_{i,t}}(\tau)$ for $\tau \in \mathcal{T}$. This stage consists of three steps. First, the standardised quantile forecasts \( Q^{\mathcal{T}}_{\widetilde{r}_{i, t}} \) from the first stage and the cross-sectional average of stock-level volatilities \( \overline{\sigma}_t \) are used as inputs to rescale the standardised quantile forecasts back to the raw stock return scale \( Q_{r_{i,t}}(\tau) = Q_{\widetilde{r}_{i,t}}(\tau) \cdot \overline{\sigma}_t \). Second, the market-wide volatility sub-network generates a second multiplicative scaling factor to reintroduce a forward-looking aspect to the market-wide volatility scalar $\overline{\sigma}_t$. This market-wide volatility scaling factor is the same for all stocks in a given period \( t \) and is used to refine the quantile forecasts rescaled in the first step. Finally, both the standardised and raw quantile forecasts are clipped at -100\% during the forward pass of the neural network.

In addition to the use of two stages, another key difference from standard feedforward neural network architecture is the inclusion of a bottleneck layer. Inspired by the autoencoder architecture, the bottleneck layer aims to capture the distribution hyperparameters. This is achieved by forcing the network to learn a compressed representation, as the bottleneck layer is limited to only four nodes.

The two-stage neural network takes three sets of inputs. The first set consists of 176 features representing stock-level characteristics based on 153 anomalies used in \cite{tobek2021does}, 18 stock-level volatility estimates, and five cross-sectional averages of stock-level mean estimates. The second set of inputs consists of cross-sectional averages of the stock-level volatility characteristics used in the first set, i.e. 18 stock-level volatility estimates. The third input is a cross-sectional average of stock-level volatilities \( \overline{\sigma}_t \) used for rescaling purposes. A detailed description of all features can be found in \autoref{subsubsec:Features}.

\subsection{Training with quantile loss function} \label{subsubsec:Training}

 In order to jointly predict multiple $\tau$-quantiles using a two-stage neural network, individual $\tau$-quantile losses must be aggregated into a single loss. Our model outputs both standardised and raw stock return quantiles, and we combine the respective losses using a simple average. Let $\mathcal{T} = \{\tau_1, \ldots, \tau_K\}$ denote a set of $K$ quantile levels, where each $\tau_k \in (0,1)$. 

\begin{equation} \label{eq:QLaggrTwoStage}
\begin{aligned}
    \mathcal{L}^{\mathcal{T}} = \frac{1}{B}\frac{1}{K}\sum_{\tau \in \mathcal{T}}\sum_{i=1}^{B} \left( \rho_{\tau}\left(r_{i,t} - \widehat{Q}_{r_{i,t}}(\tau)\right) + \rho_{\tau}\left(\widetilde{r}_{i,t} - \widehat{Q}_{\widetilde{r}_{i,t}}(\tau)\right)\right)
\end{aligned}
\end{equation}

Here, $B$ is the batch size, $K$ the number of quantiles in $\mathcal{T}$, and $\rho_{\tau}(\cdot)$ is the quantile loss function:
\begin{equation}
\rho_{\tau}(\xi) = \mathds{1}_{(\xi \geq 0)}\tau \xi + \mathds{1}_{(\xi < 0)}(\tau - 1) \xi
\end{equation} \label{eq:qlSimple}

This piecewise linear function penalises underprediction and overprediction asymmetrically based on the quantile level $\tau$, with equal weighting across stocks and quantiles in the aggregate loss.\footnote{The linear combination of losses from the two output heads uses equal weights, though in practice, the standardised return component dominates due to differences in scale. This design choice aligns with our modelling rationale, where the first stage captures most of the learning and the second stage refines the result.} Although $\rho_{\tau}(\cdot)$ is not differentiable at zero, modern autodiff frameworks such as PyTorch handle subgradients at non-differentiable points automatically.

To train the model, we minimise the loss function using stochastic gradient descent with mini-batches of size $B = 8192$ observations. Specifically, we use the Adam optimiser (\cite{kingma2014adam}) with an initial learning rate of $0.0003$ and $(\beta_1, \beta_2) = (0.9, 0.999)$. We adjust the number of epochs based on the number of observations available at each training period. The number of epochs is set to $100*(A/n)$, where $n$ is the number of observations available for training and $A$ is the adjustment constant equal to $3 000 000$ for the full sample and $1 500 000$ for the liquid sample, derived from the average number of observations in each sample over time.\footnote{See \autoref{subsec:data} for the description of the data used.} We use early stopping with a patience of 2 epochs, where 20\% of the training data is used as validation to determine the stopping point. After early stopping, the model state with the lowest validation loss is taken as the final model. We use Leaky Relu (\cite{maas2013rectifier}) as the activation function in all layers except the 
last layer in each sub-network. We use batch normalisation (\cite{ioffe2015batch}) in all layers except the last layer in both the standardised $\tau$-quantiles sub-network and the market-wide volatility sub-network. We apply dropout to all layers except the last layer in each sub-network and use a rate of $0.2$ as a result of the hyperparameter search described in \autoref{sec:appendix_hyperparameter}. We also apply an L1 penalty to the weights of the first stage in each subnet. The L1 penalty for the first layer is $0.0001$, and for the second layer, it is $0.00001$. The first layer of the market-wide volatility network is also regularised by an L2 penalty of $0.00001$. As an additional form of regularisation, we use an ensemble of 20 networks. 

The model is trained from scratch for the first training period and then fine-tuned for subsequent training periods, meaning that weights from the previous training period are used to initialise the network instead of reinitialising randomly. Fine-tuning greatly reduces the computational burden of having to re-estimate the model each period and is a natural choice given the large overlap of subsequent training samples with the widening estimation window approach. It is expected that one year of new data should not change the estimated parameters too much unless there is a novel high-information event such as the dot-com bubble or the global financial crisis, in which case fine-tuning requires more epochs to retrain.

\subsection{Approximating probability density and its moments from quantiles using B-Splines} \label{subsec:DerivingDistribution}

\begin{figure}[H]
\centering
\includegraphics[width=0.6\textwidth]{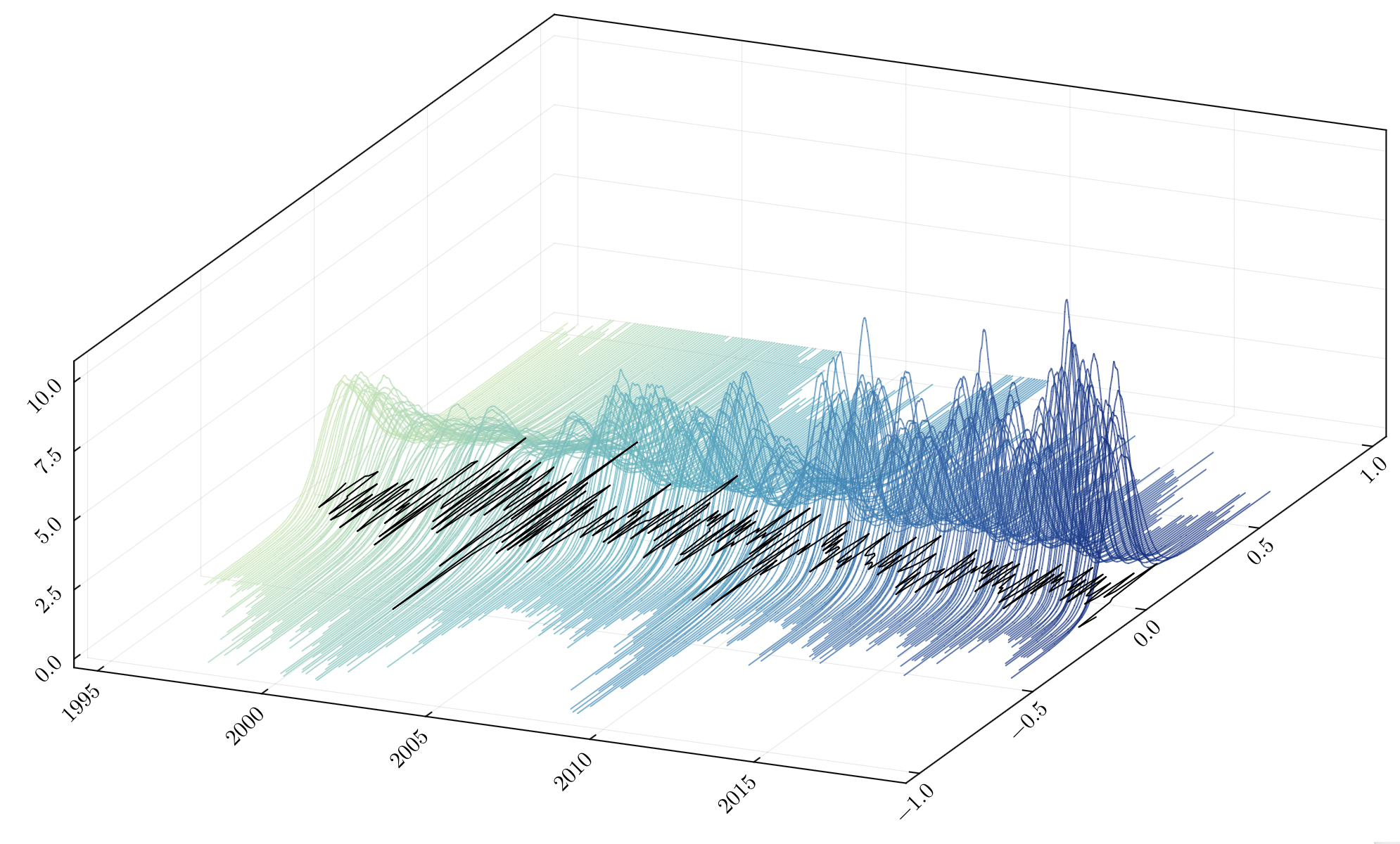}
\caption{\textbf{Example of forecasted probability density function over time.} 
This figure displays the forecasted probability density function 
for out-of-sample 22 days-ahead returns of an Microsoft stock (CUSIP 59491810) generated using our two-stage quantile neural network model.}
\label{fig:figureApple3d}
\end{figure}

To approximate the probability density, let's consider a grid of $\tau$ quantile forecasts $\{ Q_j(\tau),\tau_j \}_{j=1}^K$, where $\tau_j$ is the probability and $Q_j(\tau)$ is the corresponding quantile value, containing $K$ grid points that can be used to obtain the cumulative distribution function $F:\mathbb{R} \rightarrow [0,1]$. Since stock returns range from $-1$ to $\infty$, from a practical perspective we are interested in $F:[-1,\infty) \rightarrow [0,1]$. To get a denser representation, we use cubic B-splines\footnote{We use the splrep and BSpline functions from the Python scipy library, \cite{2020SciPy-NMeth}.} to interpolate between adjacent quantiles by constructing piecewise polynomial functions within each interval $[Q_j, Q_{j+1}]$. Instead of interpolating the predicted quantiles directly, we create a new denser $\mathcal{T}^d$ grid with one hundred equally spaced points between every two original quantile prediction points, except the largest and smallest,
i.e. the denser grid has $(K - 3) \times 100$ values, and fit the cubic B-spline to approximate the cumulative distribution function. To derive the probability density function, the derivative of the spline for the base function is calculated at each interval to obtain the new spline function representing the probability density function.
\begin{figure}[H] 
\centering
\includegraphics[width=0.6\textwidth]{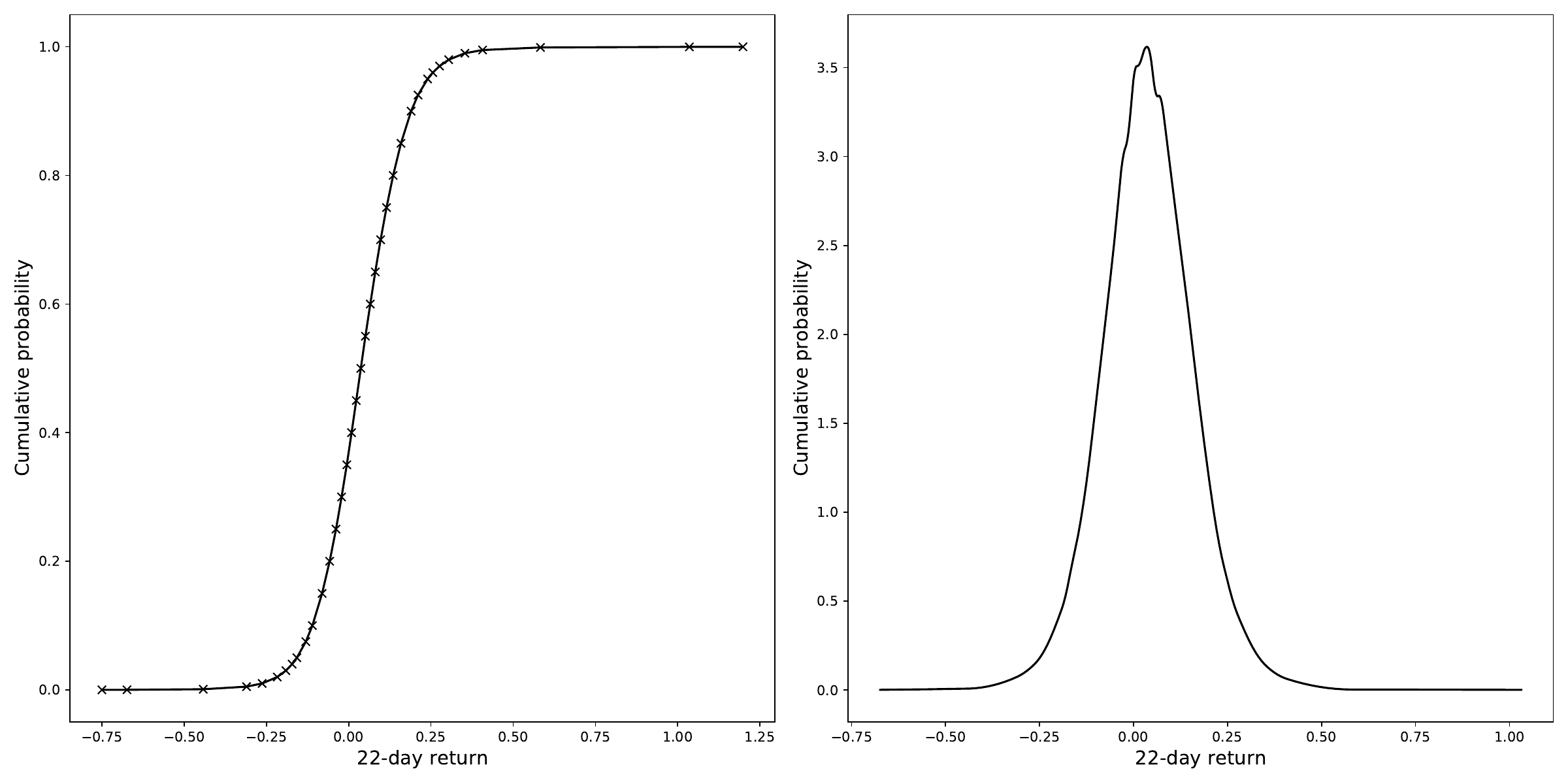}
\caption{\textbf{Interpolated cumulative distribution function and probability density function.} 
This figure shows the cumulative distribution function interpolated from the 
inverted quantile forecasts depicted as marks (left) and derived probability density function (right) for out-of-sample 22 days-ahead returns of Microsoft stock (CUSIP 59491810) in October 2008. Quantile forecasts used to obtain the cumulative distribution function are generated using the two-stage quantile neural network model.}
\label{fig:MicrosoftExampleBSplines}
\end{figure} 
Figure~\ref{fig:MicrosoftExampleBSplines} shows an example of the interpolated cumulative distribution function
on the left and the implied probability density function on the right. 
The figure is based on $\tau$-quantile forecasts of 22-day (monthly) returns for Microsoft in October 2008.
The density function is relatively close to a Student's t-distribution
with noticeably large tails during the depths of the Great Financial Crisis. Figure \ref{fig:figureApple3d} then complements this with the evolution of the forecasted distribution of the 22-day-ahead returns over the period 1996-2020.

As the interpolated density function does not cover the extreme tails of the distribution beyond the smallest and largest predicted quantile,
these tails are modelled by discrete probabilities for values at the ends of the new $\mathcal{T}^d$ grid.
We use discrete probabilities implied by the estimated spline function.
The density function is truncated for returns less than -1, and discrete probability is then considered for $-1$ returns.
Excluding the extreme tails from the density function results in a slight underestimation of the variance and a more pronounced underestimation of the kurtosis of the distribution.
Inferring the distribution in the tails from empirical data is problematic, so we chose this solution to avoid the problem to some extent.
In cases where B-spline interpolation does not produce plausible results, we use a linear B-spline function as a fallback (see Appendix \ref{app:linear_fallback} for details). A complete algorithm for approximating the probability density function from the quantile forecasts is described in detail in Appendix~\ref{app:algorithms}, Algorithm \ref{sec:appendix_density}.

Importantly, the B-spline approximation allows us to obtain moments of the cumulative probability distribution function. We first derive the $z$th non-central moment of a density function $f(x)$
\begin{equation}
m_z = \int_{-1}^{\infty} x^z f(x) \,dx
\end{equation}
using numerical integration methods and the B-spline-based density function approximation obtained above. We use a simple tractable local linear approximation algorithm, which leads to trivial integrals of polynomial functions, and the complete algorithm is detailed in Appendix~\ref{app:algorithms}, Algorithm~\ref{sec:appendix_moments}. Further, using this approximation, we calculate central moments and rescale the moments by the integrated density function to ensure they are properly defined. 

The estimated empirical quantiles from neural networks do not cover the full range of possible stock returns because it is not possible to estimate quantiles in the extreme tails of the distribution. The calculated central moments are, therefore, biased, with the bias being greater for higher moments as the tails of the distribution have a greater impact.
It is, therefore, important to test how well the algorithm captures the true moments. To do this, we compare the true moments of a theoretical distribution with the output of the algorithm and propose an adjustment to reduce the measured bias. The appendix \autoref{subsubsec:MomentsAdjustment} describes how to adjust the moments to account for bias.


\section{Empirical results} \label{sec:Results}

\subsection{Data and features}\label{subsec:data}

Our data cover U.S. and international equity markets for the periods 1963 to 2018 and 1990 to 2018, respectively. We use the merged CRSP/Compustat database from the Wharton Research Data Service (WRDS) for the US sample and LSEG Datastream\footnote{formerly called Reuters or Refinitiv Datastream.} for the international sample. Data are pre-processed using the methodology described in \cite{tobek2021does}.
There are four regions in the sample - US, Europe, Japan and Asia-Pacific, which also includes 23 developed countries.\footnote{US, Austria, Belgium, Denmark, Finland, France, Germany, Greece, Ireland, Italy, Luxembourg, Netherlands, Norway, Portugal, Spain, Sweden, Switzerland, UK, Japan, Australia, New Zealand, Hong Kong and Singapore.}
We report our results for both the \textit{full sample} and the \textit{liquid sample}, which includes only the most liquid stocks. The liquid sample includes stocks that meet the following criteria: a price greater than $1$ (or $0.1$ for Asia Pacific) at the end of the previous month, a market capitalisation within the top 95\%, and a trading volume within the top 95\%,
and a trading volume within the top 95\% of the total dollar trading volume over the previous 12 months in each region. The full sample includes stocks with a price greater than $1$ (\$0.1 for Asia Pacific) at the end of the previous month.
Market capitalisation and volume filters are not applied to the full sample.  Micro-cap stocks in the full sample therefore account for only a small fraction of the capitalisation and 
capitalisation and trading volume of the total market. \autoref{tab:n_stocks} summarizes the minimum, average, and maximum number of stocks per month observed in the cross-section of each region over the sample period, separately for the full and liquid samples. 
\begin{table}[H]
\caption{\textbf{Number of stocks in the cross-section across the regions.}
This table shows the average, minimum and maximum number of stocks in the cross-section in each region for the full and liquid samples.
The period covered is from 1973 to 2018 for the US sample and from 1990 to 2018 for the international sample.
}
\centering
\footnotesize
\begin{tabular}{lcccclcccc}
    \toprule \hline \\ [-2ex] 
        & \multicolumn{4}{c}{Full Sample} & \multicolumn{4}{c}{Liquid Sample}\\
    \cmidrule{2-5} \cmidrule{7-10}
region &   USA &  Europe &  Japan &  Asia Pacific &&   USA &  Europe &  Japan &  Asia Pacific \\
\cmidrule{2-5} \cmidrule{7-10}
min  &  3552 &    5098 &   1919 &          1157 & &  873 &     413 &    534 &           226 \\
mean &  5136 &    5746 &   3225 &          2834 & & 1167 &     690 &    749 &           430 \\
max  &  7384 &    6740 &   3795 &          4553 & & 1734 &    1044 &   1079 &           712 \\
\hline \bottomrule
\end{tabular}
\label{tab:n_stocks}
\end{table}

\subsubsection{Features}\label{subsubsec:Features}

The two-stage quantile neural network model uses both stock characteristics and market variables aggregated from the stock-level variables as features. 
There are 176 stock characteristics consisting of 153 anomalies from \cite{tobek2021does} plus an additional 18 stock-level volatility estimates. 
The 23 market-level variables consist of 18 market volatility variables and
5 market mean variables. 
This brings the total number of features to 194.

The list of 153 anomalies is given in Appendix \ref{appendix:anomalies} and includes 93 fundamental, 49 market friction and 11 I/B/E/S anomalies, mostly covered by \cite{mclean2016does}, \cite{hou2020replicating} and \cite{harvey2016and}.
Compared to the original data, we use weekly data during training. This means that the fundamental signals are updated every week with financial statement information from financial years ending at least 6 months earlier. Trade data information, such as market capitalisation, is taken as the most recent available. Missing characteristic values are imputed using cross-sectional medians \citep{gu2018empirical}.

Each sub-network in the two-stage quantile neural network model
uses its own set of features. The standardised $\tau$-quantile sub-network uses 176 features consisting of 153 anomalies-based stock characteristics together with 18 stock-level volatilities and 5 market mean returns. The 18 stock-level volatilities are calculated using daily exponentially weighted moving averages (EWMA) with $\alpha \in \{0.8, 0.9, 0.94, 0.96, 0.98, 0.99\}$ smoothing factors, EWMA of negative returns with $\alpha$ smoothing factors of $\{0. 8, 0.9, 0.94\}$ to proxy asymmetric volatility, total volatilities calculated over the past 3, 6 and 12 months, and \cite{parkinson1980extreme} high-low range-based volatility. All stock-level volatility estimates are scaled by their cross-sectional mean to remove the effects of changes in market-wide average volatility. This is done to make stock-level volatilities more consistent with the other stock-level variables, which are normalised to cross-sectional empirical quantiles.
The 5 additional market variables used as inputs to the standardised $\tau$-quantile sub-network are computed as EWMAs of cross-sectional equal-weighted averages of individual stock returns $\mu_t = \alpha \cdot \left(1/N_t \sum_{i=1}^{N_t} r_{i,t-1}\right) + (1 - \alpha) \cdot \mu_{t-1}$ over the number of stocks $N_t$ with smoothing parameters $\alpha \in \{0. 9, 0.94, 0.96, 0.99, 0.999\}$. These variables measure the average historical market-wide return and should help in forecasting when there is some persistent time variance in expected market mean returns. We rescale the mean variables by the same scalar as the predicted returns to reflect the fact that their scale over time is relevant in the estimation.

The market-wide volatility sub-network uses 18 cross-sectional averages of stock-level volatilities described above. These serve as a proxy for market-wide index volatility, as they are likely to be relevant for estimating the distribution of individual stock returns, which includes both systematic and idiosyncratic volatility components. 

\subsubsection{Validation} \label{subsubsec:validation_logic}

We use the initial period from 1973 to 1994 for hyperparameter optimization. Models
with varying hyperparameters are trained from 1973 to 1989 and validated from 1990 to
1994. The hyperparameters yielding the best validation performance, measured in terms
of average quantile loss (see \autoref{subsec:Predictability} for details), are then
applied to train models from 1995 to 2018, generating out-of-sample predictions.
Detailed hyperparameter searches for both full and liquid samples are documented in
\autoref{sec:appendix_hyperparameter} together with the validation scheme.

From 1995 to 2018, using only the U.S. sample, we retrain the model annually using an
expanding data window starting in 1973. 
To ensure that the 22-day ahead return\footnote{Our forecast target.}, ends within the current calendar year, we restrict the final forecast date each year to November 29. All features are constructed using information available up to that point. This avoids using any information from the next year, thereby preventing both label and feature leakage from future data. Separate models are trained for full and liquid samples, with monthly out-of-sample predictions for both US and international samples, using the most recently trained model at each month's end. Predictions from the full sample model are applied to the full sample and, similarly, to the liquid sample. 
Furthermore, to provide additional evidence of out-of-sample predictability robustness, we report predictability and profitability results for the period from
2019 to 2023 for the U.S. stocks in the Appendix~\ref{app:US_update}. 

\subsection{Description of the estimated moments} \label{subsec:MomementStats}

Since our main empirical results are based on the predicted quantiles and moments of the distribution of all available stocks, we first describe the estimated moments for the US data. As the rest of the international regions show similar dynamics, we present these results in Appendix \autoref{tab:correlationsInt}. 
Specifically, we predict 37 $\tau$-quantiles with $\tau \in \mathcal{T}$, where $\mathcal{T}$ is the predefined set $\mathcal{T} = \{0.00005, 0.0001, 0.001, 0.005, 0.01, \ldots, 0.05, 0.075, 0.1, 0.15, 0.2, \ldots, 0.8, 0.85, 0.9, 0.925, 0.95, \\ \ldots, 0.99, 0.995, 0.999, 0.9999, 0.99995\}$, which provides denser coverage in the tails and sparser spacing in the center of the distribution. This choice reflects a practical trade-off between statistical resolution and computational feasibility\footnote{While using more quantiles is possible, it substantially increases computational and storage costs. We also experimented with fewer quantiles, but this led to inferior interpolation performance, particularly in the tails. Nonetheless, the architecture of the two-stage quantile neural network remains flexible and can accommodate changes in both the number and placement of quantiles—allowing, for example, denser tail coverage in applications such as Value-at-Risk.}. The grid is deliberately designed to provide denser coverage in the tails of the distribution, where spline interpolation is more fragile and tail behaviour is critical for applications such as risk management while the central region is covered more sparsely (every 5\%). 
This design also helps counteract the imbalance in loss aggregation since quantile losses are typically larger in the center, uniform averaging across $\tau$ implicitly downweights the tails. Denser tail coverage helps restore balance and ensures that tail performance is better accounted for.

Further, we also predict mean, standard deviation, skewness and kurtosis of all available individual stock returns. This pattern has practical relevance for portfolio construction. In the full sample, the high correlation implies that both characteristics yield similar asset ranking and thus similar products. In contrast, in the liquid sample, correlations close to zero support a potentially more diversified portfolio construction based on higher-order moments.

\begin{figure}[H]
\centering
\includegraphics[width=1\textwidth]{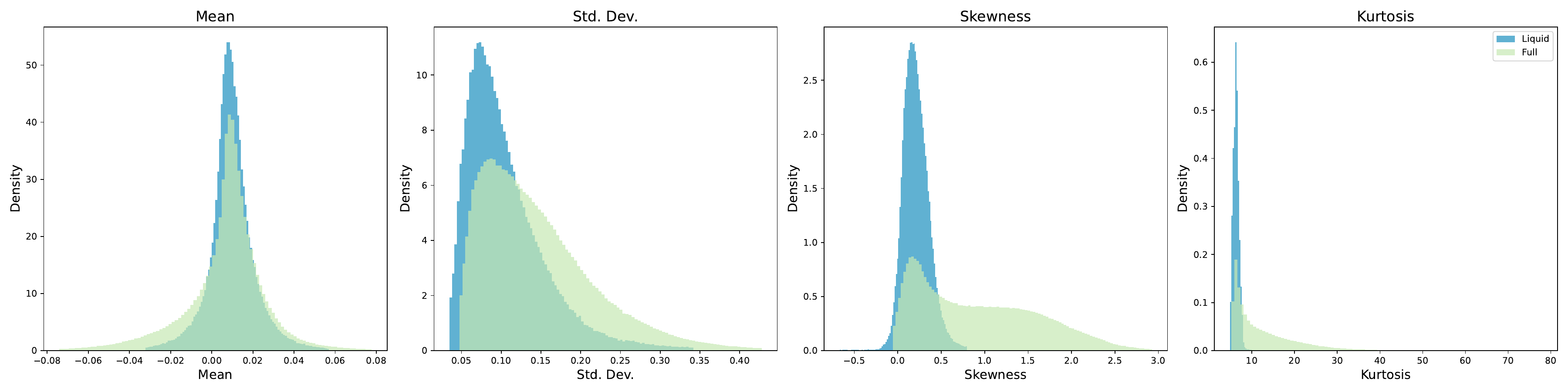}
\caption{\textbf{Histogram of Central Distribution Moments.} 
This figure shows histograms of the forecasted first four moments for the full and liquid samples in the U.S.
The Moments are forecasted using the quantiles-to-moments algorithm introduced in \autoref{subsec:DerivingDistribution} including bias adjustment described in \autoref{subsubsec:MomentsAdjustment}.
Displayed variables are trimmed to exclude the top and bottom 0.5\% of their values.
}
\label{fig:figureMomentHist}
\end{figure}

\autoref{tab:VarSummary} contains summary statistics for the predicted moments, and \autoref{fig:figureMomentHist} shows the histogram of the predicted mean, standard deviation, skewness and kurtosis for the full and liquid samples for all stocks across the US.
\begin{table}[H]
\caption{{\bf Summary statistics for the moments in the U.S.}
This table shows summary statistics for the forecasted first four moments for the full and liquid samples in the U.S. The moments are forecasted using the quantiles-to-moments algorithm introduced in \autoref{subsec:DerivingDistribution}. The table reports standard deviation, skewness, and kurtosis without bias adjustment described in  \autoref{subsubsec:MomentsAdjustment} as well as with the adjustment (Adj).
}
\centering
\footnotesize
\begin{tabular}{lccclccc}
\toprule \hline \\ [-2ex]
& \multicolumn{3}{c}{Full Sample} && \multicolumn{3}{c}{Liquid Sample}\\
\cmidrule{2-4} \cmidrule{6-8}
& Mean & Median & Std. Dev. && Mean & Median & Std. Dev.\\
\cmidrule{2-4} \cmidrule{6-8}\\ [-2ex]
Mean & 0.0068 & 0.0089 & 0.0216 &  & 0.0098 & 0.0092 & 0.0126\\
Std. Dev. & 0.1477 & 0.1317 & 0.0743 &  & 0.1071 & 0.0940 & 0.0539\\
Std. Dev. Adj & 0.1487 & 0.1325 & 0.0750 &  & 0.1074 & 0.0944 & 0.0541\\
Skewness & 0.8453 & 0.7572 & 0.6352 &  & 0.1800 & 0.1722 & 0.1993\\
Skewness Adj & 0.9340 & 0.8205 & 0.7072 &  & 0.2068 & 0.1971 & 0.1839\\
Kurtosis & 8.9784 & 7.5963 & 5.2384 &  & 5.4261 & 5.2248 & 2.3752\\
Kurtosis Adj & 13.798 & 9.9362 & 19.8491 &  & 6.8452 & 6.2727 & 8.3637\\
\hline \bottomrule
\end{tabular}
\label{tab:VarSummary}
\end{table}
We can see that the distribution of the mean for both samples is approximately centred around zero, with the liquid sample having a slightly higher density around the mean. The mean return is positive for both samples, with the liquid sample having a slightly higher mean (0.0098) than the total sample (0.0068). 
The standard deviation is higher for the total sample (0.1477) than for the liquid sample (0.1071) and shows a wider spread for the total sample than for the liquid sample. Skewness is significantly higher in the full sample (0.9340) than in the liquid sample (0.2068), reflecting more extreme positive and negative returns, while the liquid sample shows a more symmetrical distribution. Kurtosis is also higher in the full sample (13.798) than in the liquid sample (6.8452), indicating more pronounced tails compared to the liquid sample. Bias adjustment described in \autoref{subsubsec:MomentsAdjustment} has large impact only on Kurtosis, with negligible impact on standard deviation, and small impact on skewness.

\autoref{tab:VarCorrelations} also includes average cross-sectional correlations between the first four predicted moments and the median of stock returns.\footnote{The table reports results with bias adjustment described in appendix \autoref{subsubsec:MomentsAdjustment}. The adjustment has only a small impact on the reported correlations.}
There is a high correlation of $0.95$ between the forecasted mean and the median for the liquid sample, while the correlation is lower ($0.75$) for the full sample. This is consistent with the fact that the full sample contains more micro-cap stocks with higher skewness, which affects the mean but not the median. There is also a high correlation between skewness and kurtosis ($0.91$) in the full sample. In contrast, in the liquid sample, the correlation between skewness and kurtosis is close to zero. This pattern has practical relevance for portfolio construction. In the full sample, the high correlation implies that both characteristics yield similar asset ranking and thus similar products. In contrast, in the liquid sample, correlations close to zero support a potentially more diversified portfolio construction based on higher-order moments.

\begin{table}[H]
\caption{{\bf Average cross-sectional correlation between variables in the U.S.}
This table shows the average cross-sectional Spearman correlation between the forecasted first four moments and the median of the stock returns for the full and liquid samples in the U.S. Medians are forecasted using the two-stage model defined in \autoref{subsubsec:TwoStage} and moments are forecasted using the quantiles-to-density algorithm introduced in \autoref{subsec:DerivingDistribution}.
}
\resizebox{1\textwidth}{!}{
\begin{tabular}{lccccclccccc}
\toprule \hline \\ [-2ex]
& \multicolumn{5}{c}{Full Sample} && \multicolumn{5}{c}{Liquid Sample}\\
\cmidrule{2-6} \cmidrule{8-12}
& Median & Mean & Variance & Skewness & Kurtosis && Median & Mean & Variance & Skewness & Kurtosis\\
\cmidrule{2-6} \cmidrule{8-12} \\ [-2ex]
Median & 1.00 & 0.75 & -0.63 & -0.54 & -0.43 &  & 1.00 & 0.95 & -0.34 & 0.08 & 0.15\\
Mean & 0.75 & 1.00 & -0.22 & -0.08 & -0.08 &  & 0.95 & 1.00 & -0.15 & 0.27 & 0.06\\
Var & -0.63 & -0.22 & 1.00 & 0.52 & 0.22 &  & -0.34 & -0.15 & 1.00 & 0.65 & -0.39\\
Skewness & -0.54 & -0.08 & 0.52 & 1.00 & 0.91 &  & 0.08 & 0.27 & 0.65 & 1.00 & -0.16\\
Kurtosis & -0.43 & -0.08 & 0.22 & 0.91 & 1.00 &  & 0.15 & 0.06 & -0.39 & -0.16 & 1.00\\
 \hline \bottomrule
\end{tabular}
}
\label{tab:VarCorrelations}
\end{table}

\subsection{Predictability} \label{subsec:Predictability}

This section evaluates the out-of-sample predictions of the quantiles from our two-stage quantile neural network. It also evaluates the predictions of the mean and variance estimated from the quantiles. This is of particular interest as we argue that the first and second moments, estimated without any assumptions about the distribution, are more predictive of the mean and variance. To benchmark our forecasting strategy, we use a linear model (LNN), the single hidden layer neural network with 32 neurons (1hNN), the double hidden layer neural network with 128 neurons (2hNN) and the commonly used GARCH. Details of the benchmark models can be found in Appendix~\ref{subsec:NeuralNetworks} and Appendix~\ref{subsec:GARCH}. We also compare the performance of the two-stage quantile model with tree-based machine learning benchmarks in Appendix~\ref{app:US_update_tree_based}. In addition, we provide a simulation study in Appendix~\ref{app:US_update_simulations} analyzing the performance of our forecasting strategy out of sample under a known data-generating process.


To measure predictive performance across the distribution, we report the average loss, which is defined as the time series average of the cross-sectional $\tau$-quantile losses, averaged across all $\tau$s for each month. 

Specifically, we define the quantile loss for asset $i$ at time $t$ and quantile level $\tau$ as
$L_{\tau,t,i} = \rho_\tau \left( r_{i,t+1} - \hat{Q}_{\tau} \right)$ 
as in Equation~\ref{eq:qlSimple},
where \( r_{i,t+1} \) denotes the realized return, \( \hat{Q}_{\tau} \) is the predicted $\tau$-quantile from the model. 

We then compute the overall average quantile loss as
\begin{equation}\label{eq:avg_out_loss} 
    L_{\text{avg}} = \frac{1}{T} \sum_{t=1}^{T} \left( \frac{1}{N_t} \sum_{i=1}^{N_t} \left( \frac{1}{K} \sum_{\tau \in  \mathcal{T}} L_{\tau,t,i} \right) \right)
\end{equation}
where $T$ is the number of months, $N_t$ is the number of stocks available at time $t$, $K = |\mathcal{T}|$ is the number of quantiles considered, and $L_{\tau,t,i}$ is the quantile loss defined above.

\autoref{tab:avg_quantile_loss} shows the average out-of-sample $\tau$-quantile losses for the full and liquid samples across all regions. The average losses of the two-stage quantile neural network are significantly smaller than all benchmark models when looking at the full sample data. Our two-stage model also significantly outperforms the 1hNN full sample model in all regions except Europe and the 2hNN full sample model in the US and Asia Pacific. The average quantile loss is also lower in Japan, but the difference is not statistically significant at the 5\% level.

Predictive performance differences for the liquid sample mirror those observed for the full sample, although to a lesser extent and with fewer statistically significant differences. As expected, average out-of-sample quantile losses are generally lower for the liquid sample than for the full sample, reflecting the lower volatilities of the underlying stocks. The two-stage model consistently outperforms the GARCH model across all regions. However, when compared to alternative neural network specifications, the two-stage model only shows significant outperformance in the US region. Internationally, while the two-stage model achieves lower average quantile losses than the alternative neural network specifications, these differences do not reach statistical significance at the 5\% level, except for the 1hNN model in Europe. In addition to aggregated quantile loss, we also report the $\tau$-specific out-of-sample quantile losses, in Table~\ref{tab:tauSpecificQL} and continuous ranked probability score in Table~\ref{tab:scoring_rule} in Appendix~\ref{app:US_update_quantile_breakdown}.

These performance differences also highlight the role of model flexibility in capturing complex distributional patterns. In particular, the consistent improvement from LNN to 2hNN across regions reflects the benefits of modelling non-linearities in the conditional quantile functions. The two-stage model, which incorporates additional structural layers and nonlinear interactions across features and volatilities, achieves further gains—especially in the full sample—suggesting that the cross-sectional distribution of stock returns depends on non-linear effects that are not well captured by linear architectures.

\begin{table}[H]
\caption{{\bf Out-of-sample quantile forecasts evaluation.}
This table shows the average out-of-sample quantile cross-section loss for quantile forecasts from the two-stage model and alternative models. The results are obtained separately for the full and liquid samples for individual regions covering the out-of-sample period from 1995 to 2018. Neural network models are trained annually using the expanding window on US data only, while out-of-sample forecasts are generated for both the US and international samples. 
The average loss is calculated as the time series average of the cross-sectional averages of the average quantile loss across all $\tau \in \mathcal{T}$, where $\mathcal{T}$ is defined in Section~\ref{subsec:MomementStats}. The average loss is calculated as in (Eq. \ref{eq:avg_out_loss}) for the two-stage NN model, as well as benchmark neural network models (the linear model (LNN), the one-layer model (1hNN), the two-layer model (2hNN)) and the benchmark GARCH model. Note that losses are multiplied by 100.0, and the t-statistics in parentheses are computed for the average loss difference between the given model specification and the two-stage benchmark model using Newey-West standard errors with 12 lags.}
\resizebox{1\textwidth}{!}{
\begin{tabular}{llcccclcccc}
\toprule \hline \\ [-2ex]
Specification & Variable & \multicolumn{4}{c}{Full Sample} && \multicolumn{4}{c}{Liquid Sample}\\
\cmidrule{1-6} \cmidrule{8-11}
&& USA & Europe & Japan & Asia Pacific && USA & Europe & Japan & Asia Pacific\\
\cmidrule{3-6} \cmidrule{8-11} \\ [-2ex]
GARCH & $L_{avg}$ & 2.659 & 2.669 & 2.050 & 3.416 &  & 1.899 & 1.679 & 1.836 & 2.159\\
  & (t-stat) & (8.47) & (14.07) & (8.15) & (15.51) &  & (6.38) & (4.49) & (3.05) & (6.23)\\
 \cmidrule{3-6} \cmidrule{8-11} \\ [-2ex]
LNN & $L_{avg}$ & 2.592 & 2.520 & 2.020 & 3.143 &  & 1.874 & 1.658 & 1.820 & 2.115\\
 & (t-stat) & (6.05) & (3.44) & (4.43) & (4.44) &  & (3.95) & (2.07) & (1.72) & (1.53)\\
 \cmidrule{3-6} \cmidrule{8-11} \\ [-2ex]
1hNN & $L_{avg}$ & 2.569 & 2.496 & 2.004 & 3.120 &  & 1.869 & 1.655 & 1.819 & 2.121\\
 & (t-stat) & 3.81 & (0.44) & (4.03) & (3.09) &  & (3.08) & (1.69) & (1.63) & (1.57)\\
 \cmidrule{3-6} \cmidrule{8-11} \\ [-2ex]
2hNN & $L_{avg}$ & 2.564 & 2.494 & 2.001 & 3.123 &  & 1.870 & 1.654 & 1.823 & 2.128\\
 & (t-stat) & (2.22) & (0.04) & (1.71) & (2.50) &  & (2.39) & (1.10) & (1.88) & (1.73)\\
 \cmidrule{3-6} \cmidrule{8-11} \\ [-2ex]
Two stage NN & $L_{avg}$ & 2.550 & 2.494 & 1.992 & 3.106 &  & 1.858 & 1.649 & 1.813 & 2.106\\
\hline\bottomrule
\end{tabular}
}
\label{tab:avg_quantile_loss}
\end{table}

We also examine the performance of the mean and variance forecasts. To compare our two-stage model and the forecasts obtained by the algorithms described earlier, we compute the mean forecasts directly using a two-layer hidden neural network model estimated with a mean square error loss function. We follow \cite{gu2020empirical} and modify\footnote{see Appendix~\ref{app:US_update_quantile_breakdown} for the description of the testing procedure.} the calculation of the Diebold-Mariano test statistic to compare the cross-sectional average of forecast errors instead of comparing individual errors.

\autoref{tab:Mean forecast evaluation} shows the out-of-sample $R^2$ of the mean and median forecasts. Our two-stage model provides the highest out-of-sample $R^2$. For the liquid sample and across all regions, the highest $R^2$ is consistently achieved by the median forecast from the two-stage model. Forecasts using the median are not statistically different from those using the mean in the two-stage model. $R^2$ for the direct mean prediction from the 2 hidden layers neural network is negative and smaller than the two-stage predictions. In the case of the full sample, the highest $R^2$ is consistently achieved by the mean prediction derived from our model.

The results reflect a trade-off between robustness to extremes and sensitivity to the full shape of the conditional distribution. Quantile loss, by design, is less sensitive to large forecast errors, and the resulting median forecast is robust to outliers and heavy tails. However, this robustness can come at the cost of discarding potentially informative variation in the tails, which are particularly influential when forecasting the conditional mean—a non-robust functional. In the full sample, where extreme returns are more prevalent, this information appears valuable, and the mean forecast derived via numerical integration over the estimated quantile distribution outperforms both the median and the MSE-based benchmark. Notably, even the direct MSE-trained mean from the 2hNN achieves higher $R^2$ than the median in this setting, highlighting that robustness does not necessarily translate into superior predictive performance when the tails contain signal.
In contrast, the liquid sample exhibits thinner tails and fewer extreme returns. In this setting, the benefits of robustness are more pronounced, and the median forecast achieves competitive or superior performance. Importantly, since both the mean and median are derived from the same underlying quantile forecasts in our two-stage model, the comparison isolates differences attributable to the target functional rather than the model class. These findings underscore that the relative performance of robust versus non-robust functionals depends on the tail behavior of the return distribution and the presence (or absence) of influential observations.

\begin{table}[H]
\caption{{\bf Mean forecast evaluation.}
This table shows the $R^2$ of the mean and median forecasts of the next 22 days of stock returns for the full and liquid samples across all regions. The mean prediction based on the two-stage quantile model (Two-stage NN) is compared with the direct mean prediction from the two-hidden-layer neural network (2hNN MSE - Mean) and the median prediction from the two-stage model (Two-stage NN - Median). Results are obtained separately for full and liquid samples on individual regions covering the period from 1995 to 2018. Models are retrained annually using the expanding window for US data only, while out-of-sample forecasts are generated for both US and international data. Note that losses are multiplied by 100, and values in parentheses are Diebold-Mariano test statistics for the difference between the model in a row and the two-stage NN mean. Standard errors are adjusted using the Newey-West standard deviation estimator with 12 lags.}
\resizebox{1\textwidth}{!}{
\begin{tabular}{lccccclccccc}
\toprule \hline \\ [-2ex]
Specification & \multicolumn{5}{c}{Full Sample} && \multicolumn{5}{c}{Liquid Sample}\\
\cmidrule{2-6} \cmidrule{8-12}
& USA & Europe & Japan & Asia Pacific & Global && USA & Europe & Japan & Asia Pacific & Global\\
\cmidrule{2-6} \cmidrule{8-12} \\ [-2ex]
2hNN MSE - Mean & 0.65 & 0.14 & 0.85 & 0.69 & 0.51 &  & -0.68 & -0.44 & -0.48 & -0.77 & -0.54\\
& (-2.54) & (-3.56) & (-2.37) & (-3.39) & (-3.55) &  & (-1.81) & (-1.31) & (-1.84) & (-1.79) & (-1.98)\\
\cmidrule{2-6} \cmidrule{8-12}
Two stage NN - Median & 0.10 & -0.04 & 0.30 & 0.06 & 0.09 &  & 0.37 & 0.32 & 0.47 & 0.55 & 0.50\\
& (-3.50) & (-4.92) & (-2.82) & (-4.83) & (-4.93) &  & (0.23) & (0.27) & (0.98) & (0.39) & (0.52)\\
\cmidrule{2-6} \cmidrule{8-12}
Two stage NN - Mean & 1.63 & 0.87 & 1.66 & 1.63 & 1.36 &  & 0.32 & 0.22 & 0.18 & 0.40 & 0.37\\
 \hline \bottomrule
\end{tabular}
}
\label{tab:Mean forecast evaluation}
\end{table} 

Finally, to evaluate the second moment forecasts, we use the Mean Absolute Deviation (MAD) and the Root Mean Squared Error (RMSE) of the predicted 22-day volatility for the full and liquid samples for all regions. Realized volatility is calculated as the square root of the sum of squared daily returns over the 22 trading days. \autoref{tab:VolatilityForecast} shows a highly significant outperformance of the volatility forecast based on the two-stage model in terms of both MAD and RMSE for all regions and for both the full and liquid samples.

\begin{table}[H]
\caption{\textbf{Volatility forecast evaluation.}
This table shows the mean absolute deviation (MAD), root mean squared error (RMSE), and Diebold Mariano test statistic (DM) comparing the corresponding losses of the two competing models for the 22-day volatility forecasts.
Volatility forecasting based on two-stage quantile forecasting (two-stage NN) is compared with GARCH model forecasting. Results are obtained separately for full and liquid samples on individual regions covering the period from 1995 to 2018. Neural network models are retrained annually using an expanding window on US data only, while out-of-sample forecasts are generated for both US and international samples. Diebold-Mariano test statistics are computed for the difference between the GARCH and two-stage NN model volatility forecasts. Note that losses are multiplied by 100 and the standard errors used in the Diebold-Mariano test statistics are adjusted using the Newey-West standard deviation estimator with 12 lags.}
\resizebox{1\textwidth}{!}{
\begin{tabular}{llccccclccccc}
\toprule \hline \\ [-2ex]
Specification & Variable & \multicolumn{5}{c}{Full Sample} && \multicolumn{5}{c}{Liquid Sample}\\
\cmidrule{2-7} \cmidrule{9-13}
&& USA & Europe & Japan & Asia Pacific & Global && USA & Europe & Japan & Asia Pacific & Global\\
\cmidrule{2-7} \cmidrule{9-13} \\ [-2ex]
GARCH & MAD & 7.34 & 8.79 & 7.79 & 11.1 & 8.53 &  & 3.89 & 3.54 & 3.80 & 4.87 & 3.93\\
Two stage NN & MAD & 4.63 & 5.20 & 3.99 & 6.49 & 4.99 &  & 2.93 & 2.63 & 2.81 & 3.16 & 2.87\\
\cmidrule{2-7} \cmidrule{9-13} \\ [-2ex]
GARCH & RMSE & 13.14 & 16.24 & 14.63 & 18.70 & 15.50 &  & 6.93 & 7.10 & 6.72 & 9.52 & 7.36\\
Two stage NN & RMSE & 7.46 & 8.52 & 5.64 & 10.05 & 8.08 &  & 4.62 & 4.03 & 3.98 & 4.86 & 4.38\\
\cmidrule{2-7} \cmidrule{9-13}
& DM & (10.25) & (35.92) & (20.32) & (44.26) & (26.47) &  & (7.90) & (5.98) & (7.97) & (9.40) & (10.02)\\
\hline \bottomrule
\end{tabular}
}
\label{tab:VolatilityForecast}
\end{table}

\medskip

To clarify the source of improvements, we emphasise that the accuracy of the empirical moment estimates—such as the mean and variance—primarily depends on the quality of the underlying distributional forecasts. While we rely on numerical integration over interpolated quantile forecasts, this approach is not tied to our two-stage neural network architecture. In principle, any sufficiently accurate quantile forecasting method, whether traditional or machine learning-based, can be used in conjunction with this integration technique to obtain distributional moments. 
However, our empirical results show that the two-stage QNN consistently delivers more accurate quantile forecasts than alternative models, particularly in the full sample.
Overall, we demonstrate the superior distributional forecasting performance of the two-stage model and the moments derived from it compared to alternative models. The average out-of-sample quantile loss represents the forecast over the entire distribution, while the evaluation of the mean and variance forecasts provides additional evidence derived from the forecasted quantiles. 
While our evaluation focuses on distributional forecast accuracy and central moments, the underlying quantile estimates are also applicable to risk management tasks such as Value-at-Risk (VaR) forecasting. A formal backtesting exercise in this context is beyond the scope of this paper, but we acknowledge its practical relevance and view it as a promising direction for future work.

\subsection{Distribution and the cross-section of stock returns} \label{subsec:Asset pricing implications}

In addition to forecasting, we aim to investigate whether individual distributional characteristics are priced into the cross-section of returns through the profitability of long-short portfolios formed based on forecasted $\tau$-quantiles as well as the profitability of long-short portfolios formed using central moments -- mean, volatility, skewness and kurtosis.

\subsubsection{$\tau$-quantiles and the cross-section of stock returns} \label{subsubsec:QuantilesPortfolio}

\autoref{tab:QuantilePortfolios} shows the average monthly returns and annualised Sharpe ratios for decile long-short equal-weighted portfolios constructed using the $\tau$-quantile forecasts of the two-stage model. Each month, the long and short legs of the portfolios are constructed by selecting the top and bottom 10\% of stocks with the highest and lowest forecasts, respectively.

\begin{table}[H]
\caption{
{\bf Quantiles and the cross-section of stock returns in the U.S.}
This table shows the average and volatility of monthly returns and annualised Sharpe ratios 
for decile long, short and long-short equal-weighted portfolios over the period 1995 to 2018. For each $\tau$-quantile forecast, the long portfolio is formed by taking the top 10\% of stocks with the highest forecast each month,
The short portfolio is formed by taking the bottom 10\% of stocks with the lowest forecast each month. The long-short portfolio is created by taking the difference between the long and short portfolios. Values in brackets are t-statistics adjusted for Newey-West standard errors with 12 lags.
Results are obtained separately for the full and liquid samples for the US only.} \label{tab:QuantilePortfolios}
\centering
\resizebox{0.8\textwidth}{!}{
\begin{tabular}{l|ccc|ccc|ccc}
\toprule
 & \multicolumn{9}{c}{Full Sample} \\
\cmidrule{2-10}
 & \multicolumn{3}{c}{Long} & \multicolumn{3}{c}{Short} & \multicolumn{3}{c}{Long-short} \\
\cmidrule(lr){2-4} \cmidrule(lr){5-7} \cmidrule(lr){8-10}
$\tau$ & Mean & Vol & SR & Mean & Vol & SR & Mean & Vol & SR \\
\midrule
0.01 & 1.22 (5.62) & 2.53 & 1.67 & 0.14 (0.19) & 12.12 & 0.04 & 1.09 (1.52) & 12.18 & 0.31 \\
0.05 & 1.25 (5.66) & 2.46 & 1.76 & -0.09 (-0.12) & 15.59 & -0.02 & 1.34 (1.86) & 11.90 & 0.39 \\
0.1 & 1.28 (5.73) & 2.48 & 1.79 & -0.26 (-0.35) & 12.87 & -0.07 & 1.53 (2.13) & 11.78 & 0.45 \\
0.2 & 1.34 (6.04) & 2.48 & 1.87 & -0.57 (-0.78) & 12.34 & -0.16 & 1.91 (2.68) & 11.61 & 0.57 \\
0.3 & 1.46 (6.54) & 2.58 & 1.96 & -0.91 (-1.27) & 11.68 & -0.27 & 2.37 (3.36) & 11.09 & 0.74 \\
0.4 & 1.88 (7.44) & 3.16 & 2.06 & -1.34 (-1.91) & 11.05 & -0.42 & 3.22 (4.89) & 9.96 & 1.12 \\
0.5 & 2.88 (6.97) & 5.60 & 1.78 & -1.82 (-2.68) & 10.17 & -0.62 & 4.70 (8.25) & 7.17 & 2.27 \\
0.6 & 3.44 (5.68) & 8.70 & 1.37 & -2.12 (-3.51) & 8.35 & -0.88 & 5.56 (9.41) & 4.86 & 3.96 \\
0.7 & 3.48 (4.73) & 11.16 & 1.08 & -1.42 (-3.57) & 4.47 & -1.10 & 4.90 (6.57) & 8.53 & 1.99 \\
0.8 & 2.98 (3.72) & 12.59 & 0.82 & 0.44 (1.65) & 2.46 & 0.62 & 2.55 (3.16) & 11.78 & 0.75 \\
0.9 & 2.12 (2.57) & 13.35 & 0.55 & 0.75 (3.29) & 2.36 & 1.10 & 1.37 (1.69) & 12.49 & 0.38 \\
0.95 & 1.71 (2.11) & 13.16 & 0.45 & 0.86 (4.04) & 2.44 & 1.22 & 0.85 (1.07) & 12.80 & 0.23 \\
0.99 & 1.36 (1.70) & 13.09 & 0.36 & 0.97 (5.04) & 2.75 & 1.22 & 0.39 (0.50) & 12.28 & 0.11 \\
\midrule
 & \multicolumn{9}{c}{Liquid Sample} \\
\cmidrule{2-10}
 & \multicolumn{3}{c}{Long} & \multicolumn{3}{c}{Short} & \multicolumn{3}{c}{Long-short} \\
\cmidrule(lr){2-4} \cmidrule(lr){5-7} \cmidrule(lr){8-10}
$\tau$ & Mean & Vol & SR & Mean & Vol & SR & Mean & Vol & SR \\
\midrule
0.01 & 0.95 (5.26) & 2.99 & 1.10 & 0.16 (0.21) & 11.09 & 0.05 & 0.79 (1.06) & 11.40 & 0.24 \\
0.05 & 0.96 (5.63) & 2.89 & 1.15 & 0.15 (0.20) & 12.99 & 0.04 & 0.81 (1.11) & 11.22 & 0.25 \\
0.1 & 0.98 (5.72) & 2.90 & 1.17 & 0.11 (0.14) & 12.70 & 0.03 & 0.87 (1.20) & 11.59 & 0.26 \\
0.2 & 1.02 (5.89) & 2.92 & 1.21 & 0.04 (0.06) & 13.86 & 0.01 & 0.98 (1.34) & 11.32 & 0.30 \\
0.3 & 1.10 (6.09) & 3.05 & 1.25 & -0.04 (-0.06) & 13.86 & -0.01 & 1.14 (1.58) & 10.97 & 0.36 \\
0.4 & 1.39 (6.31) & 3.59 & 1.34 & -0.14 (-0.20) & 12.12 & -0.04 & 1.53 (2.23) & 9.81 & 0.54 \\
0.5 & 1.86 (5.61) & 5.86 & 1.10 & -0.35 (-0.53) & 10.10 & -0.12 & 2.21 (4.21) & 6.33 & 1.21 \\
0.6 & 1.74 (3.69) & 8.49 & 0.71 & -0.38 (-0.88) & 6.27 & -0.21 & 2.11 (5.40) & 5.45 & 1.34 \\
0.7 & 1.66 (2.86) & 10.27 & 0.56 & 0.48 (2.50) & 3.08 & 0.54 & 1.18 (2.23) & 9.51 & 0.43 \\
0.8 & 1.28 (1.94) & 11.37 & 0.39 & 0.71 (3.95) & 2.89 & 0.85 & 0.57 (0.91) & 10.97 & 0.18 \\
0.9 & 0.91 (1.24) & 11.68 & 0.27 & 0.76 (4.29) & 2.83 & 0.93 & 0.15 (0.21) & 10.39 & 0.05 \\
0.95 & 0.77 (1.03) & 11.60 & 0.23 & 0.80 (4.62) & 2.83 & 0.98 & -0.03 (-0.04) & 10.39 & -0.01 \\
0.99 & 0.60 (0.80) & 12.23 & 0.17 & 0.84 (4.72) & 2.88 & 1.01 & -0.23 (-0.32) & 11.38 & -0.07 \\
\bottomrule
\end{tabular}
}
\end{table}

The average monthly returns and Sharpe ratios are the highest for the long-short portfolios formed based on the central $\tau$-quantiles, $\tau$s of $50$\% or $60$\%, for both the full sample and the liquid sample.
Average monthly returns decrease as we move towards the tails of the distribution. For the full sample, average monthly returns are not significantly different from zero for $\tau$s greater than 80\% and less than 10\%. For the liquid sample, the average monthly returns fall even more sharply as one moves towards the tails of the distribution, becoming not 
significantly different from zero for $\tau$s greater than 70\% and less than 40\%. The highest Sharpe ratios are achieved by the long-short portfolios with the highest average monthly returns.

These results suggest that the tail $\tau$-quantiles in isolation are not priced into the cross-section of stock returns. On the contrary, the central $\tau$-quantiles are associated with the highest average returns. We also provide international evidence showing similar results in \autoref{QuantilePortfoliosInternational}. To further investigate the role of the distribution in pricing the cross-section of stock returns, we turn to the moments of the distribution in the next section.

\subsubsection{Distributional moments and the cross-section of stock returns} \label{subsubsec:MomemntsSingleSorts} 

We then examine the profitability of long-short portfolios formed on the basis of the central moments of the distribution - mean, variance, skewness and kurtosis - forecasts estimated from our two-stage quantile model. To illustrate the importance of asymmetry in the data, we first compare the profitability of portfolios formed on the mean and median forecasts. \autoref{tab:MSEvsMean} shows the average monthly returns and Sharpe ratios 
of the long-short decile portfolios based on the mean and median forecasts across all regions. 

\begin{table}[H]
\caption{{\bf Mean and median portfolios.} This table shows average monthly returns and annualised Sharpe ratios for decile long-short equal-weighted portfolios based on the median and mean predictions from the two-stage quantile NN model, the mean prediction from the two-hidden-layer neural network with multi-head quantile loss (2hNN Quantile - Mean) and the mean prediction from the two-hidden-layer neural network with MSE loss (2hNN MSE - Mean). The results are shown for the full and liquid samples and for individual regions, covering the period from 1995 to 2018.
The long portfolio is formed by taking the top 10\% of stocks with the highest forecast each month, and the short portfolio is formed by taking the bottom 10\% of stocks with the lowest forecast each month. Values in parentheses are t-statistics adjusted for Newey-West standard errors with 12 lags.
The significance of the difference in mean returns is tested using t-statistics with Newey-West standard errors with 12 lags.}
\resizebox{1\textwidth}{!}{
\begin{tabular}{llccccclccccc}
\toprule \hline \\ [-2ex]
Specification & Variable & \multicolumn{5}{c}{Full Sample} && \multicolumn{5}{c}{Liquid Sample}\\
\cmidrule{3-7} \cmidrule{9-13}
&& USA & Europe & Japan & Asia Pacific & Global && USA & Europe & Japan & Asia Pacific & Global\\
\hline \\ [-2ex]
2hNN MSE - Mean & Avg Ret & 5.00 & 3.28 & 3.11 & 5.71 & 4.27 &  & 1.62 & 1.71 & 1.67 & 2.05 & 1.76\\
 & SR & 3.47 & 2.87 & 2.52 & 3.72 & 5.50 &  & 0.87 & 1.63 & 1.52 & 1.38 & 2.01\\
 & t-stat & 4.78 & 9.26 & 5.78 & 5.79 & 9.93 &  & 1.87 & 1.53 & 2.81 & 2.30 & 2.54\\
\hline \\ [-2ex]
2hNN Quantile - Mean & Avg Ret & 5.59 & 4.17 & 3.44 & 6.43 & 4.91 &  & 1.67 & 1.86 & 1.68 & 2.22 & 1.86\\
 & SR & 4.10 & 3.70 & 2.86 & 4.28 & 6.35 &  & 0.86 & 1.80 & 1.64 & 1.51 & 2.09\\
& t-stat & 3.70 & 6.01 & 4.20 & 6.26 & 7.92 &  & 1.71 & 0.68 & 2.09 & 1.90 & 1.93\\
\hline \\ [-2ex]
Two stage NN - Median & Avg Ret & 4.70 & 3.35 & 3.10 & 5.07 & 4.05 &  & 2.21 & 2.00 & 2.25 & 2.95 & 2.35\\
 & SR & 2.27 & 2.48 & 1.84 & 3.06 & 3.53 &  & 1.21 & 1.52 & 1.51 & 1.82 & 2.03\\
 & t-stat & 4.26 & 7.14 & 5.62 & 8.75 & 9.16 &  & 0.84 & 0.19 & 0.10 & 1.07 & 0.75\\
 \hline \\ [-2ex]
Two stage NN - Mean & Avg Ret & 6.44 & 5.29 & 4.03 & 8.05 & 5.95 &  & 2.35 & 2.02 & 2.27 & 3.12 & 2.44\\
 & SR & 4.34 & 4.49 & 3.31 & 4.99 & 6.92 &  & 1.70 & 1.88 & 1.92 & 2.13 & 2.83\\

\hline\bottomrule
\end{tabular}
}
\label{tab:MSEvsMean}
\end{table}

To benchmark our forecasts, we use the mean forecasts obtained directly from the feed-forward neural network model estimated with the mean squared error loss function, as well as its distributional forecast equivalent where the mean forecast is derived from predicted quantiles. The long-short portfolio based on the mean forecast of our two-stage quantile model leads to significantly higher average monthly returns and Sharpe ratios compared to the portfolios based on the mean forecasts obtained directly from the simple neural network for the full sample across all regions. This is consistent with the $R^2$ results in \autoref{tab:Mean forecast evaluation}, as our two-stage model outperforms the direct mean forecast in terms of $R^2$ across all regions and also leads to higher corresponding profitability. In the case of the full sample, the average monthly returns of our mean forecast are economically higher than the direct mean forecast in all regions and significantly higher for Japan, Asia-Pacific and globally. 

Next, the average monthly returns of the mean- and median-based long-short portfolios are statistically indistinguishable for the liquid sample. This is also consistent with the $R^2$ results in \autoref{tab:Mean forecast evaluation}, as there are no significant differences in the quality of the mean and median forecasts for the liquid sample. However, the results are different for the full sample. The difference between the mean and the median is associated with distributional asymmetry, which is more pronounced for stocks in the full sample. Mean-based long-short portfolios have significantly higher average monthly returns than median-based long-short portfolios. This is true for all regions and also globally. The difference between mean and median pricing for the full sample indicates the effect of forecast distributional asymmetry in the cross-section of stock returns.

\begin{table}
\caption{{\bf Moments decile portfolios.}
This table shows average monthly returns for individual deciles (1-10) and long-short equal-weighted portfolios built at the predicted moments.
Results are reported for the full and liquid samples, as well as for individual regions, covering the period from 1995 to 2018. Values in parentheses are t-statistics adjusted for Newey-West standard errors with 12 lags.}
\label{tab:SingleSorts}
\centering
\scriptsize
\resizebox{0.9\textwidth}{!}{
\begin{tabular}{lcccclcccc}
\toprule \hline \\ [-2ex]
\textbf{Mean} & \multicolumn{4}{c}{Full Sample} && \multicolumn{4}{c}{Liquid Sample}\\
\cmidrule{2-5} \cmidrule{7-10}
Decile & USA & Europe & Japan & Asia Pacific && USA & Europe & Japan & Asia Pacific \\
\cmidrule{2-5} \cmidrule{7-10} \\ [-2ex]
1 & -2.37 (-3.87) & -1.06 (-1.87) & -1.24 (-1.99) & -1.69 (-2.40) &  & -0.48 (-0.81) & -0.52 (-0.90) & -1.13 (-1.99) & -1.23 (-1.73)\\
2 & -0.46 (-0.97) & -0.29 (-0.63) & -0.40 (-0.77) & -0.31 (-0.53) &  & 0.33 (0.80) & 0.18 (0.44) & -0.32 (-0.65) & -0.00 (-0.00)\\
3 & 0.33 (0.90) & 0.04 (0.10) & -0.19 (-0.41) & 0.13 (0.23) &  & 0.64 (2.00) & 0.51 (1.32) & -0.01 (-0.02) & 0.48 (0.97)\\
4 & 0.68 (2.07) & 0.35 (0.93) & 0.10 (0.24) & 0.46 (0.87) &  & 0.75 (2.56) & 0.71 (1.97) & 0.16 (0.38) & 0.58 (1.25)\\
5 & 1.01 (3.39) & 0.57 (1.64) & 0.38 (0.94) & 0.87 (1.69) &  & 0.90 (3.21) & 0.72 (2.03) & 0.20 (0.50) & 0.68 (1.53)\\
6 & 1.28 (4.37) & 0.92 (2.67) & 0.70 (1.82) & 1.30 (2.57) &  & 1.00 (3.77) & 0.91 (2.70) & 0.34 (0.96) & 1.04 (2.59)\\
7 & 1.46 (4.94) & 1.22 (3.38) & 0.88 (2.27) & 1.68 (3.27) &  & 1.17 (4.17) & 0.94 (2.57) & 0.47 (1.27) & 1.10 (2.61)\\
8 & 1.82 (5.50) & 1.51 (4.05) & 1.27 (3.05) & 2.30 (4.30) &  & 1.20 (3.77) & 1.13 (3.06) & 0.65 (1.74) & 1.26 (2.83)\\
9 & 2.30 (5.92) & 2.05 (5.18) & 1.63 (3.68) & 3.05 (5.02) &  & 1.44 (4.49) & 1.37 (3.50) & 0.91 (2.45) & 1.52 (3.18)\\
10 & 4.07 (6.72) & 4.23 (8.55) & 2.79 (5.44) & 6.36 (8.66) &  & 1.87 (5.03) & 1.50 (3.55) & 1.14 (2.89) & 1.88 (3.68)\\
\cmidrule{1-5} \cmidrule{7-10} \\ [-2ex]
10-1 & 6.44 (9.76) & 5.29 (12.36) & 4.03 (10.26) & 8.05 (18.56) &  & 2.35 (5.16) & 2.02 (5.51) & 2.27 (6.90) & 3.11 (7.89)\\
\cmidrule{1-5} \cmidrule{7-10} \\ [-2ex]
 \textbf{Median} \\
\cmidrule{1-5} \cmidrule{7-10} \\ [-2ex]
1 & -1.82 (-2.68) & -0.42 (-0.62) & -0.96 (-1.40) & -0.99 (-1.21) &  & -0.35 (-0.54) & -0.51 (-0.80) & -1.10 (-1.76) & -1.09 (-1.43)\\
2 & 0.16 (0.29) & 0.52 (1.00) & 0.08 (0.13) & 1.01 (1.42) &  & 0.34 (0.73) & 0.23 (0.51) & -0.33 (-0.63) & -0.10 (-0.18)\\
3 & 0.65 (1.48) & 0.29 (0.66) & 0.25 (0.53) & 1.28 (2.01) &  & 0.61 (1.75) & 0.50 (1.28) & -0.03 (-0.07) & 0.50 (1.01)\\
4 & 0.99 (2.68) & 0.50 (1.31) & 0.28 (0.64) & 1.14 (1.99) &  & 0.75 (2.44) & 0.69 (1.84) & 0.17 (0.40) & 0.48 (1.00)\\
5 & 1.19 (3.74) & 0.69 (1.99) & 0.36 (0.92) & 1.14 (2.23) &  & 0.83 (2.93) & 0.77 (2.19) & 0.29 (0.78) & 0.65 (1.46)\\
6 & 1.29 (4.45) & 0.91 (2.82) & 0.53 (1.44) & 1.08 (2.24) &  & 1.08 (3.91) & 0.91 (2.67) & 0.22 (0.66) & 0.93 (2.20)\\
7 & 1.44 (5.51) & 1.09 (3.38) & 0.82 (2.18) & 1.37 (2.95) &  & 1.13 (4.29) & 0.95 (2.76) & 0.46 (1.25) & 1.16 (2.81)\\
8 & 1.52 (5.26) & 1.36 (3.92) & 1.04 (2.72) & 1.82 (3.83) &  & 1.18 (4.07) & 1.03 (2.83) & 0.65 (1.87) & 1.39 (3.31)\\
9 & 1.80 (6.00) & 1.65 (4.58) & 1.34 (3.54) & 2.22 (4.40) &  & 1.41 (4.81) & 1.39 (3.80) & 0.92 (2.61) & 1.53 (3.53)\\
10 & 2.88 (6.97) & 2.94 (6.73) & 2.14 (4.62) & 4.08 (6.56) &  & 1.86 (5.61) & 1.49 (3.73) & 1.16 (3.17) & 1.86 (3.81)\\
\cmidrule{1-5} \cmidrule{7-10} \\ [-2ex]
10-1 & 4.70 (8.25) & 3.35 (6.64) & 3.10 (6.76) & 5.07 (10.76) &  & 2.21 (4.22) & 2.00 (4.65) & 2.25 (5.88) & 2.96 (6.27)\\
\cmidrule{1-5} \cmidrule{7-10} \\ [-2ex]
\textbf{Volatility}\\
\cmidrule{1-5} \cmidrule{7-10} \\ [-2ex]
1 & 1.02 (4.96) & 0.75 (3.23) & 0.25 (0.98) & 0.76 (2.50) &  & 0.88 (5.13) & 0.81 (3.54) & 0.33 (1.57) & 1.00 (3.14)\\
2 & 1.11 (4.28) & 0.83 (2.87) & 0.36 (1.18) & 1.11 (2.89) &  & 0.98 (4.19) & 0.84 (2.88) & 0.26 (0.90) & 0.95 (2.80)\\
3 & 1.08 (3.97) & 0.92 (2.85) & 0.49 (1.44) & 1.07 (2.40) &  & 1.02 (4.02) & 0.85 (2.84) & 0.32 (0.95) & 0.91 (2.22)\\
4 & 1.09 (3.82) & 0.85 (2.38) & 0.59 (1.55) & 0.98 (1.93) &  & 1.10 (4.09) & 0.82 (2.50) & 0.27 (0.80) & 0.97 (2.20)\\
5 & 1.05 (3.24) & 0.85 (2.30) & 0.60 (1.44) & 0.95 (1.73) &  & 0.98 (3.49) & 0.81 (2.16) & 0.50 (1.38) & 0.97 (2.16)\\
6 & 1.05 (2.89) & 0.66 (1.65) & 0.62 (1.39) & 0.96 (1.70) &  & 0.96 (3.25) & 0.82 (2.11) & 0.36 (0.95) & 0.79 (1.52)\\
7 & 1.05 (2.44) & 0.57 (1.25) & 0.65 (1.31) & 1.08 (1.72) &  & 0.91 (2.41) & 0.74 (1.79) & 0.26 (0.56) & 0.62 (1.24)\\
8 & 0.86 (1.65) & 0.37 (0.73) & 0.83 (1.48) & 1.00 (1.37) &  & 0.84 (1.89) & 0.84 (1.82) & 0.31 (0.64) & 0.71 (1.24)\\
9 & 0.74 (1.16) & 0.62 (1.04) & 0.89 (1.41) & 1.61 (1.94) &  & 0.72 (1.29) & 0.71 (1.33) & 0.13 (0.22) & 0.62 (0.90)\\
10 & 1.05 (1.33) & 3.12 (4.21) & 0.61 (0.85) & 4.63 (4.99) &  & 0.44 (0.58) & 0.23 (0.32) & -0.34 (-0.45) & -0.24 (-0.31)\\
\cmidrule{1-5} \cmidrule{7-10} \\ [-2ex]
10-1 & 0.03 (0.04) & 2.36 (3.35) & 0.35 (0.62) & 3.86 (5.10) &  & -0.44 (-0.60) & -0.57 (-0.96) & -0.68 (-1.05) & -1.24 (-1.81)\\
\cmidrule{1-5} \cmidrule{7-10} \\ [-2ex]
\textbf{Skewness}\\
\cmidrule{1-5} \cmidrule{7-10} \\ [-2ex]
1 & 0.85 (3.00) & 0.73 (2.07) & 0.31 (0.96) & 0.69 (1.53) &  & 0.61 (2.65) & 0.30 (0.90) & -0.06 (-0.19) & 0.44 (1.19)\\
2 & 1.08 (4.01) & 0.88 (2.27) & 0.37 (1.04) & 0.84 (1.76) &  & 0.75 (2.82) & 0.45 (1.20) & -0.04 (-0.11) & 0.46 (1.07)\\
3 & 1.05 (3.50) & 0.83 (1.99) & 0.48 (1.12) & 0.77 (1.50) &  & 0.77 (2.74) & 0.58 (1.66) & 0.06 (0.14) & 0.59 (1.38)\\
4 & 1.03 (3.13) & 0.61 (1.45) & 0.50 (1.08) & 0.67 (1.20) &  & 0.81 (2.93) & 0.79 (2.26) & 0.19 (0.48) & 0.70 (1.55)\\
5 & 1.04 (3.04) & 0.51 (1.21) & 0.55 (1.13) & 0.97 (1.70) &  & 0.84 (2.94) & 0.78 (2.18) & 0.20 (0.48) & 0.86 (1.85)\\
6 & 0.88 (2.51) & 0.59 (1.50) & 0.56 (1.12) & 1.31 (2.26) &  & 1.02 (3.23) & 0.84 (2.47) & 0.42 (1.03) & 0.74 (1.52)\\
7 & 0.90 (2.27) & 0.71 (1.74) & 0.67 (1.38) & 1.67 (2.70) &  & 1.11 (3.45) & 1.00 (2.79) & 0.38 (0.85) & 1.02 (2.15)\\
8 & 0.96 (2.18) & 0.99 (2.32) & 0.87 (1.74) & 2.20 (3.02) &  & 1.09 (2.95) & 0.97 (2.35) & 0.52 (1.20) & 1.01 (2.05)\\
9 & 1.23 (2.45) & 1.65 (3.83) & 0.98 (1.92) & 2.67 (3.65) &  & 1.03 (2.34) & 1.04 (2.26) & 0.54 (1.25) & 0.95 (1.54)\\
10 & 1.09 (1.92) & 2.03 (4.92) & 0.61 (1.30) & 2.37 (3.63) &  & 0.80 (1.10) & 0.72 (1.16) & 0.19 (0.35) & 0.54 (0.75)\\
\cmidrule{1-5} \cmidrule{7-10} \\ [-2ex]
10-1 & 0.24 (0.57) & 1.30 (4.07) & 0.30 (1.08) & 1.68 (4.06) &  & 0.18 (0.28) & 0.42 (1.02) & 0.25 (0.72) & 0.11 (0.20)\\
\cmidrule{1-5} \cmidrule{7-10} \\ [-2ex]
\textbf{Kurtosis}\\
\cmidrule{1-5} \cmidrule{7-10} \\ [-2ex]
1 & 1.08 (2.69) & 0.82 (1.90) & 0.48 (1.01) & 0.80 (1.46) &  & 0.76 (1.26) & 0.39 (0.61) & 0.03 (0.04) & 0.47 (0.70)\\
2 & 0.99 (3.32) & 0.93 (2.28) & 0.47 (1.26) & 0.68 (1.34) &  & 0.70 (1.44) & 0.67 (1.34) & -0.17 (-0.32) & 0.45 (0.70)\\
3 & 1.00 (3.35) & 0.76 (1.76) & 0.51 (1.25) & 0.69 (1.28) &  & 0.70 (1.85) & 0.63 (1.44) & 0.11 (0.24) & 0.34 (0.61)\\
4 & 1.05 (3.40) & 0.69 (1.62) & 0.49 (1.11) & 0.90 (1.58) &  & 0.80 (2.43) & 0.73 (1.98) & 0.23 (0.57) & 0.72 (1.57)\\
5 & 0.96 (2.76) & 0.54 (1.29) & 0.57 (1.19) & 1.07 (1.93) &  & 0.78 (2.49) & 0.90 (2.48) & 0.37 (0.99) & 0.76 (1.68)\\
6 & 0.99 (2.64) & 0.72 (1.68) & 0.70 (1.37) & 1.78 (2.83) &  & 0.91 (3.19) & 0.78 (2.27) & 0.43 (1.32) & 0.99 (2.34)\\
7 & 0.99 (2.27) & 0.84 (2.05) & 0.74 (1.51) & 1.94 (2.91) &  & 1.13 (4.45) & 0.90 (2.71) & 0.50 (1.46) & 1.00 (2.30)\\
8 & 1.02 (2.30) & 1.31 (3.09) & 0.91 (1.82) & 2.36 (3.28) &  & 1.07 (3.98) & 0.90 (2.67) & 0.37 (1.10) & 0.88 (2.12)\\
9 & 1.19 (2.42) & 1.49 (4.03) & 0.76 (1.62) & 2.41 (3.68) &  & 1.07 (3.77) & 0.94 (2.85) & 0.37 (1.07) & 1.02 (2.29)\\
10 & 0.83 (2.16) & 1.42 (4.54) & 0.27 (0.64) & 1.52 (2.95) &  & 0.92 (3.37) & 0.62 (1.90) & 0.16 (0.47) & 0.69 (1.63)\\
\cmidrule{1-5} \cmidrule{7-10} \\ [-2ex]
10-1 & -0.25 (-0.77) & 0.60 (1.82) & -0.21 (-0.66) & 0.73 (1.80) &  & 0.15 (0.33) & 0.23 (0.62) & 0.14 (0.33) & 0.22 (0.57)\\
\midrule \bottomrule
\end{tabular}
}
\end{table}

In addition to examining the role of individual $\tau$-quantiles and the mean, we also look at the decile portfolios and the corresponding long-short portfolios formed on the basis of the higher moments.  The higher moments are obtained by the quantiles-to-moments algorithm using the two-stage model quantile forecasts as described in \autoref{subsec:DerivingDistribution}. \autoref{tab:SingleSorts} shows the average monthly returns and Sharpe ratios of the equally weighted decile long-short portfolios based on the forecast mean, median, volatility, skewness and kurtosis. 

Overall, none of the higher moments appear to be monotonically priced in the cross-section of stock returns in the liquid or full sample. Average monthly returns of the long-short portfolios formed on the higher moments are mostly not significantly different from zero. There are some regional exceptions for the full sample, where volatility and skewness-based long-short portfolios have significantly positive average monthly returns in Europe and Asia Pacific. Results for the value-weighted long-short portfolios are mostly consistent with the equal-weighted portfolios and are shown in \autoref{tab:SingleSortsVW} in Appendix.

The relationship between predicted volatility and future returns is negative in the liquid sample but statistically insignificant, reflecting the overall ambiguity documented in prior studies on the pricing of volatility risk \citep{ang2006cross,bollerslev2020good}. This result is consistent with the focus on large-cap stocks, where volatility effects are typically weaker. In the full sample, microstructure-related distortions in equal-weighted portfolios further obscure the relationship, although the pattern becomes somewhat clearer with value weighting.

The results for skewness are inconclusive for a monotonic relationship with future returns. However, there is a noticeable hill-shaped effect; stocks with low skewness have lower future returns relative to stocks with more average skewness. The low returns of the high-skewness portfolio (10) lead to a non-monotonic relationship across the decile portfolios and thus also affects the 10-1 long-short portfolio. Skewness would be more convincingly priced if 9-1 long-short portfolios were used instead.

\clearpage
\section{Conclusion}

We develop a novel machine learning approach to forecasting the distribution of stock returns, focusing on both US and international markets. Our method uses a multi-head quantile neural network to predict a set of $\tau$ quantiles based on stock characteristics and market variables. We interpolate these quantiles using cubic B-splines to derive the distribution and its moments, such as mean, volatility, skewness and kurtosis. This approach outperforms traditional parametric models and simpler neural networks in forecasting stock returns, providing more accurate out-of-sample mean and variance forecasts.

We also examine the pricing of distributional characteristics in the cross-section of stock returns using a comprehensive dataset of US and international stocks. We find that only the central $\tau$-quantiles are priced, with significant profitability of long-short portfolios based on these predictions. In addition, our results show that the mean obtained by numerical integration of the interpolated distribution leads to higher out-of-sample returns than direct mean forecasts from other models.
We find no evidence that other moments, such as variance, skewness and kurtosis, are priced in the cross-section, adding to the ongoing debate in the asset pricing literature.

\bibliographystyle{chicago}
\bibliography{references}

\clearpage

\appendix

\setcounter{table}{0}

\begin{center}
    \Large \textbf{Appendix for} 
\end{center}
\begin{center}
    \Large
    ``Predicting the distributions of stock returns around the globe in the era of big data and learning''
\end{center}

\section{List of the anomalies}\label{appendix:anomalies}
\renewcommand{\thetable}{A\arabic{table}}
\counterwithin{table}{section} 
\counterwithin{figure}{section}

{\footnotesize
\begin{longtable}{p{.50\textwidth} p{.50\textwidth}}
\hline
\endfoot
\caption{ \bf{List of anomalies}}
\label{table14} \\
\toprule

\small \textbf{Fundamental}    &                                                  \\ 
\midrule
\textbf{Accruals}    &         \\ 
Accruals & \cite{sloan1996create}\\Change in Common Equity & \cite{richardson2006implications}\\Change in Current Operating Assets & \cite{richardson2006implications}\\Change in Current Operating Liabilities & \cite{richardson2006implications}\\Change in Financial Liabilities & \cite{richardson2006implications}\\Change in Long-Term Investments & \cite{richardson2006implications}\\Change in Net Financial Assets & \cite{richardson2006implications}\\Change in Net Non-Cash Working Capital & \cite{richardson2006implications}\\Change in Net Non-Current Operating Assets & \cite{richardson2006implications}\\Change in Non-Current Operating Assets & \cite{richardson2006implications}\\Change in Non-Current Operating Liabilities & \cite{richardson2006implications}\\Change in Short-Term Investments & \cite{richardson2006implications}\\Discretionary Accruals & \cite{dechow1995detecting}\\Growth in Inventory & \cite{thomas2002inventory}\\Inventory Change & \cite{thomas2002inventory}\\Inventory Growth & \cite{belo2011inventory}\\M/B and Accruals & \cite{bartov2004risk}\\Net Working Capital Changes & \cite{soliman2008use}\\Percent Operating Accrual & \cite{hafzalla2011percent}\\Percent Total Accrual & \cite{hafzalla2011percent}\\Total Accruals & \cite{richardson2006implications}\\
\textbf{Intangibles} &                                                   \\
$\bigtriangleup$ Gross Margin - $\bigtriangleup$ Sales & \cite{abarbanell1998abnormal}\\$\bigtriangleup$ Sales - $\bigtriangleup$ Accounts Receivable & \cite{abarbanell1998abnormal}\\$\bigtriangleup$ Sales - $\bigtriangleup$ Inventory & \cite{abarbanell1998abnormal}\\$\bigtriangleup$ Sales - $\bigtriangleup$ SG and A & \cite{abarbanell1998abnormal}\\Asset Liquidity & \cite{ortiz2014real}\\Asset Liquidity II & \cite{ortiz2014real}\\Cash-to-assets & \cite{palazzo2012cash}\\Earnings Conservatism & \cite{francis2004costs}\\Earnings Persistence & \cite{francis2004costs}\\Earnings Predictability & \cite{francis2004costs}\\Earnings Smoothness & \cite{francis2004costs}\\Earnings Timeliness & \cite{francis2004costs}\\Herfindahl Index & \cite{hou2006industry}\\Hiring rate & \cite{belo2014labor}\\Industry Concentration Assets & \cite{hou2006industry}\\Industry Concentration Book Equity & \cite{hou2006industry}\\Industry-adjusted Organizational Capital-to-Assets & \cite{eisfeldt2013organization}\\Industry-adjusted Real Estate Ratio & \cite{tuzel2010corporate}\\Org. Capital & \cite{eisfeldt2013organization}\\RD / Market Equity & \cite{chan2001stock}\\RD Capital-to-assets & \cite{li2011financial}\\RD Expenses-to-sales & \cite{chan2001stock}\\Tangibility & \cite{hahn2009financial}\\Unexpected RD Increases & \cite{eberhart2004examination}\\Whited-Wu Index & \cite{whited2006financial}\\
\textbf{Investment} &                                                     \\ 
$\bigtriangleup$ CAPEX - $\bigtriangleup$ Industry CAPEX & \cite{abarbanell1998abnormal}\\Asset Growth & \cite{cooper2008asset}\\Change Net Operating Assets & \cite{hirshleifer2004investors}\\Changes in PPE and Inventory-to-Assets & \cite{lyandres2007new}\\Composite Debt Issuance & \cite{lyandres2007new}\\Composite Equity Issuance (5-Year) & \cite{daniel2006market}\\Debt Issuance & \cite{spiess1995underperformance}\\Growth in LTNOA & \cite{fairfield2003accrued}\\Investment & \cite{titman2004capital}\\Net Debt Finance & \cite{bradshaw2006relation}\\Net Equity Finance & \cite{bradshaw2006relation}\\Net Operating Assets & \cite{hirshleifer2004investors}\\Noncurrent Operating Assets Changes & \cite{soliman2008use}\\Share Repurchases & \cite{ikenberry1995market}\\Total XFIN & \cite{bradshaw2006relation}\\
\textbf{Profitability} &                                                    \\ 
Asset Turnover & \cite{soliman2008use}\\Capital Turnover & \cite{haugen1996commonality}\\Cash-based Operating Profitability & \cite{ball2016accruals}\\Change in Asset Turnover & \cite{soliman2008use}\\Change in Profit Margin & \cite{soliman2008use}\\Earnings / Price & \cite{basu1977investment}\\Earnings Consistency & \cite{alwathainani2009consistency}\\F-Score & \cite{piotroski2000value}\\Gross Profitability & \cite{novy2013other}\\Labor Force Efficiency & \cite{abarbanell1998abnormal}\\Leverage & \cite{bhandari1988debt}\\O-Score (More Financial Distress) & \cite{dichev1998risk}\\Operating Profits to Assets & \cite{ball2016accruals}\\Operating Profits to Equity & \cite{FAMA20151}\\Profit Margin & \cite{soliman2008use}\\Return on Net Operating Assets & \cite{soliman2008use}\\Return-on-Equity & \cite{haugen1996commonality}\\Z-Score (Less Financial Distress) & \cite{dichev1998risk}\\
\textbf{Value} &                                                  \\
Assets-to-Market & \cite{fama1992cross}\\Book Equity / Market Equity & \cite{fama1992cross}\\Cash Flow / Market Equity & \cite{lakonishok1994contrarian}\\Duration of Equity & \cite{dechow2004implied}\\Enterprise Component of Book/Price & \cite{penman2007book}\\Enterprise Multiple & \cite{loughran2011new}\\Intangible Return & \cite{daniel2006market}\\Leverage Component of Book/Price & \cite{penman2007book}\\Net Payout Yield & \cite{boudoukh2007importance}\\Operating Leverage & \cite{novy2010operating}\\Payout Yield & \cite{boudoukh2007importance}\\Sales Growth & \cite{lakonishok1994contrarian}\\Sales/Price & \cite{barbee1996sales}\\Sustainable Growth & \cite{lockwood2010sustainable}\\
\midrule
\small  \textbf{Market Friction}    &                                                 \\ 
\midrule
11-Month Residual Momentum & \cite{blitz2011residual}\\52-Week High & \cite{george200452}\\Amihud's Measure (Illiquidity) & \cite{amihud2002illiquidity}\\Beta & \cite{fama1973risk}\\Betting against Beta & \cite{frazzini2014betting}\\Bid-Ask Spread & \cite{amihud1986asset}\\Cash Flow Variance & \cite{haugen1996commonality}\\Coefficient of Variation of Share Turnover & \cite{chordia2001trading}\\Coskewness & \cite{harvey2000conditional}\\Downside Beta & \cite{ang2006downside}\\Earnings Forecast-to-Price & \cite{elgers2001delayed}\\Firm Age & \cite{barry1984differential}\\Firm Age-Momentum & \cite{zhang2006information}\\Idiosyncratic Risk & \cite{ang2006cross}\\Industry Momentum & \cite{moskowitz1999industries}\\Lagged Momentum & \cite{novy2012momentum}\\Liquidity Beta 1 & \cite{acharya2005asset}\\Liquidity Beta 2 & \cite{acharya2005asset}\\Liquidity Beta 3 & \cite{acharya2005asset}\\Liquidity Beta 4 & \cite{acharya2005asset}\\Liquidity Beta 5 & \cite{acharya2005asset}\\Liquidity Shocks & \cite{bali2013liquidity}\\Long-Term Reversal & \cite{bondt1985does}\\Max & \cite{bali2011maxing}\\Momentum & \cite{jegadeesh1993returns}\\Momentum and LT Reversal & \cite{kot2006can}\\Momentum-Reversal & \cite{jegadeesh1993returns}\\Momentum-Volume & \cite{lee2000price}\\Price & \cite{blume1973price}\\Seasonality & \cite{heston2008seasonality}\\Seasonality 1 A & \cite{heston2008seasonality}\\Seasonality 1 N & \cite{heston2008seasonality}\\Seasonality 11-15 A & \cite{heston2008seasonality}\\Seasonality 11-15 N & \cite{heston2008seasonality}\\Seasonality 16-20 A & \cite{heston2008seasonality}\\Seasonality 16-20 N & \cite{heston2008seasonality}\\Seasonality 2-5 A & \cite{heston2008seasonality}\\Seasonality 2-5 N & \cite{heston2008seasonality}\\Seasonality 6-10 A & \cite{heston2008seasonality}\\Seasonality 6-10 N & \cite{heston2008seasonality}\\Share Issuance (1-Year) & \cite{pontiff2008share}\\Share Turnover & \cite{datar1998liquidity}\\Short-Term Reversal & \cite{jegadeesh1990evidence}\\Size & \cite{banz1981relationship}\\Tail Risk & \cite{kelly2014tail}\\Total Volatility & \cite{ang2006cross}\\Volume / Market Value of Equity & \cite{haugen1996commonality}\\Volume Trend & \cite{haugen1996commonality}\\Volume Variance & \cite{chordia2001trading}\\    
\midrule
\small  \textbf{I/B/E/S}    &                                               \\ 
\midrule
Analyst Value & \cite{frankel1998accounting}\\Analysts Coverage & \cite{elgers2001delayed}\\Change in Forecast + Accrual & \cite{barth2004analyst}\\Change in Recommendation & \cite{jegadeesh2004analyzing}\\Changes in Analyst Earnings Forecasts & \cite{hawkins1984earnings}\\Disparity between LT and ST Earnings Growth Forecasts & \cite{da2011disparity}\\Dispersion in Analyst LT Growth Forecasts & \cite{anderson2005heterogeneous}\\Down Forecast & \cite{barber2001can}\\Forecast Dispersion & \cite{diether2002differences}\\Long-Term Growth Forecasts & \cite{la1996expectations}\\Up Forecast & \cite{barber2001can}\\
\end{longtable}
}

\section{Alternative models} \label{sec:AppendixAlternativeModels}
\renewcommand{\thetable}{B\arabic{table}}
\counterwithin{table}{section} 
\counterwithin{figure}{section}

We describe the benchmark models used to evaluate the performance of our two-stage
model and the accuracy of our distributional moment forecasts. We compare the
forecasting performance of the two-stage model with the GARCH model and
three different neural network architectures, as outlined in \autoref{tab:avg_quantile_loss}.
Detailed descriptions of the neural networks and the GARCH model are provided in
Sections \ref{subsec:NeuralNetworks} and \ref{subsec:GARCH}, respectively. \\

\subsection{Feed-forward neural networks} \label{subsec:NeuralNetworks}
\renewcommand{\thetable}{B\arabic{table}}

We use multi-head feed-forward neural networks with varying architectures as benchmarks for the two-stage model 
quantile forecasts. We consider a simple multi-output linear model without any hidden layers, 
neural networks with 1 hidden layer, and neural networks with 2 hidden layers
as benchmarks for the two-stage model. Each model has 176 features and 37 outputs.
The input features are the same as those for the standardised $\tau$-quantiles network in 
the two-stage model. The 37 outputs are the $\tau$-quantile forecasts of raw future stock returns for the
same $\tau$ values as in the two-stage model. See \autoref{subsubsec:Features} for the feature specification and \autoref{subsubsec:TwoStage} for the main 
model introduction.

Each neural network is trained using the aggregated multi-$\tau$-quantile loss.

\begin{equation} \label{eq:QLaggr} 
\begin{aligned}
    \mathcal{L}^{\mathcal{T}}_{raw} = \frac{1}{B}\frac{1}{K}\sum_{\tau \in \mathcal{T}}\sum_{i=1}^{B}  \rho_{\tau}\left(r_{i,t} - \widehat{Q}_{r_{i,t}}(\tau)\right)
\end{aligned}
\end{equation}
where $\widehat{Q}_{r_{i,t}}(\tau)$ is the forecast $\tau$-quantile of the raw return of stock $i$ at period $t$, $B$ is the lot size and $K = 37$ is the number of $\tau$ in the $\mathcal{T}$ set.

We also use a single-head feed-forward neural network with 2 hidden layers and mean squared error loss function as a benchmark for 
mean forecasts comparison as in \autoref{tab:Mean forecast evaluation} and \autoref{tab:MSEvsMean}. This model also uses
the same 176 features as the networks above. Architecture schemas for all alternative neural network models are shown in \autoref{fig:NeuralNetworks}. \\

\begin{figure}[H]
\centering
\begin{subfigure}[b]{0.45\textwidth}
    \centering
    \begin{tikzpicture}[shorten >=1pt,draw=black!50, node distance=\layersep, scale=0.7] 
        \tikzstyle{every pin edge}=[<-,shorten <=1pt]
        \tikzstyle{neuron}=[circle, draw=none,
    fill=black!15, minimum size=17pt, inner sep=0pt]
        \tikzstyle{neuronin}=[circle, draw=none,
    fill=color1, minimum size=17pt, inner sep=0pt]
    \tikzstyle{neuronout}=[circle, draw=none,
    fill=color2, minimum size=17pt, inner sep=0pt]
        \tikzstyle{annot} = [text width=4em, text centered]

        \path[yshift=0.5cm]
            node[neuronin] (N1-L1-1) at (0cm, -1cm) {};
        \path[yshift=0.5cm]
            node (N1-L1-2) at (0cm,-2cm) {\vdots};
        \path[yshift=0.5cm]
            node[neuronin] (N1-L1-3)at (0cm, -3cm) {};
        \path[yshift=0.5cm]
            node (N1-L1-4) at (0cm,-4cm) {\vdots};
        \path[yshift=0.5cm]
            node[neuronin] (N1-L1-5) at (0cm, -5cm) {};

        \path[yshift=0.5cm]
            node[neuronout] (NQ-1) at (3cm, -1cm) {};
        \path[yshift=0.5cm]
            node (NQ-2) at (3cm,-2cm) {\vdots};
        \path[yshift=0.5cm]
            node[neuronout] (NQ-3)at (3cm, -3cm) {};
        \path[yshift=0.5cm]
            node (NQ-4) at (3cm,-4cm) {\vdots};
        \path[yshift=0.5cm]
            node[neuronout] (NQ-5) at (3cm, -5cm) {};

        \foreach \i in {1,3,5}
            \foreach \j in {1,3,5}
                \path (N1-L1-\i) edge (NQ-\j);

        \node[annot, above=0.5cm of N1-L1-1] (input-desc) {(176)};
        
        \node[annot, above=0.5cm of NQ-1] (output-desc) {\footnotesize{$Q_{r}(\tau)$} \\ \footnotesize{(37)}};

    \end{tikzpicture}
    \caption{\centering Linear model (LNN)}
    \label{fig:LinearModel}
\end{subfigure}
\hfill
\begin{subfigure}[b]{0.45\textwidth}
    \centering
    \begin{tikzpicture}[shorten >=1pt,draw=black!50, node distance=\layersep, scale=0.7] 
        \tikzstyle{every pin edge}=[<-,shorten <=1pt]
        \tikzstyle{neuron}=[circle, draw=none,
    fill=black!15, minimum size=17pt, inner sep=0pt]
        \tikzstyle{neuronin}=[circle, draw=none,
    fill=color1, minimum size=17pt, inner sep=0pt]
    \tikzstyle{neuronout}=[circle, draw=none,
    fill=color2, minimum size=17pt, inner sep=0pt]
        \tikzstyle{annot} = [text width=4em, text centered]

        \path[yshift=0.5cm]
            node[neuronin] (N1-L1-1) at (0cm, -1cm) {};
        \path[yshift=0.5cm]
            node (N1-L1-2) at (0cm,-2cm) {\vdots};
        \path[yshift=0.5cm]
            node[neuronin] (N1-L1-3)at (0cm, -3cm) {};
        \path[yshift=0.5cm]
            node (N1-L1-4) at (0cm,-4cm) {\vdots};
        \path[yshift=0.5cm]
            node[neuronin] (N1-L1-5) at (0cm, -5cm) {};

        \path[yshift=0.5cm]
            node[neuron] (Hidden-1) at (2cm, -1cm) {};
        \path[yshift=0.5cm]
            node (Hidden-2) at (2cm,-2cm) {\vdots};
        \path[yshift=0.5cm]
            node[neuron] (Hidden-3) at (2cm,-3cm) {};
        \path[yshift=0.5cm]
            node (Hidden-4) at (2cm,-4cm) {\vdots};
        \path[yshift=0.5cm]
            node[neuron] (Hidden-5) at (2cm, -5cm) {};

        \path[yshift=0.5cm]
            node[neuronout] (NQ-1) at (4cm, -1cm) {};
        \path[yshift=0.5cm]
            node (NQ-2) at (4cm,-2cm) {\vdots};
        \path[yshift=0.5cm]
            node[neuronout] (NQ-3)at (4cm, -3cm) {};
        \path[yshift=0.5cm]
            node (NQ-4) at (4cm,-4cm) {\vdots};
        \path[yshift=0.5cm]
            node[neuronout] (NQ-5) at (4cm, -5cm) {};

        \foreach \i in {1,3,5}
            \foreach \j in {1,3,5}
                \path (N1-L1-\i) edge (Hidden-\j);

        \foreach \i in {1,3,5}
            \foreach \j in {1,3,5}
                \path (Hidden-\i) edge (NQ-\j);

        \node[annot, above=0.5cm of N1-L1-1] (input-desc) {(176)};
        
        \node[annot, above=0.5cm of Hidden-1] (hidden-desc) {(128)};

        \node[annot, above=0.5cm of NQ-1] (output-desc) {\footnotesize{$Q_{r}(\tau)$} \\ \footnotesize{(37)}};

    \end{tikzpicture}
    \caption{One hidden layer neural network (1hNN)}
    \label{fig:OneHiddenLayer}
\end{subfigure}
\vskip\baselineskip
\begin{subfigure}[b]{0.45\textwidth}
    \centering
    \begin{tikzpicture}[shorten >=1pt,draw=black!50, node distance=\layersep, scale = 0.7] 
        \tikzstyle{every pin edge}=[<-,shorten <=1pt]
        \tikzstyle{neuron}=[circle, draw=none,
    fill=black!15, minimum size=17pt, inner sep=0pt]
        \tikzstyle{neuronin}=[circle, draw=none,
    fill=color1, minimum size=17pt, inner sep=0pt]
    \tikzstyle{neuronout}=[circle, draw=none,
    fill=color2, minimum size=17pt, inner sep=0pt]
        \tikzstyle{annot} = [text width=4em, text centered]

        \path[yshift=0.5cm]
            node[neuronin] (N1-L1-1) at (0cm, -1cm) {};
        \path[yshift=0.5cm]
            node (N1-L1-2) at (0cm,-2cm) {\vdots};
        \path[yshift=0.5cm]
            node[neuronin] (N1-L1-3)at (0cm, -3cm) {};
        \path[yshift=0.5cm]
            node (N1-L1-4) at (0cm,-4cm) {\vdots};
        \path[yshift=0.5cm]
            node[neuronin] (N1-L1-5) at (0cm, -5cm) {};

        \path[yshift=0.5cm]
            node[neuron] (Hidden1-1) at (2cm, -1cm) {};
        \path[yshift=0.5cm]
            node (Hidden1-2) at (2cm,-2cm) {\vdots};
        \path[yshift=0.5cm]
            node[neuron] (Hidden1-3) at (2cm,-3cm) {};
        \path[yshift=0.5cm]
            node (Hidden1-4) at (2cm,-4cm) {\vdots};
        \path[yshift=0.5cm]
            node[neuron] (Hidden1-5) at (2cm, -5cm) {};

        \path[yshift=0.5cm]
            node[neuron] (Hidden2-1) at (4cm, -1cm) {};
        \path[yshift=0.5cm]
            node (Hidden2-2) at (4cm,-2cm) {\vdots};
        \path[yshift=0.5cm]
            node[neuron] (Hidden2-3) at (4cm,-3cm) {};
        \path[yshift=0.5cm]
            node (Hidden2-4) at (4cm,-4cm) {\vdots};
        \path[yshift=0.5cm]
            node[neuron] (Hidden2-5) at (4cm, -5cm) {};

        \path[yshift=0.5cm]
            node[neuronout] (NQ-1) at (6cm, -1cm) {};
        \path[yshift=0.5cm]
            node (NQ-2) at (6cm,-2cm) {\vdots};
        \path[yshift=0.5cm]
            node[neuronout] (NQ-3)at (6cm, -3cm) {};
        \path[yshift=0.5cm]
            node (NQ-4) at (6cm,-4cm) {\vdots};
        \path[yshift=0.5cm]
            node[neuronout] (NQ-5) at (6cm, -5cm) {};

        \foreach \i in {1,3,5}
            \foreach \j in {1,3,5}
                \path (N1-L1-\i) edge (Hidden1-\j);

        \foreach \i in {1,3,5}
            \foreach \j in {1,3,5}
                \path (Hidden1-\i) edge (Hidden2-\j);

        \foreach \i in {1,3,5}
            \foreach \j in {1,3,5}
                \path (Hidden2-\i) edge (NQ-\j);

        \node[annot, above=0.5cm of N1-L1-1] (input-desc) {(176)};
        
        \node[annot, above=0.5cm of Hidden1-1] (hidden1-desc) {(128)};

        \node[annot, above=0.5cm of Hidden2-1] (hidden2-desc) {(128)};

        \node[annot, above=0.5cm of NQ-1] (output-desc) {\footnotesize{$Q_{r}(\tau)$} \\ \footnotesize{(37)}};

    \end{tikzpicture}
    \caption{ \centering Two hidden layers neural network (2hNN)
    \newline }
    \label{fig:TwoHiddenLayers}
\end{subfigure}
\begin{subfigure}[b]{0.45\textwidth}
    \centering
    \begin{tikzpicture}[shorten >=1pt,draw=black!50, node distance=\layersep, scale = 0.7] 
        \tikzstyle{every pin edge}=[<-,shorten <=1pt]
        \tikzstyle{neuron}=[circle, draw=none,
    fill=black!15, minimum size=17pt, inner sep=0pt]
         \tikzstyle{neuronin}=[circle, draw=none,
    fill=color1, minimum size=17pt, inner sep=0pt]
    \tikzstyle{neuronout}=[circle, draw=none,
    fill=color2, minimum size=17pt, inner sep=0pt]
        \tikzstyle{annot} = [text width=4em, text centered]

        \path[yshift=0.5cm]
            node[neuronin] (N1-L1-1) at (0cm, -1cm) {};
        \path[yshift=0.5cm]
            node (N1-L1-2) at (0cm,-2cm) {\vdots};
        \path[yshift=0.5cm]
            node[neuronin] (N1-L1-3)at (0cm, -3cm) {};
        \path[yshift=0.5cm]
            node (N1-L1-4) at (0cm,-4cm) {\vdots};
        \path[yshift=0.5cm]
            node[neuronin] (N1-L1-5) at (0cm, -5cm) {};

        \path[yshift=0.5cm]
            node[neuron] (Hidden1-1) at (2cm, -1cm) {};
        \path[yshift=0.5cm]
            node (Hidden1-2) at (2cm,-2cm) {\vdots};
        \path[yshift=0.5cm]
            node[neuron] (Hidden1-3) at (2cm,-3cm) {};
        \path[yshift=0.5cm]
            node (Hidden1-4) at (2cm,-4cm) {\vdots};
        \path[yshift=0.5cm]
            node[neuron] (Hidden1-5) at (2cm, -5cm) {};

        \path[yshift=0.5cm]
            node[neuron] (Hidden2-1) at (4cm, -1cm) {};
        \path[yshift=0.5cm]
            node (Hidden2-2) at (4cm,-2cm) {\vdots};
        \path[yshift=0.5cm]
            node[neuron] (Hidden2-3) at (4cm,-3cm) {};
        \path[yshift=0.5cm]
            node (Hidden2-4) at (4cm,-4cm) {\vdots};
        \path[yshift=0.5cm]
            node[neuron] (Hidden2-5) at (4cm, -5cm) {};

        \path[yshift=0.5cm]
            node[neuronout] (NQ-1) at (6cm, -3cm) {};

        \foreach \i in {1,3,5}
            \foreach \j in {1,3,5}
                \path (N1-L1-\i) edge (Hidden1-\j);

        \foreach \i in {1,3,5}
            \foreach \j in {1,3,5}
                \path (Hidden1-\i) edge (Hidden2-\j);

        \foreach \i in {1,3,5}
            \path (Hidden2-\i) edge (NQ-1);

        \node[annot, above=0.5cm of N1-L1-1] (input-desc) {(176)};
        
        \node[annot, above=0.5cm of Hidden1-1] (hidden1-desc) {(128)};

        \node[annot, above=0.5cm of Hidden2-1] (hidden2-desc) {(128)};

        \node[annot, above=0.5cm of NQ-1] (output-desc) {\footnotesize{$\mu$} \\ \footnotesize{(1)}};
    \end{tikzpicture}
    \caption{ \centering Single output neural network with two hidden layers (2hNN MSE)}
    \label{fig:TwoHiddenLayersSingleOutput}
\end{subfigure}

\caption{\textbf{Architectures of benchmark multi-head feed-forward neural networks.} \\
\autoref{fig:LinearModel} depicts the linear model (LNN),
\autoref{fig:OneHiddenLayer} illustrates the neural network with one hidden layer (1hNN),
\autoref{fig:TwoHiddenLayers} shows the neural network with two hidden layers (2hNN),
and \autoref{fig:TwoHiddenLayersSingleOutput} presents the two hidden layers neural network with a single output node (2hNN MSE),
which is used to generate mean forecasts directly using the mean squared error loss function.
All networks have 176 input features; multi-output networks have 37 outputs, varying in the number of hidden layers and neurons.
}
\label{fig:NeuralNetworks}
\end{figure}

Most of the hyper-parameters for all four benchmark neural networks are the same as for the two-stage model and are available in \autoref{FixedHyperparameters}. Dropout rate is set to 0.2 and learning rate to 0.0003, in line with the selected hyper-parameter for the two-stage model. Hyper-parameters that differ are the architecture in terms of the number of layers and neurons, the loss function, and other model-specific settings.

\subsection{GARCH benchmark} \label{subsec:GARCH}

In order to provide a valid benchmark for the two-stage model,
we examine various specifications of the GARCH model. We utilize daily returns from all regions for the period from 1995 to 2018, corresponding to the time period and stock universe used for the main results in this paper.
We base our estimations on daily returns using a rolling window of 36 months. 
Results are compared for both the full and liquid samples.

We consider several variants of GARCH(1,1) model with the standardised residuals distributed according to normal and t-distributions. Mean $\mu$ is either estimated or set equal to the risk-free rate at the end of the estimation window plus 5\% divided by 252. Assuming these distributions, we simulate 100,000 paths of residuals for 22 days ahead and consequently derive the corresponding return paths. During this process, volatility is simulated using \autoref{eq:GARCH_volatility}

\begin{equation} 
\sigma_{i,j}^2 = \omega + \alpha \epsilon_{i-1,j}^2 + \beta \sigma_{i-1,j}^2
\label{eq:GARCH_volatility}
\end{equation} 

where $\omega$, $\alpha$, and $\beta$ are the GARCH parameters, $\epsilon_{i-1,j}$ represents the previous day's residual,
and $\sigma_{i-1,j}^2$ is the previous day's volatility.
The cumulative return for the $j$-th simulated path is then calculated according to \autoref{eq:GARCH_cumulative_return}.

\begin{equation}
r_j = \left(\prod_{i=1}^{22} \left(1 + \mu + \epsilon_{i,j} \sqrt{\sigma_{i,j}^2}\right)\right) - 1
\label{eq:GARCH_cumulative_return}
\end{equation}

Finally, the variance forecast is taken as the sample standard deviation of all the cumulative returns over the 22 days, 
and the quantile forecast is obtained as the sample quantile of all the cumulative returns over the 22 days.

\begin{table}[H]
\caption{{\bf Performance of Various GARCH Specifications} This table shows 
out-of-sample average quantile losses for different specifications of the GARCH model for both the full and liquid samples.
The significance of the difference in average quantile losses between the GARCH(1,1) with Student's t-distribution and other specifications 
is captured by t-statistics adjusted for Newey-West standard errors with 12 lags.
All models, unless specified, use the mean as a risk-free rate at the end of the estimation window plus 5\%.
GARCH(1,1) t-dist uses a t-distribution with an estimated number of degrees of freedom, and GARCH(1,1) normal dist uses a normal distribution. 
GARCH(1,1) fixed, t-dist models have the number of degrees of freedom fixed at 3, 4, or 5, and their auto-regressive parameters are fixed to $\alpha = 0.06$ and $\beta = 0.94$.
Daily risk-free rate data are taken from the Fama-French data library.}
\centering
\resizebox{0.7\textwidth}{!}{
\begin{tabular}{llll}
\toprule \hline \\ [-2ex]
Specification & Variable & Full Sample & Liquid Sample\\\\
\hline \\ [-2ex]
GARCH(1, 1), t-dist & Avg Loss & 0.02642 & 0.01887\\
GARCH(1, 1) fixed, t-dist 3 df & Avg Loss & 0.02636 & 0.01906\\
 & t-stat & -0.96 & 4.97\\
GARCH(1, 1) fixed, t-dist 4 df & Avg Loss & 0.02641 & 0.01900\\
 & t-stat & -0.17 & 4.25\\
GARCH(1, 1) fixed, t-dist 5 df & Avg Loss & 0.02646 & 0.01901\\
 & t-stat & 1.06 & 4.53\\
GARCH(1, 1), normal dist & Avg Loss & 0.02681 & 0.01899\\
 & t-stat & 15.25 & 10.30\\
GARCH(1, 1), t-dist, $\mu$ estimated & Avg Loss & 0.02769 & 0.01930\\
 & t-stat & 5.22 & 4.23\\
GARCH(2, 2), t-dist & Avg Loss & 0.02649 & 0.01893\\
 & t-stat & 2.10 & 1.26\\
EGARCH(1, 1, 1), t-dist & Avg Loss & 0.02681 & 0.01913\\
 & t-stat & 3.38 & 2.25\\
\hline \bottomrule
\end{tabular}
}
\end{table} \label{tab:GARCH_comparison}

For the t-distribution, besides estimating the number of degrees of freedom from the data, 
we also consider fixing the number of degrees of freedom to 3, 4, and 5. 
Furthermore, we try GARCH(2, 2) and EGARCH(1, 1, 1) specifications also with t-distributed standardised residuals, in both cases estimating degrees of freedom from the data. 

Table \ref{tab:GARCH_comparison} shows the out-of-sample average quantile losses for the considered GARCH models for both the full sample and the liquid sample. 
The best performance for the full sample is achieved by the GARCH(1, 1) model with t-distributed standardised residuals fixed to 4 degrees of freedom and auto-regressive parameters fixed to $\alpha = 0.06$ and $\beta = 0.94$.\footnote{This specification of GARCH is widely used in practice and was coined by J.P Morgan in 1996.}
GARCH(1, 1) specification without fixing the parameters performs similarly and achieves the best result for the liquid sample. Based on these results, we choose the GARCH(1, 1) with t-distributed 
standardised residuals and estimated number of degrees of freedom to serve as the benchmark for the main empirical study in this paper.\footnote{Note that the estimation of GARCH process can fail or it can generate explosive forecasts. GARCH(1, 1) fixed, t-dist 4 df is used for these cases as it can always be computed. The results in \autoref{tab:GARCH_comparison} already reflect this for all the models.}  \\

A more recent method for quantile prediction is MIDAS quantile regression introduced in \cite{ghysels2016invest}. MIDAS quantile regressions, in their simplest form, predict quantile as $Q = \alpha + \beta f(\gamma)$. $\alpha$, $\beta$, and $\gamma$ are three parameters to be estimated via non-linear OLS. $f(\gamma)$ is a weighted average of past absolute daily returns where the parameter $\gamma$ determines the speed of weight decay for less recent returns. The simplest specification is quite similar to GARCH(1, 1) with its auto-regressive component. In contrast to GARCH, MIDAS quantile regression does not rely on any distributional assumptions, which, however, also means that it cannot be applied to the extreme quantiles being estimated in the present study. The range of quantities that can be estimated via MIDAS is limited by a number of observations for one stock. Three years of data for one stock allows, at most, an estimation of 0.0014 quantile, which is much larger than the minimum quantiles estimated by neural networks 0.00005.

\section{Additional empirical results} \label{sec:AppendixResults}
\counterwithin{table}{section} 
\counterwithin{figure}{section}
\renewcommand{\thetable}{C\arabic{table}}

\begin{table}[H] 
\caption{{\bf Average cross-sectional correlation between moments and median internationally.}
This table presents the average cross-sectional Spearman correlation between the forecasted first four
moments and the median of the stock returns for the full and liquid samples in Europe, Japan, and Asia Pacific.
Medians are forecasted using the two-stage model defined in \autoref{subsubsec:TwoStage}, and
moments are forecasted using the quantiles-to-moments algorithm defined in \autoref{subsec:DerivingDistribution}.
The predictions span the time period from 1995 to 2018.}

\resizebox{1\textwidth}{!}{
\begin{tabular}{lccccclccccc}

\toprule \hline \\ [-2ex]
& \multicolumn{11}{c}{Europe} \\
\hline \\ [-2ex]
& \multicolumn{5}{c}{Full Sample} && \multicolumn{5}{c}{Liquid Sample}\\
\cmidrule{2-6} \cmidrule{8-12}
{} &    Median &      Mean &  Variance &  Skewness &  Kurtosis & &   Median &      Mean &  Variance &  Skewness &  Kurtosis \\
\midrule
Median & 1.0 & 0.81 & -0.58 & -0.49 & -0.43 &  & 1.0 & 0.97 & -0.2 & 0.32 & 0.18\\
Mean & 0.81 & 1.0 & -0.25 & -0.13 & -0.16 &  & 0.97 & 1.0 & -0.05 & 0.46 & 0.11\\
Variance & -0.58 & -0.25 & 1.0 & 0.43 & 0.2 &  & -0.2 & -0.05 & 1.0 & 0.49 & -0.47\\
Skewness & -0.49 & -0.13 & 0.43 & 1.0 & 0.94 &  & 0.32 & 0.46 & 0.49 & 1.0 & -0.1\\
Kurtosis & -0.43 & -0.16 & 0.2 & 0.94 & 1.0 &  & 0.18 & 0.11 & -0.47 & -0.1 & 1.0\\
        
\hline \\ [-2ex]
& \multicolumn{11}{c}{Japan} \\
\hline \\ [-2ex]
& \multicolumn{5}{c}{Full Sample} && \multicolumn{5}{c}{Liquid Sample}\\
\cmidrule{2-6} \cmidrule{8-12}
{} &    Median &      Mean &  Variance &  Skewness &  Kurtosis & &   Median &      Mean &  Variance &  Skewness &  Kurtosis \\
\midrule
Median & 1.0 & 0.83 & -0.51 & -0.39 & -0.32 &  & 1.0 & 0.97 & -0.27 & 0.27 & 0.23\\
Mean & 0.83 & 1.0 & -0.19 & -0.02 & -0.04 &  & 0.97 & 1.0 & -0.12 & 0.42 & 0.17\\
Variance & -0.51 & -0.19 & 1.0 & 0.31 & 0.04 &  & -0.27 & -0.12 & 1.0 & 0.48 & -0.44\\
Skewness & -0.39 & -0.02 & 0.31 & 1.0 & 0.93 &  & 0.27 & 0.42 & 0.48 & 1.0 & 0.0\\
Kurtosis & -0.32 & -0.04 & 0.04 & 0.93 & 1.0 &  & 0.23 & 0.17 & -0.44 & 0.0 & 1.0\\
        
\hline \\ [-2ex]
& \multicolumn{11}{c}{Asia Pacific} \\
\hline \\ [-2ex]
& \multicolumn{5}{c}{Full Sample} && \multicolumn{5}{c}{Liquid Sample}\\
\cmidrule{2-6} \cmidrule{8-12}
{} &    Median &      Mean &  Variance &  Skewness &  Kurtosis & &   Median &      Mean &  Variance &  Skewness &  Kurtosis \\
\midrule
Median & 1.0 & 0.8 & -0.65 & -0.54 & -0.43 &  & 1.0 & 0.96 & -0.24 & 0.19 & 0.18\\
Mean & 0.8 & 1.0 & -0.29 & -0.18 & -0.16 &  & 0.96 & 1.0 & -0.07 & 0.36 & 0.1\\
Variance & -0.65 & -0.29 & 1.0 & 0.52 & 0.26 &  & -0.24 & -0.07 & 1.0 & 0.61 & -0.42\\
Skewness & -0.54 & -0.18 & 0.52 & 1.0 & 0.92 &  & 0.19 & 0.36 & 0.61 & 1.0 & -0.1\\
Kurtosis & -0.43 & -0.16 & 0.26 & 0.92 & 1.0 &  & 0.18 & 0.1 & -0.42 & -0.1 & 1.0\\
\hline \bottomrule
\end{tabular}
}
\label{tab:correlationsInt}
\end{table}

\begin{table}[H]
\caption{{\bf Quantiles and long-short portfolios internationally.}
This table displays average monthly returns and annualized Sharpe ratios (SR)
for decile long, short, and long-short equal-weighted portfolios over the period from 1995 to 2018.
For each $\tau$-quantile forecast, the long portfolio is formed by selecting the top 10\% of stocks with the 
highest forecasts each month,
the short portfolio by selecting the bottom 10\% of stocks with the lowest forecasts,
and the long-short portfolio by taking the difference between the long and short portfolios.
Values in brackets are t-statistics adjusted for Newey-West standard errors with 12 lags.
Results are obtained separately for full and liquid samples for Europe, Japan, and Asia Pacific.}
\centering
\resizebox{1\textwidth}{!}{
\begin{tabular}{l|cc|cc|cclcc|cc|cc}
\toprule \hline \\ [-2ex]
 & \multicolumn{13}{c}{Europe}\\
\hline \\ [-2ex]
 & \multicolumn{6}{c}{Full Sample} && \multicolumn{6}{c}{Liquid Sample}\\
\cmidrule{2-7} \cmidrule{9-14}
 & \multicolumn{2}{c}{Long} & \multicolumn{2}{c}{Short} & \multicolumn{2}{c}{Long-Short} & & \multicolumn{2}{c}{Long} & \multicolumn{2}{c}{Short} & \multicolumn{2}{c}{Long-Short}\\
$\tau$& Mean & SR & Mean & SR & Mean & SR && Mean & SR & Mean & SR & Mean & SR\\
\hline \\ [-2ex]
0.01 & 0.91 (3.92) & 1.04 & 2.57 (3.61) & 1.07 & -1.66 (-2.41) & -0.77 &  & 0.98 (4.06) & 0.99 & 0.06 (0.09) & 0.02 & 0.92 (1.51) & 0.39\\
0.05 & 0.86 (3.79) & 1.00 & 2.32 (3.22) & 0.96 & -1.46 (-2.10) & -0.66 &  & 1.00 (4.31) & 1.03 & -0.01 (-0.01) & -0.00 & 1.01 (1.63) & 0.42\\
0.1 & 0.88 (3.89) & 1.02 & 2.13 (2.94) & 0.88 & -1.25 (-1.80) & -0.57 &  & 1.03 (4.53) & 1.05 & -0.05 (-0.07) & -0.02 & 1.09 (1.79) & 0.46\\
0.2 & 0.95 (4.21) & 1.11 & 1.70 (2.37) & 0.71 & -0.75 (-1.10) & -0.35 &  & 1.09 (4.53) & 1.09 & -0.14 (-0.20) & -0.05 & 1.23 (2.03) & 0.55\\
0.3 & 1.08 (4.73) & 1.23 & 1.19 (1.67) & 0.51 & -0.10 (-0.16) & -0.05 &  & 1.22 (4.76) & 1.15 & -0.28 (-0.38) & -0.10 & 1.50 (2.54) & 0.70\\
0.4 & 1.69 (5.95) & 1.59 & 0.35 (0.50) & 0.16 & 1.35 (2.20) & 0.74 &  & 1.39 (4.76) & 1.12 & -0.33 (-0.47) & -0.13 & 1.71 (3.12) & 0.91\\
0.5 & 2.94 (6.73) & 1.80 & -0.42 (-0.62) & -0.20 & 3.35 (6.63) & 2.48 &  & 1.49 (3.73) & 0.88 & -0.50 (-0.80) & -0.22 & 2.00 (4.65) & 1.52\\
0.6 & 3.65 (6.89) & 1.82 & -0.67 (-1.17) & -0.37 & 4.32 (10.29) & 3.48 &  & 1.49 (3.05) & 0.70 & -0.31 (-0.65) & -0.19 & 1.80 (5.33) & 1.47\\
0.7 & 3.98 (6.39) & 1.70 & -0.57 (-1.69) & -0.51 & 4.55 (9.80) & 2.55 &  & 1.29 (2.21) & 0.52 & 0.23 (0.78) & 0.21 & 1.06 (2.63) & 0.59\\
0.8 & 4.12 (5.82) & 1.63 & 0.17 (0.74) & 0.20 & 3.95 (6.10) & 1.74 &  & 0.97 (1.51) & 0.36 & 0.54 (2.23) & 0.53 & 0.43 (0.85) & 0.20\\
0.9 & 3.70 (5.07) & 1.46 & 0.45 (1.93) & 0.54 & 3.25 (4.70) & 1.41 &  & 0.70 (1.01) & 0.25 & 0.64 (2.66) & 0.64 & 0.06 (0.11) & 0.03\\
0.95 & 3.54 (4.78) & 1.40 & 0.57 (2.46) & 0.66 & 2.98 (4.27) & 1.31 &  & 0.54 (0.76) & 0.19 & 0.67 (2.80) & 0.68 & -0.13 (-0.23) & -0.06\\
0.99 & 3.40 (4.69) & 1.39 & 0.75 (2.98) & 0.78 & 2.65 (3.97) & 1.22 &  & 0.42 (0.58) & 0.15 & 0.70 (2.96) & 0.71 & -0.28 (-0.48) & -0.12\\
\hline \\ [-2ex]
 & \multicolumn{13}{c}{Japan}\\
\hline \\ [-2ex]
 & \multicolumn{6}{c}{Full Sample} && \multicolumn{6}{c}{Liquid Sample}\\
\cmidrule{2-7} \cmidrule{9-14}
 & \multicolumn{2}{c}{Long} & \multicolumn{2}{c}{Short} & \multicolumn{2}{c}{Long-Short} & & \multicolumn{2}{c}{Long} & \multicolumn{2}{c}{Short} & \multicolumn{2}{c}{Long-Short}\\
$\tau$& Mean & SR & Mean & SR & Mean & SR && Mean & SR & Mean & SR & Mean & SR\\
\hline \\ [-2ex]
0.01 & 0.62 (2.24) & 0.61 & -0.04 (-0.06) & -0.01 & 0.66 (1.21) & 0.25 &  & 0.50 (2.44) & 0.43 & -0.57 (-0.75) & -0.19 & 1.07 (1.68) & 0.40\\
0.05 & 0.63 (2.22) & 0.62 & -0.16 (-0.22) & -0.05 & 0.79 (1.45) & 0.30 &  & 0.53 (2.62) & 0.46 & -0.60 (-0.78) & -0.20 & 1.13 (1.76) & 0.43\\
0.1 & 0.67 (2.31) & 0.65 & -0.25 (-0.35) & -0.08 & 0.92 (1.73) & 0.35 &  & 0.61 (2.91) & 0.53 & -0.63 (-0.84) & -0.21 & 1.24 (1.96) & 0.47\\
0.2 & 0.79 (2.63) & 0.76 & -0.39 (-0.55) & -0.13 & 1.18 (2.29) & 0.46 &  & 0.71 (3.32) & 0.60 & -0.64 (-0.85) & -0.22 & 1.35 (2.16) & 0.53\\
0.3 & 1.01 (3.27) & 0.93 & -0.53 (-0.75) & -0.18 & 1.54 (3.13) & 0.63 &  & 0.88 (3.76) & 0.71 & -0.77 (-1.04) & -0.26 & 1.65 (2.78) & 0.68\\
0.4 & 1.49 (4.33) & 1.22 & -0.72 (-1.03) & -0.25 & 2.21 (4.77) & 1.01 &  & 1.08 (4.04) & 0.76 & -0.88 (-1.24) & -0.31 & 1.96 (3.72) & 0.93\\
0.5 & 2.14 (4.62) & 1.29 & -0.96 (-1.40) & -0.35 & 3.10 (6.76) & 1.84 &  & 1.16 (3.17) & 0.67 & -1.10 (-1.76) & -0.44 & 2.25 (5.87) & 1.51\\
0.6 & 2.29 (4.09) & 1.08 & -1.16 (-1.92) & -0.50 & 3.45 (7.98) & 2.96 &  & 1.05 (2.12) & 0.50 & -0.98 (-2.07) & -0.53 & 2.03 (6.33) & 1.83\\
0.7 & 2.26 (3.58) & 0.90 & -1.03 (-2.25) & -0.70 & 3.28 (6.44) & 1.98 &  & 0.74 (1.19) & 0.30 & -0.31 (-0.99) & -0.24 & 1.05 (2.16) & 0.55\\
0.8 & 1.86 (2.69) & 0.66 & -0.44 (-1.30) & -0.40 & 2.30 (3.95) & 0.98 &  & 0.34 (0.48) & 0.12 & 0.02 (0.08) & 0.02 & 0.32 (0.55) & 0.13\\
0.9 & 1.34 (1.87) & 0.44 & -0.14 (-0.50) & -0.14 & 1.49 (2.49) & 0.57 &  & 0.08 (0.11) & 0.03 & 0.16 (0.69) & 0.14 & -0.08 (-0.14) & -0.03\\
0.95 & 1.11 (1.57) & 0.36 & -0.01 (-0.04) & -0.01 & 1.13 (1.94) & 0.42 &  & -0.04 (-0.05) & -0.01 & 0.20 (0.89) & 0.18 & -0.24 (-0.39) & -0.09\\
0.99 & 0.92 (1.33) & 0.30 & 0.18 (0.73) & 0.17 & 0.75 (1.34) & 0.29 &  & -0.15 (-0.20) & -0.05 & 0.24 (1.08) & 0.21 & -0.39 (-0.61) & -0.14\\
\hline \\ [-2ex]
 & \multicolumn{13}{c}{Asia Pacific}\\
\hline \\ [-2ex]
 & \multicolumn{6}{c}{Full Sample} && \multicolumn{6}{c}{Liquid Sample}\\
\cmidrule{2-7} \cmidrule{9-14}
 & \multicolumn{2}{c}{Long} & \multicolumn{2}{c}{Short} & \multicolumn{2}{c}{Long-Short} & & \multicolumn{2}{c}{Long} & \multicolumn{2}{c}{Short} & \multicolumn{2}{c}{Long-Short}\\
$\tau$& Mean & SR & Mean & SR & Mean & SR && Mean & SR & Mean & SR & Mean & SR\\
\hline \\ [-2ex]
0.01 & 1.01 (3.33) & 0.95 & 3.57 (4.15) & 1.11 & -2.56 (-3.63) & -0.99 &  & 1.18 (3.77) & 0.79 & -0.50 (-0.64) & -0.14 & 1.68 (2.36) & 0.54\\
0.05 & 1.02 (3.34) & 0.98 & 2.94 (3.42) & 0.92 & -1.92 (-2.82) & -0.75 &  & 1.18 (3.74) & 0.79 & -0.61 (-0.77) & -0.17 & 1.79 (2.51) & 0.58\\
0.1 & 1.05 (3.42) & 1.01 & 2.47 (2.87) & 0.78 & -1.41 (-2.11) & -0.56 &  & 1.23 (3.92) & 0.83 & -0.67 (-0.84) & -0.19 & 1.90 (2.66) & 0.62\\
0.2 & 1.10 (3.51) & 1.05 & 1.64 (1.90) & 0.53 & -0.54 (-0.82) & -0.22 &  & 1.31 (4.20) & 0.87 & -0.70 (-0.88) & -0.20 & 2.01 (2.86) & 0.68\\
0.3 & 1.28 (3.91) & 1.16 & 0.71 (0.84) & 0.23 & 0.57 (0.91) & 0.25 &  & 1.52 (4.56) & 0.96 & -0.83 (-1.04) & -0.24 & 2.35 (3.39) & 0.84\\
0.4 & 2.30 (5.37) & 1.55 & -0.18 (-0.21) & -0.06 & 2.48 (4.54) & 1.24 &  & 1.81 (5.39) & 1.07 & -0.90 (-1.15) & -0.27 & 2.71 (4.65) & 1.23\\
0.5 & 4.08 (6.56) & 1.75 & -0.99 (-1.21) & -0.35 & 5.07 (10.75) & 3.06 &  & 1.86 (3.81) & 0.76 & -1.09 (-1.43) & -0.36 & 2.95 (6.28) & 1.82\\
0.6 & 5.31 (6.81) & 1.74 & -1.38 (-1.82) & -0.55 & 6.69 (14.28) & 3.55 &  & 1.54 (2.62) & 0.51 & -0.59 (-1.04) & -0.27 & 2.13 (7.48) & 1.10\\
0.7 & 6.04 (7.17) & 1.75 & -1.33 (-2.43) & -0.82 & 7.37 (13.30) & 2.84 &  & 1.29 (2.05) & 0.39 & 0.43 (1.23) & 0.28 & 0.86 (2.02) & 0.31\\
0.8 & 6.02 (6.81) & 1.67 & -0.01 (-0.02) & -0.01 & 6.03 (8.91) & 1.97 &  & 0.83 (1.21) & 0.24 & 0.71 (2.16) & 0.49 & 0.12 (0.22) & 0.04\\
0.9 & 5.69 (6.24) & 1.58 & 0.41 (1.30) & 0.39 & 5.29 (7.41) & 1.75 &  & 0.48 (0.68) & 0.13 & 0.79 (2.46) & 0.55 & -0.30 (-0.51) & -0.09\\
0.95 & 5.41 (5.98) & 1.55 & 0.56 (1.88) & 0.53 & 4.85 (6.68) & 1.67 &  & 0.27 (0.36) & 0.07 & 0.85 (2.74) & 0.59 & -0.58 (-0.91) & -0.18\\
0.99 & 5.18 (5.54) & 1.52 & 0.71 (2.19) & 0.61 & 4.47 (5.83) & 1.62 &  & 0.17 (0.22) & 0.04 & 0.95 (3.06) & 0.65 & -0.78 (-1.17) & -0.24\\
\hline \bottomrule
\end{tabular}
}
\label{QuantilePortfoliosInternational}
\end{table}

\begin{table}[H]
\caption{{\bf Moments value-weighted decile portfolios.}
This table shows average monthly returns for individual decile (1-10) and long-short value-weighted portfolios
formed on the forecasted mean, median, volatility, skewness, and kurtosis. Results are reported for 
the full and liquid samples as well as individual regions, covering the period from 1995 to 2018.
Values in brackets are t-statistics adjusted for Newey-West standard errors with 12 lags.
}
\centering
\scriptsize
\resizebox{1\textwidth}{!}{
\begin{tabular}{lcccclcccc}
\toprule \hline \\ [-2ex]
 & \multicolumn{9}{c}{\textbf{Mean}}\\
\hline \\ [-2ex]
& \multicolumn{4}{c}{Full Sample} && \multicolumn{4}{c}{Liquid Sample}\\
\cmidrule{2-5} \cmidrule{7-10}
Decile & USA & Europe & Japan & Asia Pacific && USA & Europe & Japan & Asia Pacific \\
\hline \\ [-2ex]
1 & -1.84 (-2.37) & -2.18 (-3.20) & -1.46 (-2.31) & -3.27 (-4.24) &  & -0.65 (-1.00) & -0.40 (-0.88) & -0.80 (-1.49) & -0.88 (-1.37)\\
2 & -0.74 (-1.24) & -0.87 (-1.54) & -0.64 (-1.15) & -1.64 (-2.44) &  & 0.28 (0.75) & 0.37 (0.96) & -0.24 (-0.47) & 0.17 (0.34)\\
3 & 0.10 (0.22) & -0.40 (-0.80) & -0.37 (-0.77) & -1.08 (-1.65) &  & 0.60 (2.01) & 0.44 (1.13) & -0.09 (-0.21) & 0.34 (0.67)\\
4 & 0.49 (1.42) & 0.48 (1.17) & -0.08 (-0.19) & -0.59 (-0.95) &  & 0.57 (1.96) & 0.66 (1.92) & 0.09 (0.22) & 0.50 (1.21)\\
5 & 0.74 (2.22) & 0.73 (1.91) & 0.28 (0.68) & 0.23 (0.46) &  & 0.86 (3.26) & 0.69 (2.02) & 0.19 (0.48) & 0.72 (1.74)\\
6 & 0.89 (2.84) & 0.94 (2.74) & 0.31 (0.76) & 0.73 (1.41) &  & 0.91 (3.45) & 0.75 (2.17) & 0.31 (0.90) & 1.10 (2.78)\\
7 & 1.28 (3.91) & 1.25 (3.43) & 0.62 (1.52) & 1.30 (2.45) &  & 1.05 (3.71) & 0.83 (2.43) & 0.45 (1.26) & 1.26 (3.87)\\
8 & 1.38 (4.07) & 1.46 (3.84) & 0.91 (2.31) & 1.68 (3.34) &  & 1.14 (3.83) & 1.09 (3.22) & 0.57 (1.55) & 1.32 (3.25)\\
9 & 1.84 (4.51) & 1.89 (4.35) & 1.24 (2.59) & 2.16 (3.91) &  & 1.17 (3.51) & 1.27 (3.17) & 1.08 (3.09) & 1.50 (3.73)\\
10 & 2.46 (5.05) & 2.36 (4.86) & 1.88 (3.46) & 3.50 (5.42) &  & 2.11 (5.32) & 1.45 (3.51) & 1.10 (2.50) & 1.85 (4.01)\\
10-1 & 4.30 (6.63) & 4.54 (7.38) & 3.34 (8.10) & 6.77 (15.25) &  & 2.76 (5.02) & 1.85 (5.26) & 1.90 (6.07) & 2.73 (6.18)\\
\hline \\ [-2ex]
 & \multicolumn{9}{c}{\textbf{Volatility}}\\
\hline \\ [-2ex]
& \multicolumn{4}{c}{Full Sample} && \multicolumn{4}{c}{Liquid Sample}\\
\cmidrule{2-5} \cmidrule{7-10}
Decile & USA & Europe & Japan & Asia Pacific && USA & Europe & Japan & Asia Pacific \\
\hline \\ [-2ex]
1 & 0.84 (4.03) & 0.78 (3.24) & 0.29 (1.41) & 0.83 (3.03) &  & 0.81 (4.01) & 0.72 (3.17) & 0.23 (0.97) & 0.69 (2.50)\\
2 & 0.97 (3.61) & 0.72 (2.45) & 0.22 (0.73) & 1.03 (2.46) &  & 0.90 (3.75) & 0.73 (2.47) & 0.22 (0.68) & 0.97 (2.50)\\
3 & 0.93 (3.03) & 0.74 (2.25) & 0.28 (0.77) & 1.15 (2.60) &  & 0.87 (3.18) & 0.75 (2.37) & 0.25 (0.57) & 1.05 (2.47)\\
4 & 1.15 (3.44) & 0.71 (1.85) & 0.28 (0.75) & 0.84 (1.55) &  & 1.02 (3.20) & 0.71 (1.94) & 0.04 (0.09) & 1.24 (2.51)\\
5 & 0.83 (1.88) & 0.77 (1.77) & 0.20 (0.53) & 0.86 (1.56) &  & 0.98 (3.00) & 0.66 (1.61) & 0.41 (1.18) & 1.02 (2.12)\\
6 & 1.10 (2.24) & 0.74 (1.50) & 0.32 (0.65) & 0.81 (1.28) &  & 0.97 (2.38) & 0.67 (1.66) & 0.05 (0.13) & 0.73 (1.44)\\
7 & 0.83 (1.56) & 0.80 (1.46) & 0.41 (0.76) & 0.21 (0.30) &  & 0.80 (1.67) & 0.54 (1.19) & 0.13 (0.25) & 0.58 (1.04)\\
8 & 0.71 (1.09) & 0.01 (0.01) & 0.41 (0.69) & -0.01 (-0.02) &  & 1.00 (1.83) & 0.58 (1.08) & 0.46 (0.87) & 0.79 (1.27)\\
9 & 0.64 (0.81) & 0.05 (0.07) & 0.66 (0.82) & 0.06 (0.08) &  & 0.69 (1.13) & 0.81 (1.36) & 0.53 (0.74) & 0.78 (0.99)\\
10 & 0.16 (0.18) & -0.87 (-0.93) & -0.07 (-0.08) & 1.07 (1.04) &  & 0.32 (0.38) & 0.13 (0.18) & -0.22 (-0.25) & -0.04 (-0.05)\\
10-1 & -0.67 (-0.76) & -1.65 (-1.93) & -0.35 (-0.44) & 0.24 (0.28) &  & -0.49 (-0.63) & -0.59 (-1.00) & -0.45 (-0.59) & -0.73 (-1.14)\\
\hline \\ [-2ex]
 & \multicolumn{9}{c}{\textbf{Skewness}}\\
\hline \\ [-2ex]
& \multicolumn{4}{c}{Full Sample} && \multicolumn{4}{c}{Liquid Sample}\\
\cmidrule{2-5} \cmidrule{7-10}
Decile & USA & Europe & Japan & Asia Pacific && USA & Europe & Japan & Asia Pacific \\
\hline \\ [-2ex]
1 & 0.79 (2.56) & 0.65 (1.90) & 0.22 (0.63) & 0.78 (1.88) &  & 0.55 (2.36) & 0.33 (1.05) & 0.06 (0.19) & 0.78 (2.37)\\
2 & 1.01 (4.11) & 1.05 (3.06) & 0.35 (0.88) & 1.13 (3.31) &  & 0.72 (3.02) & 0.44 (1.16) & -0.23 (-0.59) & 0.59 (1.48)\\
3 & 0.96 (2.95) & 0.97 (2.81) & 0.39 (0.82) & 0.82 (1.74) &  & 0.80 (2.88) & 0.61 (1.87) & 0.14 (0.33) & 0.88 (1.98)\\
4 & 0.81 (1.85) & 0.48 (1.37) & 0.19 (0.38) & 0.70 (1.32) &  & 0.77 (2.58) & 0.74 (2.22) & 0.23 (0.53) & 0.87 (1.92)\\
5 & 0.82 (1.70) & 0.60 (1.37) & 0.50 (0.96) & 0.40 (0.72) &  & 0.86 (2.62) & 0.78 (2.09) & 0.32 (0.76) & 1.14 (2.86)\\
6 & 0.80 (1.56) & 0.50 (0.99) & 0.08 (0.15) & 0.79 (1.31) &  & 0.99 (2.63) & 0.72 (2.29) & 0.61 (1.44) & 1.09 (2.41)\\
7 & 0.95 (1.68) & 0.11 (0.20) & 0.40 (0.74) & 1.90 (1.22) &  & 1.15 (2.98) & 0.88 (2.23) & 0.56 (1.31) & 1.26 (2.46)\\
8 & 0.81 (1.66) & 0.50 (1.07) & 0.39 (0.74) & 0.46 (0.70) &  & 1.18 (3.03) & 0.78 (1.95) & 0.74 (1.71) & 1.18 (2.62)\\
9 & 0.85 (1.61) & -0.39 (-0.70) & 0.42 (0.74) & 0.32 (0.50) &  & 1.31 (2.60) & 1.10 (2.63) & 0.60 (1.23) & 1.20 (2.08)\\
10 & 0.89 (1.73) & -0.26 (-0.50) & -0.10 (-0.21) & 0.35 (0.52) &  & 0.75 (0.96) & 0.71 (1.14) & 0.31 (0.53) & 0.83 (1.23)\\
10-1 & 0.10 (0.23) & -0.91 (-2.32) & -0.32 (-0.89) & -0.44 (-0.95) &  & 0.19 (0.27) & 0.38 (0.78) & 0.25 (0.60) & 0.05 (0.09)\\
\hline \\ [-2ex]
 & \multicolumn{9}{c}{\textbf{Kurtosis}}\\
\hline \\ [-2ex]
& \multicolumn{4}{c}{Full Sample} && \multicolumn{4}{c}{Liquid Sample}\\
\cmidrule{2-5} \cmidrule{7-10}
Decile & USA & Europe & Japan & Asia Pacific && USA & Europe & Japan & Asia Pacific \\
\hline \\ [-2ex]
1 & 1.05 (2.54) & 0.77 (2.08) & 0.44 (0.95) & 0.85 (1.92) &  & 0.50 (0.90) & 0.38 (0.61) & 0.11 (0.15) & 0.57 (0.92)\\
2 & 0.92 (3.29) & 0.80 (2.31) & 0.28 (0.84) & 0.88 (2.18) &  & 0.80 (1.97) & 0.71 (1.81) & 0.38 (0.85) & 0.85 (1.13)\\
3 & 0.74 (2.70) & 0.85 (2.86) & 0.25 (0.75) & 0.71 (1.99) &  & 0.81 (2.99) & 0.48 (1.24) & 0.30 (0.66) & 0.71 (1.78)\\
4 & 0.82 (3.10) & 0.68 (2.38) & 0.06 (0.16) & 0.66 (1.58) &  & 0.75 (2.67) & 0.60 (1.68) & 0.12 (0.32) & 1.03 (2.56)\\
5 & 0.80 (2.54) & 0.61 (1.52) & 0.26 (0.64) & 0.50 (1.17) &  & 0.93 (3.30) & 0.69 (1.92) & 0.06 (0.15) & 0.60 (1.34)\\
6 & 0.83 (2.48) & 0.42 (1.11) & 0.37 (0.74) & 1.10 (1.93) &  & 0.91 (3.12) & 0.76 (2.32) & 0.26 (0.78) & 0.85 (2.13)\\
7 & 0.77 (2.05) & 0.82 (1.89) & 0.28 (0.67) & 0.87 (1.69) &  & 0.90 (3.01) & 0.71 (2.02) & 0.14 (0.39) & 1.01 (2.49)\\
8 & 0.93 (2.70) & 0.16 (0.56) & 0.35 (0.80) & 0.79 (1.57) &  & 1.01 (3.02) & 0.76 (2.27) & 0.18 (0.46) & 0.82 (1.95)\\
9 & 1.05 (3.06) & -0.07 (-0.18) & 0.23 (0.50) & 0.75 (1.39) &  & 1.11 (3.47) & 0.81 (2.42) & 0.40 (1.01) & 1.21 (2.79)\\
10 & 0.94 (4.72) & -0.24 (-0.62) & -0.19 (-0.53) & 0.51 (0.46) &  & 0.80 (2.45) & 0.55 (1.62) & 0.22 (0.57) & 1.02 (2.20)\\
10-1 & -0.11 (-0.31) & -1.01 (-2.72) & -0.62 (-1.68) & -0.34 (-0.29) &  & 0.29 (0.76) & 0.17 (0.44) & 0.11 (0.24) & 0.46 (1.40)\\
\hline \bottomrule
\end{tabular}
}
\label{tab:SingleSortsVW}
\end{table}

\begin{table}[H]
\caption{
{\bf Quantiles and the cross-section of stock returns in the U.S.: 2hNN benchmark}
This table shows the average and volatility of monthly returns and annualised Sharpe ratios 
for decile long, short and long-short equal-weighted portfolios over the period 1995 to 2018. 
For each $\tau$-quantile forecast, the long portfolio is formed by taking the top 10\% of stocks with the highest forecast each month. The quantile forecast are from neural network with two hidden layers (2hNN). The short portfolio is formed by taking the bottom 10\% of stocks with the lowest forecast each month. The long-short portfolio is created by taking the difference between the long and short portfolios. Values in brackets are t-statistics adjusted for Newey-West standard errors with 12 lags. Results are obtained separately for the full and liquid samples for the US only.} \label{tab:QuantilePortfoliosBenchmark}
\centering
\resizebox{0.8\textwidth}{!}{
\begin{tabular}{l|ccc|ccc|ccc}
\toprule
 & \multicolumn{9}{c}{Full Sample} \\
\cmidrule{2-10}
 & \multicolumn{3}{c}{Long} & \multicolumn{3}{c}{Short} & \multicolumn{3}{c}{Long-short} \\
\cmidrule(lr){2-4} \cmidrule(lr){5-7} \cmidrule(lr){8-10}
$\tau$ & Mean & Vol & SR & Mean & Vol & SR & Mean & Vol & SR \\
\midrule
0.01 & 1.16 (4.91) & 2.61 & 1.54 & 0.26 (0.35) & 12.87 & 0.07 & 0.89 (1.23) & 12.33 & 0.25 \\
0.05 & 1.22 (5.25) & 2.52 & 1.68 & -0.03 (-0.04) & 10.39 & -0.01 & 1.25 (1.73) & 12.03 & 0.36 \\
0.1 & 1.27 (5.52) & 2.50 & 1.76 & -0.22 (-0.30) & 12.70 & -0.06 & 1.50 (2.05) & 11.81 & 0.44 \\
0.2 & 1.37 (6.02) & 2.52 & 1.88 & -0.56 (-0.75) & 12.12 & -0.16 & 1.94 (2.60) & 11.39 & 0.59 \\
0.3 & 1.58 (6.80) & 2.76 & 1.98 & -0.96 (-1.25) & 11.88 & -0.28 & 2.53 (3.31) & 10.82 & 0.81 \\
0.4 & 2.01 (7.51) & 3.55 & 1.96 & -1.37 (-1.81) & 11.04 & -0.43 & 3.37 (4.49) & 9.65 & 1.21 \\
0.5 & 2.73 (7.11) & 5.63 & 1.68 & -1.66 (-2.27) & 10.27 & -0.56 & 4.39 (6.59) & 6.98 & 2.18 \\
0.6 & 3.11 (6.06) & 7.98 & 1.35 & -1.79 (-2.80) & 8.61 & -0.72 & 4.90 (9.48) & 4.66 & 3.64 \\
0.7 & 3.01 (4.81) & 10.22 & 1.02 & -0.74 (-2.14) & 4.01 & -0.64 & 3.75 (6.33) & 7.51 & 1.73 \\
0.8 & 2.51 (3.37) & 12.08 & 0.72 & 0.58 (2.18) & 2.51 & 0.80 & 1.93 (2.58) & 11.14 & 0.60 \\
0.9 & 1.90 (2.35) & 13.16 & 0.50 & 0.81 (3.57) & 2.38 & 1.18 & 1.09 (1.35) & 12.59 & 0.30 \\
0.95 & 1.53 (1.90) & 13.25 & 0.40 & 0.89 (4.18) & 2.47 & 1.25 & 0.64 (0.80) & 12.32 & 0.18 \\
0.99 & 1.23 (1.57) & 12.91 & 0.33 & 0.95 (4.68) & 2.74 & 1.20 & 0.28 (0.37) & 12.12 & 0.08 \\
\midrule
 & \multicolumn{9}{c}{Liquid Sample} \\
\cmidrule{2-10}
 & \multicolumn{3}{c}{Long} & \multicolumn{3}{c}{Short} & \multicolumn{3}{c}{Long-short} \\
\cmidrule(lr){2-4} \cmidrule(lr){5-7} \cmidrule(lr){8-10}
$\tau$ & Mean & Vol & SR & Mean & Vol & SR & Mean & Vol & SR \\
\midrule
0.01 & 0.95 (5.25) & 2.96 & 1.11 & 0.19 (0.25) & 13.16 & 0.05 & 0.76 (1.03) & 11.45 & 0.23 \\
0.05 & 0.95 (5.37) & 2.91 & 1.13 & 0.12 (0.15) & 13.86 & 0.03 & 0.84 (1.14) & 11.64 & 0.25 \\
0.1 & 0.99 (5.60) & 2.96 & 1.16 & 0.10 (0.14) & 11.55 & 0.03 & 0.88 (1.20) & 11.29 & 0.27 \\
0.2 & 1.10 (6.18) & 3.05 & 1.25 & 0.04 (0.06) & 13.86 & 0.01 & 1.06 (1.48) & 11.13 & 0.33 \\
0.3 & 1.25 (6.49) & 3.36 & 1.29 & -0.01 (-0.02) & 0.00 & -0.00 & 1.26 (1.85) & 10.39 & 0.42 \\
0.4 & 1.42 (5.97) & 4.10 & 1.20 & -0.06 (-0.08) & 10.39 & -0.02 & 1.47 (2.39) & 8.93 & 0.57 \\
0.5 & 1.68 (4.81) & 6.40 & 0.91 & -0.10 (-0.17) & 8.66 & -0.04 & 1.78 (4.59) & 6.42 & 0.96 \\
0.6 & 1.54 (3.43) & 8.60 & 0.62 & 0.11 (0.29) & 6.35 & 0.06 & 1.42 (5.92) & 6.65 & 0.74 \\
0.7 & 1.24 (2.21) & 10.23 & 0.42 & 0.46 (2.29) & 3.39 & 0.47 & 0.78 (1.63) & 9.01 & 0.30 \\
0.8 & 1.03 (1.54) & 11.51 & 0.31 & 0.65 (3.41) & 2.92 & 0.77 & 0.38 (0.60) & 10.97 & 0.12 \\
0.9 & 0.77 (1.07) & 12.12 & 0.22 & 0.74 (3.98) & 2.88 & 0.89 & 0.03 (0.04) & 10.39 & 0.01 \\
0.95 & 0.61 (0.81) & 12.43 & 0.17 & 0.75 (3.94) & 2.89 & 0.90 & -0.14 (-0.19) & 12.12 & -0.04 \\
0.99 & 0.50 (0.67) & 12.37 & 0.14 & 0.80 (4.23) & 2.92 & 0.95 & -0.30 (-0.41) & 11.55 & -0.09 \\
\bottomrule
\end{tabular}
}
\end{table}

\begin{table}
\caption{{\bf Moments decile portfolios: 2hNN benchmark.}
This table shows average monthly returns for individual deciles (1-10) and long-short equal-weighted portfolios built at the predicted moments. The quantile forecast are from neural network with two hidden layers (2hNN). Results are reported for the full and liquid samples, as well as for individual regions, covering the period from 1995 to 2018. Values in parentheses are t-statistics adjusted for Newey-West standard errors with 12 lags.}
\label{tab:SingleSorts2hNN}

\centering
\scriptsize
\resizebox{1\textwidth}{!}{
\begin{tabular}{lcccclcccc}
\toprule \hline \\ [-2ex]
\textbf{Mean} & \multicolumn{4}{c}{Full Sample} && \multicolumn{4}{c}{Liquid Sample}\\
\cmidrule{2-5} \cmidrule{7-10}
Decile & USA & Europe & Japan & Asia Pacific && USA & Europe & Japan & Asia Pacific \\
\cmidrule{2-5} \cmidrule{7-10} \\ [-2ex]
1 & -1.98 (-3.18) & -0.52 (-0.88) & -1.07 (-1.64) & -1.19 (-1.57) &  & -0.01 (-0.01) & -0.41 (-0.77) & -0.77 (-1.45) & -0.67 (-1.07)\\
2 & -0.13 (-0.29) & -0.16 (-0.33) & -0.18 (-0.35) & -0.05 (-0.08) &  & 0.52 (1.41) & 0.27 (0.66) & -0.28 (-0.61) & 0.26 (0.58)\\
3 & 0.51 (1.39) & 0.07 (0.15) & -0.01 (-0.03) & 0.39 (0.70) &  & 0.54 (1.73) & 0.43 (1.09) & -0.14 (-0.32) & 0.47 (1.10)\\
4 & 0.76 (2.40) & 0.36 (0.95) & 0.17 (0.41) & 0.74 (1.39) &  & 0.74 (2.49) & 0.63 (1.71) & 0.10 (0.25) & 0.72 (1.58)\\
5 & 0.93 (3.18) & 0.67 (1.92) & 0.38 (0.94) & 0.92 (1.91) &  & 0.95 (3.44) & 0.78 (2.28) & 0.31 (0.80) & 0.81 (1.86)\\
6 & 1.23 (4.32) & 0.95 (2.80) & 0.60 (1.61) & 1.36 (2.95) &  & 0.97 (3.55) & 1.03 (2.94) & 0.33 (0.87) & 1.02 (2.45)\\
7 & 1.38 (4.51) & 1.18 (3.38) & 0.89 (2.29) & 1.78 (3.51) &  & 1.11 (3.77) & 0.97 (2.71) & 0.57 (1.50) & 1.09 (2.37)\\
8 & 1.73 (5.20) & 1.47 (4.09) & 1.16 (2.96) & 2.23 (4.13) &  & 1.06 (3.25) & 1.06 (2.93) & 0.62 (1.75) & 1.01 (1.99)\\
9 & 2.08 (5.77) & 1.85 (4.96) & 1.58 (3.57) & 2.74 (4.60) &  & 1.29 (3.60) & 1.26 (3.22) & 0.74 (1.94) & 1.06 (1.97)\\
10 & 3.60 (6.70) & 3.65 (7.86) & 2.37 (4.94) & 5.24 (7.26) &  & 1.66 (4.00) & 1.45 (3.23) & 0.92 (2.08) & 1.54 (2.62)\\
\cmidrule{1-5} \cmidrule{7-10} \\ [-2ex]
10-1 & 5.59 (10.48) & 4.17 (10.22) & 3.44 (9.24) & 6.43 (12.91) &  & 1.67 (5.79) & 1.86 (7.77) & 1.68 (5.78) & 2.22 (5.83)\\
\cmidrule{1-5} \cmidrule{7-10} \\ [-2ex]
\textbf{Volatility}\\
\cmidrule{1-5} \cmidrule{7-10} \\ [-2ex]
1 & 1.02 (4.71) & 0.74 (3.06) & 0.22 (0.85) & 0.77 (2.58) &  & 0.87 (4.77) & 0.80 (3.46) & 0.31 (1.41) & 0.94 (3.01)\\
2 & 1.07 (4.12) & 0.85 (2.95) & 0.34 (1.06) & 0.98 (2.49) &  & 0.93 (4.00) & 0.84 (2.92) & 0.24 (0.77) & 0.91 (2.51)\\
3 & 1.06 (3.93) & 0.86 (2.74) & 0.46 (1.33) & 1.07 (2.38) &  & 1.08 (4.39) & 0.78 (2.47) & 0.28 (0.85) & 1.03 (2.47)\\
4 & 1.06 (3.66) & 0.89 (2.54) & 0.57 (1.53) & 0.98 (1.99) &  & 1.05 (4.02) & 0.84 (2.49) & 0.35 (1.00) & 1.00 (2.39)\\
5 & 1.11 (3.47) & 0.81 (2.17) & 0.64 (1.53) & 0.93 (1.68) &  & 1.01 (3.49) & 0.80 (2.24) & 0.37 (1.04) & 0.77 (1.70)\\
6 & 1.14 (3.11) & 0.76 (1.86) & 0.67 (1.51) & 1.01 (1.72) &  & 0.98 (3.37) & 0.81 (2.09) & 0.35 (0.90) & 0.89 (1.77)\\
7 & 1.07 (2.46) & 0.56 (1.24) & 0.74 (1.49) & 1.17 (1.76) &  & 0.94 (2.53) & 0.74 (1.89) & 0.32 (0.73) & 0.78 (1.42)\\
8 & 0.85 (1.64) & 0.35 (0.70) & 0.77 (1.39) & 1.14 (1.57) &  & 0.85 (1.86) & 0.90 (1.94) & 0.35 (0.72) & 0.73 (1.31)\\
9 & 0.80 (1.25) & 0.70 (1.18) & 0.95 (1.52) & 1.92 (2.30) &  & 0.72 (1.27) & 0.70 (1.28) & 0.09 (0.17) & 0.33 (0.51)\\
10 & 0.93 (1.19) & 3.02 (4.00) & 0.52 (0.74) & 4.18 (4.70) &  & 0.41 (0.54) & 0.25 (0.36) & -0.25 (-0.33) & -0.07 (-0.08)\\
\cmidrule{1-5} \cmidrule{7-10} \\ [-2ex]
10-1 & -0.09 (-0.12) & 2.28 (3.16) & 0.30 (0.53) & 3.41 (4.64) &  & -0.45 (-0.61) & -0.55 (-0.93) & -0.56 (-0.88) & -1.00 (-1.44)\\
\cmidrule{1-5} \cmidrule{7-10} \\ [-2ex]
\textbf{Skewness}\\
\cmidrule{1-5} \cmidrule{7-10} \\ [-2ex]
1 & 1.00 (3.70) & 0.77 (2.17) & 0.35 (1.01) & 0.73 (1.70) &  & 0.84 (3.41) & 0.69 (2.15) & 0.24 (0.68) & 0.80 (2.37)\\
2 & 1.05 (3.79) & 0.77 (1.99) & 0.36 (0.92) & 0.83 (1.74) &  & 0.92 (3.58) & 0.76 (2.27) & 0.24 (0.71) & 0.86 (2.21)\\
3 & 1.02 (3.45) & 0.68 (1.68) & 0.40 (0.98) & 0.82 (1.62) &  & 1.02 (4.14) & 0.81 (2.48) & 0.35 (1.12) & 0.75 (1.75)\\
4 & 1.07 (3.36) & 0.58 (1.44) & 0.45 (1.01) & 0.83 (1.60) &  & 1.05 (4.03) & 0.75 (2.28) & 0.33 (0.87) & 0.80 (1.78)\\
5 & 1.02 (3.12) & 0.56 (1.39) & 0.59 (1.24) & 0.98 (1.85) &  & 1.09 (3.98) & 0.83 (2.25) & 0.38 (0.99) & 0.56 (1.24)\\
6 & 0.98 (2.84) & 0.61 (1.58) & 0.64 (1.31) & 1.46 (2.48) &  & 1.01 (3.20) & 0.81 (2.16) & 0.22 (0.54) & 0.82 (1.87)\\
7 & 0.97 (2.54) & 0.78 (2.00) & 0.78 (1.63) & 1.65 (2.72) &  & 0.88 (2.56) & 0.75 (1.86) & 0.37 (0.81) & 0.71 (1.40)\\
8 & 1.19 (2.66) & 1.04 (2.71) & 0.88 (1.83) & 2.27 (3.34) &  & 0.74 (1.83) & 0.85 (1.94) & 0.36 (0.79) & 0.70 (1.23)\\
9 & 1.10 (2.05) & 1.63 (3.85) & 0.92 (1.85) & 2.77 (3.60) &  & 0.71 (1.39) & 0.71 (1.57) & 0.04 (0.08) & 0.70 (1.11)\\
10 & 0.69 (1.11) & 2.10 (4.13) & 0.52 (1.01) & 1.82 (2.24) &  & 0.59 (0.86) & 0.51 (0.82) & -0.13 (-0.22) & 0.60 (0.81)\\
\cmidrule{1-5} \cmidrule{7-10} \\ [-2ex]
10-1 & -0.30 (-0.62) & 1.33 (3.66) & 0.17 (0.54) & 1.09 (2.00) &  & -0.25 (-0.42) & -0.18 (-0.43) & -0.37 (-0.98) & -0.20 (-0.37)\\
\cmidrule{1-5} \cmidrule{7-10} \\ [-2ex]
\textbf{Kurtosis}\\
\cmidrule{1-5} \cmidrule{7-10} \\ [-2ex]
1 & 1.06 (3.08) & 0.71 (1.56) & 0.33 (0.70) & 0.61 (1.14) &  & 1.15 (3.44) & 0.89 (2.09) & 0.51 (1.17) & 0.86 (1.74)\\
2 & 1.04 (3.23) & 0.77 (1.92) & 0.43 (1.07) & 0.73 (1.48) &  & 1.04 (3.07) & 0.87 (2.09) & 0.31 (0.71) & 0.73 (1.60)\\
3 & 1.02 (3.28) & 0.75 (1.89) & 0.44 (1.12) & 0.85 (1.74) &  & 0.90 (2.66) & 0.79 (2.08) & 0.32 (0.79) & 0.81 (1.58)\\
4 & 1.09 (3.62) & 0.66 (1.61) & 0.48 (1.13) & 0.92 (1.78) &  & 1.01 (3.27) & 0.76 (2.01) & 0.30 (0.73) & 0.78 (1.57)\\
5 & 1.04 (3.23) & 0.71 (1.77) & 0.66 (1.46) & 1.03 (1.84) &  & 0.98 (3.27) & 0.65 (1.72) & 0.25 (0.64) & 0.63 (1.28)\\
6 & 0.96 (2.79) & 0.69 (1.72) & 0.70 (1.49) & 1.62 (2.72) &  & 0.88 (2.90) & 0.72 (1.90) & 0.27 (0.67) & 0.76 (1.66)\\
7 & 0.98 (2.53) & 0.81 (2.14) & 0.85 (1.79) & 1.88 (3.04) &  & 0.80 (2.34) & 0.82 (2.24) & 0.23 (0.58) & 0.75 (1.68)\\
8 & 1.07 (2.41) & 1.09 (2.83) & 0.86 (1.80) & 2.45 (3.43) &  & 0.74 (2.13) & 0.83 (2.18) & 0.15 (0.37) & 0.86 (1.74)\\
9 & 1.05 (2.11) & 1.48 (3.64) & 0.77 (1.60) & 2.46 (3.51) &  & 0.69 (1.63) & 0.59 (1.46) & 0.20 (0.48) & 0.63 (1.24)\\
10 & 0.79 (1.52) & 1.87 (4.57) & 0.35 (0.76) & 1.59 (2.35) &  & 0.64 (1.35) & 0.54 (1.29) & -0.13 (-0.30) & 0.49 (0.83)\\
\cmidrule{1-5} \cmidrule{7-10} \\ [-2ex]
10-1 & -0.27 (-0.75) & 1.16 (3.80) & 0.01 (0.05) & 0.98 (2.29) &  & -0.51 (-1.28) & -0.35 (-1.50) & -0.63 (-2.40) & -0.37 (-0.93)\\
\midrule \bottomrule
\end{tabular}
}
\end{table}

\begin{table}[H]
\caption{{\bf Mean forecast evaluation: 2hNN benchmark.}
This table shows the $R^2$ of the mean and median forecasts of the next 22 days of stock returns for the full and liquid samples across all regions. The mean prediction based on the two-stage quantile model (Two-stage NN) is compared with the direct mean prediction from the two-hidden-layer neural network with MSE loss (2hNN MSE - Mean), derived mean predictions from the two-hidden-layer neural network with multi-head quantile loss (2hNN Quantile - Mean), and the median prediction from the two-stage model (Two-stage NN - Median). Results are obtained separately for full and liquid samples on individual regions covering the period from 1995 to 2018. Models are retrained annually using the expanding window for US data only, while out-of-sample forecasts are generated for both US and international data. Note that losses are multiplied by 100, and values in parentheses are Diebold-Mariano test statistics for the difference between the model in a row and the two-stage NN mean. Standard errors are adjusted using the Newey-West standard deviation estimator with 12 lags.}
\resizebox{1\textwidth}{!}{
\begin{tabular}{lccccclccccc}
\toprule \hline \\ [-2ex]
Specification & \multicolumn{5}{c}{Full Sample} && \multicolumn{5}{c}{Liquid Sample}\\
\cmidrule{2-6} \cmidrule{8-12}
& USA & Europe & Japan & Asia Pacific & Global && USA & Europe & Japan & Asia Pacific & Global\\
\cmidrule{2-6} \cmidrule{8-12} \\ [-2ex]
2hNN MSE - Mean & 0.65 & 0.14 & 0.85 & 0.69 & 0.51 &  & -0.68 & -0.44 & -0.48 & -0.77 & -0.54\\
& (-2.54) & (-3.56) & (-2.37) & (-3.39) & (-3.55) &  & (-1.81) & (-1.31) & (-1.84) & (-1.79) & (-1.98)\\
\cmidrule{2-6} \cmidrule{8-12}
2hNN Quantile - Mean & 1.00 & 0.45 & 1.21 & 1.01 & 0.79 &  & -0.98 & -0.77 & -0.89 & -0.93 & -0.92\\
& (-1.95) & (-2.54) & (-1.53) & (-3.02) & (-2.68) &  & (-1.86) & (-1.78) & (-2.24) & (-1.86) & (-2.21)\\
\cmidrule{2-6} \cmidrule{8-12}
Two stage NN - Median & 0.10 & -0.04 & 0.30 & 0.06 & 0.09 &  & 0.37 & 0.32 & 0.47 & 0.55 & 0.50\\
& (-3.50) & (-4.92) & (-2.82) & (-4.83) & (-4.93) &  & (0.23) & (0.27) & (0.98) & (0.39) & (0.52)\\
\cmidrule{2-6} \cmidrule{8-12}
Two stage NN - Mean & 1.63 & 0.87 & 1.66 & 1.63 & 1.36 &  & 0.32 & 0.22 & 0.18 & 0.40 & 0.37\\
 \hline \bottomrule
\end{tabular}
}
\label{tab:Mean forecast evaluation NN2}
\end{table}

\begin{table}[H]
\caption{\textbf{Volatility forecast evaluation: 2hNN benchmark.}
This table shows the mean absolute deviation (MAD), root mean squared error (RMSE), and Diebold Mariano test statistic (DM) comparing the corresponding losses of the three competing models for the 22-day volatility forecasts.
Volatility forecasting based on two-stage quantile forecasting (two-stage NN) is compared with GARCH model forecasting and volatility derived from quantile forecast from neural network with two hidden layers (2hNN). Results are obtained separately for full and liquid samples on individual regions covering the period from 1995 to 2018. Neural network models are retrained annually using an expanding window on US data only, while out-of-sample forecasts are generated for both US and international samples. Diebold-Mariano test statistics are computed for the difference between the two-stage NN model volatility forecasts and the two competing models - GARCH and 2hNN. Note that losses are multiplied by 100 and the standard errors used in the Diebold-Mariano test statistics are adjusted using the Newey-West standard deviation estimator with 12 lags.}
\resizebox{1\textwidth}{!}{
\begin{tabular}{llccccclccccc}
\toprule \hline \\ [-2ex]
Specification & Variable & \multicolumn{5}{c}{Full Sample} && \multicolumn{5}{c}{Liquid Sample}\\
\cmidrule{2-7} \cmidrule{9-13}
&& USA & Europe & Japan & Asia Pacific & Global && USA & Europe & Japan & Asia Pacific & Global\\
\cmidrule{2-7} \cmidrule{9-13} \\ [-2ex]
GARCH & MAD & 7.34 & 8.79 & 7.79 & 11.1 & 8.53 &  & 3.89 & 3.54 & 3.80 & 4.87 & 3.93\\
2hNN & MAD & 4.66 & 5.32 & 3.91 & 6.41 & 5.01 &  & 3.04 & 2.76 & 2.85 & 3.18 & 2.95\\
Two stage NN & MAD & 4.63 & 5.20 & 3.99 & 6.49 & 4.99 &  & 2.93 & 2.63 & 2.81 & 3.16 & 2.87\\
\cmidrule{2-7} \cmidrule{9-13} \\ [-2ex]
GARCH & RMSE & 13.14 & 16.24 & 14.63 & 18.70 & 15.50 &  & 6.93 & 7.10 & 6.72 & 9.52 & 7.36\\
2hNN & RMSE & 7.57 & 8.48 & 5.58 & 10.43 & 8.08 &  & 4.73 & 4.12 & 4.05 & 4.95 & 4.48\\
Two stage NN & RMSE & 7.46 & 8.52 & 5.64 & 10.05 & 8.08 &  & 4.62 & 4.03 & 3.98 & 4.86 & 4.38\\
\cmidrule{2-7} \cmidrule{9-13}
GARCH & DM & (10.25) & (35.92) & (20.32) & (44.26) & (26.47) &  & (7.90) & (5.98) & (7.97) & (9.40) & (10.02)\\
2hNN & DM & (1.56) & (-1.12) & (-1.69) & (-1.03) & (-0.18) &  & (2.38) & (1.59) & (1.60) & (1.41) & (2.10)\\
\hline \bottomrule
\end{tabular}
}
\label{tab:VolatilityForecastNN2}
\end{table}

\newpage
\section{Hyperparameter search and validation} 
\label{sec:appendix_hyperparameter}
\counterwithin{table}{section} 
\counterwithin{figure}{section}
\renewcommand{\thetable}{D\arabic{table}}

We conduct a hyperparameter search using U.S. data from the period 1973 to 1994, employing weekly sampling and 22-day ahead returns, consistent with the setup for the main results in the paper.
The period from 1973 to 1989 is utilized for training models with various hyperparameter values we are exploring.
The period from 1990 to 1994 is used to evaluate the trained models and select the best hyperparameter values.
Forecasts for the validation period are generated using the best hyperparameter values identified in the previous step.
Hyperparameters are optimized once on the early period and held fixed throughout the evaluation horizon to limit the computational burden associated with repeated tuning.
To maintain consistency with the train-test split used for the main results of the paper, we retrain the model annually during the validation period and use the most recent model for forecasting.
Specifically, given a set of hyperparameter values, the model is initially trained on the 1973-1989 training sample and then rolled forward by one year four times, always using the model for the subsequent year's forecasts.
We utilize the full sample given the greater number of available observations compared to the liquid sample, which provides more stable results. Lengthening the validation sample would result in a shorter training sample, which is undesirable. The training sample should cover at least one severe market-wide negative shock in equities, which occurred in 1987. This approach suggests there is some potential room for improvement in finding better hyperparameters for the liquid sample and/or for international regions.

It is not feasible to perform a full hyperparameter search for all possible hyperparameters of the neural networks. Therefore, we divide the hyperparameter search into two sequential steps and focus only on a subset of hyperparameters. The rest of the hyperparameters are fixed and are based on the optimal hyperparameters found previously in the literature, e.g., \cite{gu2020empirical} or \cite{tobek2021does}. \autoref{FixedHyperparameters} shows the fixed hyperparameters and their values.

\begin{table}[H]
\caption{\textbf{Fixed Hyperparameters.}
This table lists the hyperparameters that are fixed during the hyperparameter search.
The number of epochs is set to $100*(A/n)$, where $n$ is the number of observations available for
training and $A$ is an adjustment constant equal to $3{\small,}000{\small,}000$ for the full sample
and $1{\small,}500{\small,}000$ for the liquid sample, derived from the average number of observations 
in individual samples over time.}
\centering
\begin{tabular}{c|c}
Hyperparameter & Value \\
\hline
 Activation function & LeakyReLU \\
 Batch size & 8192 \\
 Batch normalization & True \\
 Epochs & $100*(A/n)$\\
 Early-stopping patience & 2 epochs\\
 Early-stopping validation size & 20\% \\
 Optimizer & Adam \\
 Adam momentum & (0.9, 0.999) \\
 Decay factor & 1.0 \\
 Decay step size & 100 \\
 L1 penalty for first layer & 0.0001\\
 Num. of networks in ensemble & 20 liquid,  10 full sample \\
\end{tabular}
\label{FixedHyperparameters}
\end{table}

We start with a grid search for optimization algorithm hyperparameters, namely the learning rate (LR) and dropout rate (DR).
We test values of $0.1$, $0.001$, and $0.0001$ for LR and $0$, $0.2$, and $0.4$ for DR, evaluating the average quantile loss for the validation sample.
We use a two-stage feedforward neural network with two hidden layers, each with $128$ neurons, as a starting point to perform the first round of hyperparameter search.
The architecture was chosen to provide slightly higher capacity than neural networks used in \cite{gu2020empirical} or \cite{tobek2021does}.
Our evaluation criterion is the average quantile loss during the validation period from 1990 to 1994.
\autoref{HyperparameterSearch1} displays the results of the hyperparameter search.

\begin{table}[H]
\caption{{\bf Hyperparameter Search for Learning Rate and Dropout.}
We conduct a hyperparameter search for the period from 1973 to 1994 using U.S. data only.
We search for the optimal learning rate and dropout rate for the two-stage neural network.
The rest of the hyperparameters are fixed as shown in \autoref{FixedHyperparameters}. 
Loss is calculated as the average quantile loss for the validation period from 1990 to 1994.}
\centering
\resizebox{0.45\textwidth}{!}{
\begin{tabular}{ccc}
\toprule \hline \\ [-2ex]
Learning Rate & Dropout Rate & Loss\\
\hline \\ [-2ex]
0.01 & 0.0 & 0.022827\\
0.01 & 0.2 & 0.022982\\
0.01 & 0.4 & 0.023135\\
0.001 & 0.0 & 0.022958\\
0.001 & 0.2 & 0.022821\\
0.001 & 0.4 & 0.022881\\
0.0001 & 0.0 & 0.022936\\
0.0001 & 0.2 & 0.022835\\
0.0001 & 0.4 & 0.022844\\
0.0003 & 0.2 & \textbf{0.022778}\\
0.003 & 0.2 & 0.022914\\
0.0003 & 0.3 & 0.022815\\
0.0003 & 0.1 & 0.022829\\
\hline \bottomrule
\end{tabular}
}

\label{HyperparameterSearch1}
\end{table}

The lowest average quantile loss is achieved for the learning rate of $0.0003$ and dropout rate of $0.2$, although dropout rates of $0.1$ and $0.3$ are not significantly worse than $0.2$. Fixing the learning rate and dropout rate at the optimal values achieved in the first part of the hyperparameter search, we perform the second part of the hyperparameter search to find an optimal architecture for the two-stage feedforward neural network, where we try different numbers of neurons and layers in the individual stages of the network.
\autoref{HyperparameterSearch2} shows the results of the architecture hyperparameter search.

\begin{table}[H]
\caption{{\bf Hyperparameter Search for Architecture.}
We search for the optimal number of neurons and layers in the individual stages of the two-stage feed-forward neural network.
The loss function is the average quantile loss achieved during the validation period from 1990 to 1994.}
\centering
\resizebox{0.4\textwidth}{!}{
\begin{tabular}{ccc}
\toprule \hline \\ [-2ex]
Stage1 (block) & Stage2 & Loss \\
\hline \\ [-2ex]
2x128 & 8x1 & \textbf{0.022778}\\
2x128 & 0x1 & 0.022815\\
2x128 & 16x1 & 0.022822\\
2x64 & 8x1 & 0.022829\\
2x256 & 8x1 & 0.022786\\
3x128 & 8x1 & 0.022810\\
1x128 & 8x1 & 0.022816\\
4x64 & 8x1 & 0.022872\\
\hline \bottomrule
\end{tabular}
}

\label{HyperparameterSearch2}
\end{table}

A smaller or greater number of neurons than $8$ in the second stage leads to worse results. A smaller number of neurons is preferred to prevent potential overfitting and to speed up learning.
The network already needs to be heavily regularized to achieve satisfactory performance; it, therefore, should not require more learning capacity by adding more neurons.

The L1 and L2 penalties were selected in isolation as it is not computationally possible to add them as another hyper-parameter in the search. The chosen values proved to be extremely robust and to work well across all the network architectures. The L1 penalty for the first layer of the first stage and L1/L2 penalty for the first layer of the second stage were selected based on inspection of estimated weights on the 1973-1994 training sample. The L1 penalty for the first layer of the first stage was selected using a one layer network and inspecting how many weights are non-negative.\footnote{Note that one layer neural network is just a standard quantile linear regression with L1 penalty on its weights that is estimated using stochastic gradient descent.} The goal was to induce some selection of input variables and to limit the effects of collinearity. The L1 penalty for the first stage is $0.0001$, and it forces weights of about $100-150$ of the input variables to zero.\footnote{The L1 penalty for the first stage is changed to $0.00001$ for neural network specification, where raw returns are directly predicted from the first stage to reflect the smaller scale of the predicted variable.}

The L1 and L2 penalties for the second stage were chosen solely to limit collinearity issues without any focus on variable selection. The penalty was chosen the smallest possible while still leading to no obvious collinearlity issues.\footnote{The goal was to not have a mix of very large negative and positive weights, which are a symptom of the collinearity.} The L1 and L2 penalties for the second stage are the same at $0.00001$. The L1 penalty for second stage is approximately $10$ times smaller than the L1 penalty for the first stage. The scale of the standardised stock returns, serving as a label for the first stage, is close to $1$. The scale of the raw stock returns, serving as a label for the second stage, is roughly $10$ times smaller. The L1 penalties for both stages are, therefore, consistent given the scale of the variables used as an input for the loss function.

\subsection{Validation}

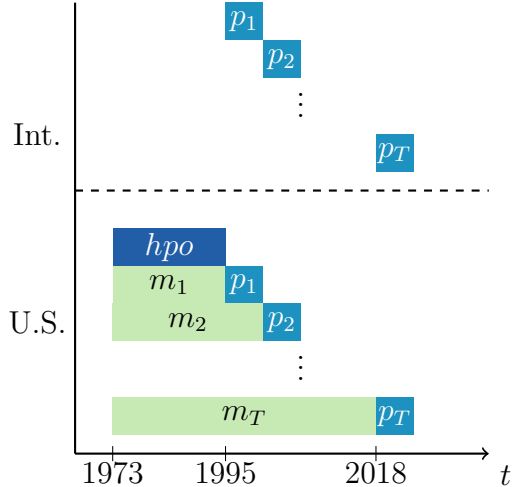
\begin{figure}[H]
\centering    
\begin{tikzpicture}[
    scale=0.5,
    node distance=2cm,
    block/.style={rectangle, draw, fill=blue!20, text width=5em, text centered, rounded corners, minimum height=4em},
    line/.style={draw, -latex'}
]

\draw[thick,->] (0,0) -- (11,0) node[anchor=north west] {$t$};
\draw[thick] (0,0) -- (0,12) node[anchor=south east] {};

\draw (1,-0.2) -- (1,0.2);
\draw (4,-0.2) -- (4,0.2);
\draw (8,-0.2) -- (8,0.2);

\node at (1,-0.5) {1973};
\node at (4,-0.5) {1995};
\node at (8,-0.5) {2018};

\node at (-1, 3.5) {U.S.};
\node at (-1, 8.5) {Int.};

\draw[thick, dashed] (0,7) -- (11,7);

\fill[blue!20] (1,5) rectangle (4,6);
\node at (2.5,5.5) {$hpo$}; 

\fill[red!20] (1,4) rectangle (4,5);
\node at (2.5,4.5) {$m_1$}; 

\fill[green!20] (4,4) rectangle (5,5);
\node at (4.5,4.5) {$p_1$}; 

\fill[red!20] (1,3) rectangle (5,4);
\node at (3,3.5) {$m_2$}; 

\fill[green!20] (5,3) rectangle (6,4);
\node at (5.5,3.5) {$p_2$}; 

\node at (6,2.5) {$\vdots$};

\fill[red!20] (1,0.5) rectangle (8,1.5);
\node at (4.5,1) {$m_{T}$}; 

\fill[green!20] (8,0.5) rectangle (9,1.5);
\node at (8.5,1) {$p_{T}$}; 

\fill[green!20] (4,11) rectangle (5,12);
\node at (4.5,11.5) {$p_{1}$}; 

\fill[green!20] (5,10) rectangle (6,11);
\node at (5.5,10.5) {$p_{2}$}; 

\fill[green!20] (8,7.5) rectangle (9,8.5);
\node at (8.5,8) {$p_{T}$}; 

\node at (6,9.5) {$\vdots$};
\end{tikzpicture}

\caption{\textbf{Validation sample-splitting schema.} 
    This figure shows how the data is split into training ($m_i$), validation ($hpo$) and testing ($p_i$)
    sets for the U.S. and international samples. 
    All models are trained on the U.S. data and then used to generate out-of-sample predictions
    for both the U.S. and the international sample. 
    Models are retrained each year using expanding window of data reaching back to 1973.
    Predictions are generated monthly, always using the most up-to-date model available at the end of each month, e.g.
    $p_1$ predictions represent all monthly predictions during the period of year $1$ and
    are generated using the model trained on $m_1$.}
\label{fig:validation_sample_splitting}
\end{figure}

\autoref{fig:validation_sample_splitting} illustrates the data split into training,
validation, and testing sets for the U.S. and international samples. Hyperparameter
optimization and model training are conducted solely on U.S. data, using international data only for out-of-sample evaluation.

In order to maximize the number of observations available for training, future returns are computed each week along with all the required inputs to the neural network.
Future returns cover the next 22 business days\footnote{The count of business days here also includes holidays for simplicity.}, approximately one month of trading.
This leads to approximately 4.3 times more partially overlapping observations available for training compared to using non-overlapping monthly data only\footnote{Using only monthly data during the training leads to inferior models because it can miss some reverting rare price changes.}. Weekly sampled future returns are used only during the model training. Out-of-sample predictions are generated and evaluated using regular calendar months.

\section{Algorithms for approximating probability distribution function and moments from quantiles}
\label{app:algorithms}
\counterwithin{table}{section} 
\counterwithin{figure}{section}
\renewcommand{\thetable}{E\arabic{table}}

Algorithm \ref{sec:appendix_density} contains the algorithm for deriving the probability density function from the quantiles. 
The algorithm is referred to as \textit{quantiles-to-density} in the paper. Algorithm
\ref{sec:appendix_moments} contains the algorithm for deriving the moments of the cumulative probaility distribution function.
The algorithm is referred to as \textit{density-to-moments} in the paper, and the algorithm below
does not contain the adjustment of the moments described in \autoref{subsubsec:MomentsAdjustment}.

\subsection{Fallback to linear B-Splines} \label{app:linear_fallback}

The liquid universe of stocks generally exhibits well-behaved density functions.
Therefore, the fallback to the linear approximation of the density is rarely needed.
However, it is sometimes necessary for the full sample due to the presence of infrequently traded stocks with small capitalization.
There can be many days with zero trading activity for micro-cap stocks, especially in the historical period.
Days with zero trading are a well-documented proxy for liquidity in the finance literature, formally introduced by \cite{lesmond1999new}.
See \cite{goyenko2009liquidity} for an overview of other related liquidity proxies along with a comparison of their relative performance.
This method could also be applied in bond returns forecasting where high illiquidity and zero-return observations are common.
The presence of zero-return days can lead to a large probability mass around zero and, in extreme cases, to a discrete probability of a zero return.
The continuous density function implied from the cumulative distribution function is then not well-behaved, and the fallback to linear approximation becomes necessary.

\begin{figure}[H]
\centering
\includegraphics[width=1\textwidth]{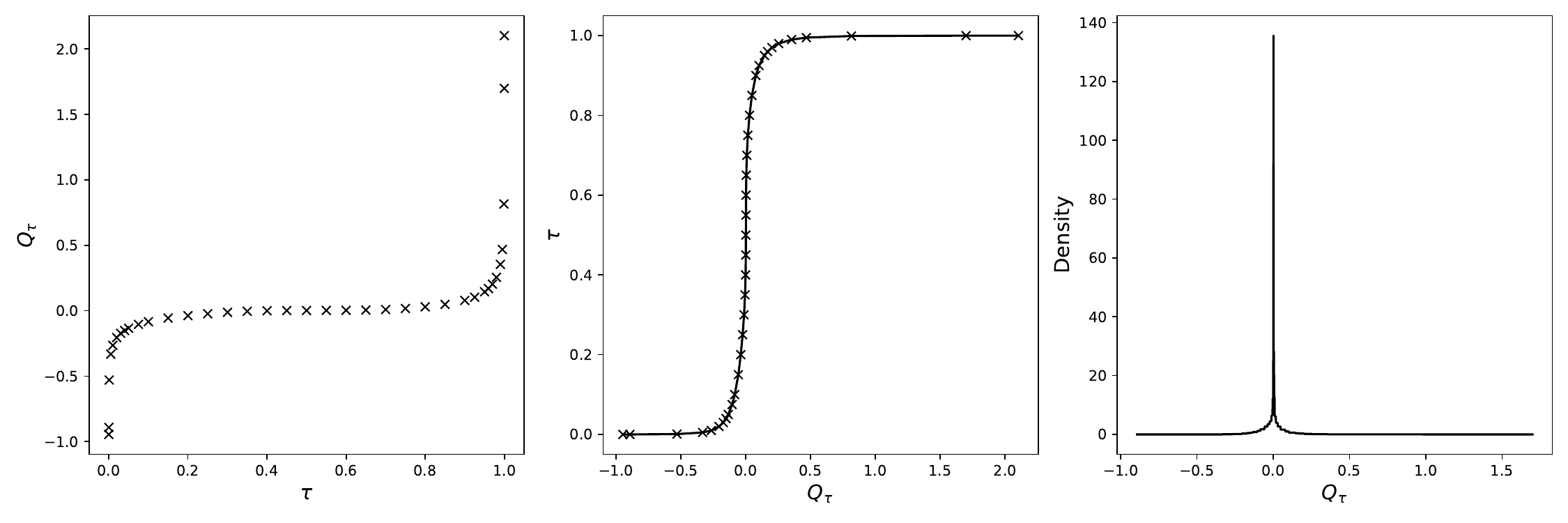}
\caption{\textbf{Example of density with a large mass around zero.}
This figure shows the derived probability density function for monthly return predictions for Idenix Pharmaceuticals in October 2014.
}
\label{figureDirichletExample}
\end{figure}

For larger cap stocks where the density fallback was applied, the cause was manually checked. 
The density always exhibited a large mass around zero, indicating that the model was predicting a discrete probability of zero return in the following month.
Fallback predictions for large cap stocks were often associated with a corporate event, such as being close to official delisting or a recent trading halt.
One such case is detailed in Figure~\ref{figureDirichletExample}, depicting forecasts for October 2014 for Idenix Pharmaceuticals, which was delisted just five days later.
Idenix Pharmaceuticals received a tender takeover offer from Merck on June 9, 2014\footnote{The announcement of the acquisition is available here \url{https://www.merck.com/news/merck-to-acquire-idenix/}.}.

\begin{algorithm}
	\caption{PDF} 
	\label{sec:appendix_density}
	\footnotesize
\begin{algorithmic}[1]
\Require $\textbf{x}, \textbf{y}$
\State $gp = 100$
\Comment{Density of grid points}
\State $d_{min} = 1e-5$
\Comment{Minimum value for density function}
\State $eps = 1e-4$
\Comment{Epsilon}
\State $z = 0$
\Comment{Keep only the highest quantile with -1 return}
\For{$i$ in 1 to length(x) - 1}
    \If {$x[i + 1] = -1$}
        \State $z = z + 1$
    \EndIf
\EndFor
\State  $x = x[z:]$
\State  $y = y[z:]$
\If {$min(y) < 0.05$}
\Comment{Get 0.9 and 0.1 quantiles for outlier detection}
    \State $x_{Q10} = x[y = 0.1]$
\Else
     \State $x_{Q10} = x[3]$   
\EndIf
\State $x_{Q90} = x[y = 0.9]$
\For{$i$ in 1 to length(x) - 1}
\Comment{Fix inconsistent quantile predictions}
    \If {$x[i + 1] < x[i] + eps$}
        \State $x[i + 1] = x[i] + eps$
    \EndIf
\EndFor
\State allocate empty vector $\Tilde{\textbf{x}}$
\Comment{Create grid of denser x values}
\For{$i$ in 1 to length(x) - 3}
    \State $x_l = x[i + 1]$
    \State $x_u = x[i + 2]$
    \State add $range(x_l, x_u, (x_u - x_l) / gp)$ to $\Tilde{\textbf{x}}$
\EndFor
\State drop duplicates in $\Tilde{\textbf{x}}$
\State sort values of $\Tilde{\textbf{x}}$ from the smallest to the largest
\State fit cubic B-spline interpolation $S$ on $\textbf{x}$ and $\textbf{y}$
\State use $S$ to fit its first derivative on $\Tilde{\textbf{x}}$: $\textbf{d} = S'(\Tilde{\textbf{x}})$
\If {$min(\textbf{d}[(\Tilde{\textbf{x}} \ge x_{Q10}) \& (\Tilde{\textbf{x}} \leq x_{Q90})]) < d_{min}$}
\Comment{Density fallback}
    \State fit linear B-spline interpolation $S$ on $\textbf{x}$ and $\textbf{y}$
    \State use $S$ to fit its first derivative on $\Tilde{\textbf{x}}$: $\textbf{d} = S'(\Tilde{\textbf{x}})$
\EndIf
\State $\textbf{d}[\textbf{d} < d_{min}] = d_{min}$
\Comment{Enforce minimum density}
\State $w = min(\textbf{d}[(\Tilde{\textbf{x}} <= x_{Q10})])$
\Comment{Enforce monotonicity in the tails}
\State $v = arg min(\textbf{d}[(\Tilde{\textbf{x}} <= x_{Q10})])$
\State $\textbf{d}[:v] = w$
\State $w = min(\textbf{d}[(\Tilde{\textbf{x}} >= x_{Q90})])$
\State $v = arg min(\textbf{d}[(\Tilde{\textbf{x}} >= x_{Q90})])$
\State $\textbf{d}[v:] = w$
\State $x_{min} = min(x)$
\Comment{get max and min of x}
\State $x_{max} = max(x)$
\If {$min(\Tilde{\textbf{x}}) < -1$}
\Comment{Truncate the distribution at -1 if needed}
    \State $\textbf{d} = \textbf{d}[\Tilde{\textbf{x}} \ge -1]$
    \State $\Tilde{\textbf{x}} = \Tilde{\textbf{x}}[\Tilde{\textbf{x}} \ge -1]$
    \State $x_{min} = -1$
\EndIf
\State $P_l = 1 - S(x_{max})$
\Comment{Compute discrete probabilities at tails}
\State $P_u = S(x_{min})$
\end{algorithmic}
\end{algorithm}

\subsection{Central Moments}

We can derive central moments from the non-central moments using the \autoref{eq:centralMoments}.

\begin{equation} \label{eq:centralMoments}
\begin{aligned}
\sigma &= m_2 - m_1^2 \\
\text{skewness} &= \frac{m_3 - 3 m_1 \sigma - m_1^3}{\sigma^{3 / 2}} \\
\text{kurtosis} &= \frac{m_4 - 4 m_1 m_3 + 6 m_1^2 m_2 - 3 m_1^4}{\sigma^2} \\
\end{aligned}
\end{equation}

where $m_i$ for $i \in {1,2,3,4}$ are non-central moments and $\sigma$ is the variance.

\subsection{Adjustment of moments} \label{subsubsec:MomentsAdjustment}

The required moment adjustments are estimated using simulated data with known precise theoretical moments.
One million non-central Student's t distributions are generated with various degrees of freedom and noncentrality parameter.
\footnote{Non-central Student's t distribution is defined as $\frac{Y + nc}{\sqrt{V/df}}$ where $Y$ is a standard normal distribution, $V$ is an independent chi-square random variable, $nc$ is noncentrality parameter, and $df$ are degrees of freedom.
See \href{https://docs.scipy.org/doc/scipy/reference/generated/scipy.stats.nct.html}{scipy.stats.nct} documentation for more details on its implementation.}
Degrees of freedom are randomly selected from \{5, 6, 7, 8, 9, 10, 11, 12, 13, 14, 15, 16, 17, 18, 19, 20, 30, 40, 50, 60, 70, 80, 90, 100, 1000, 10000\} with replacement and equal probability of each outcome.
The noncentrality parameter is randomly selected using uniform distribution of $[-0.5, 5]$ interval focusing mostly on positive skewness, as is observed in the empirical equities data.
The scale of the generated distributions is set to 0.1 to limit the effects of the fact that stock returns cannot be smaller than $-1$,
which is reflected in the density-to-moments algorithm but not in the theoretical distributions.
Precise theoretical moments of the distribution are saved, and quantiles of the distribution are fed into the ``naive`` density-to-moments algorithm to compute predicted moments.
Generated distributions with theoretical kurtosis larger than $20$ are discarded as they would create strong leverage points and skew the estimated parameters to capture these extremes.

The values of the parameters are then estimated using OLS, where precise theoretical moments are predicted using
moments produced by the ``naive`` density-to-moments algorithm.
The simple linear regression is preferred here as it provides a satisfactory fit while being easy to interpret and more likely to generalize to other distributions.\footnote{It is possible to further improve the fit by employing neural networks or higher order polynomials and interactions between the exogenous variables if required.}

The following adjustment is proposed to decrease bias in the moments estimated from the ``naive`` density-to-moments algorithm.

\begin{equation}\label{eq:mom_adj}
\begin{aligned}
k_e & = k - 3 \\
\Tilde{v} & = v \times (1.0023 - 0.0021 \times s + 0.0022 \times k_e) \\
\Tilde{s} & = 0.9950 \times s + 0.0261 \times s^2 + 0.0107 \times k_e \\
\Tilde{k} & = 3 + 1.4185 \times k_e + 0.0466 \times k_e^2 - 0.7395 \times s
\end{aligned}
\end{equation}

Where $v, s, k$ are variance, skewness, and kurtosis from the ``naive`` density-to-moments algorithm, $\Tilde{v}, \Tilde{s}, \Tilde{k}$ are their adjusted counterparts, and $k_e$ is excess kurtosis.
All of the estimated parameters are significant with t-distribution over $100$ in absolute terms. The fit is high for all the regressions with $R^2$ over $0.9$.

Variance $v$ requires only a small adjustment, which mainly accounts for the fact that it is underestimated for heavy-tailed distributions with high kurtosis.
The adjustment for skewness $s$ is larger to account for its underestimation when either skewness or kurtosis of the distribution is high.
Finally, as expected, kurtosis requires the largest adjustment due to the difficulty of capturing it with the ``naive`` density-to-moments algorithm.

The adjusted density-to-moments algorithm is next tested on probability distributions that are close to those actually observed in the predictions of the future return distribution.
The predicted densities from neural networks have positive skewness and are heavy-tailed with positive excess kurtosis.
A range of distributions is tested in Table \ref{tab:MomentAdjustment}.
Normal distribution is the starting point due to its wide use in quantitative finance.
t-distribution with small degrees of freedom then serves as a benchmark for a more heavy-tailed distribution.
Finally, non-central Student’s t-distribution is the closest match to the distribution actually observed in the data.

\begin{table}[H]
\caption{{\bf Adjustment fit of density-to-moments algorithm.} 
$m, v, s, k$ denote mean, variance, skewness, and kurtosis estimated using the ``naive`` (non-adjusted) density-to-moments algorithm defined in \autoref{subsubsec:MomentsAdjustment}, $m_t, v_t, s_t, k_t$ corresponding theoretical values
and $\Tilde{v}, \Tilde{s}, \Tilde{k}$ are generaged using adjusted density-to-moments algorithm, i.e. values are adjusted using equations \ref{eq:mom_adj}.
}
\resizebox{1\textwidth}{!}{
\begin{tabular}{lllllllllllll}
\toprule \hline \\ [-2ex]
Dist & $m_t$ & $m$ & $v_t$ & $v$ & $\Tilde{v}$ & $s_t$ & $s$ & $\Tilde{s}$ & $k_t$ & $k$ & $\Tilde{k}$ \\
\hline \\ [-2ex]
Normal & 0.000 & -0.000 & 1.000 & 0.998 & 1.001 & 0.000 & -0.000 & -0.000 & 3.000 & 2.980 & 2.971\\
t: df=10 & 0.000 & -0.000 & 1.250 & 1.244 & 1.250 & 0.000 & -0.000 & 0.009 & 4.000 & 3.852 & 4.242\\
t: df=6 & 0.000 & -0.000 & 1.500 & 1.487 & 1.497 & 0.000 & -0.000 & 0.023 & 6.000 & 5.161 & 6.282\\
t: df=5 & 0.000 & -0.000 & 1.667 & 1.646 & 1.662 & 0.000 & -0.001 & 0.035 & 9.000 & 6.360 & 8.293\\
nct: df=5, nc=1 & 0.119 & 0.119 & 1.919 & 1.892 & 1.914 & 1.266 & 1.116 & 1.200 & 13.321 & 8.385 & 11.164\\
nct: df=6, nc=3 & 0.345 & 0.345 & 3.072 & 3.035 & 3.076 & 1.832 & 1.686 & 1.823 & 12.991 & 9.691 & 13.330\\
nct: df=5, nc=4 & 0.476 & 0.476 & 5.698 & 5.593 & 5.736 & 2.718 & 2.372 & 2.644 & 29.832 & 15.827 & 27.104\\
\hline \bottomrule
\end{tabular}
}
\label{tab:MomentAdjustment}
\end{table}

Table \ref{tab:MomentAdjustment} documents a good fit for lower moments and even for kurtosis post-adjustment.
The computed mean is almost identical to the theoretical mean, as is the computed variance.
Skewness benefits from the moment adjustment for heavy-tailed positively skewed distribution while exhibiting reasonable fit even without the adjustment.
Kurtosis is the hardest to fit precisely, but the adjustment is quite precise even for the cases with the highest theoretical kurtosis.
This section has, therefore, confirmed that the proposed moment estimation technique from the quantiles of the distribution yields reasonable results for empirical analysis.

\begin{algorithm}
	\caption{Moments} 
	\label{sec:appendix_moments}
	\footnotesize
\begin{algorithmic}[1]
\Require $\textbf{x}, \textbf{y}, P_l, P_u, x_{min}, x_{max}$
\State $m_0, m_1, m_2, m_3, m_4 = 0, 0, 0, 0, 0$
\For{$i$ in 1 to length(x) - 1}
    \State $x_1 = \textbf{x}[i]$
    \State $x_2 = \textbf{x}[i + 1]$
    \State $y_1 = \textbf{y}[i]$
    \State $y_2 = \textbf{y}[i + 1]$
    \State $b = (y_2 - y_1) / (x_2 - x_1)$
    \State $a = y_1 - b x_1$
    \State $m_0 = m_0 + a x_2 + 1 / 2 b x_2^2 - a x_1 - 1 / 2 b x_1^2$
    \State $m_1 = m_1 + 1 / 2 a x_2^2 + 1 / 3 b x_2^3 - 1 / 2 a x_1^2 - 1 / 3 b x_1^3$
    \State $m_2 = m_2 + 1 / 3 a x_2^3 + 1 / 4 b x_2^4 - 1 / 3 a x_1^3 - 1 / 4 b x_1^4$
    \State $m_3 = m_3 + 1 / 4 a x_2^4 + 1 / 5 b x_2^5 - 1 / 4 a x_1^4 - 1 / 5 b x_1^5$
    \State $m_4 = m_4 + 1 / 5 a x_2^5 + 1 / 6 b x_2^6 - 1 / 5 a x_1^5 - 1 / 6 b x_1^6$
\EndFor
\State $m_1 = m_1 + P_l x_{min} + P_u x_{max}$
\Comment{Add discrete probabilities in the tails}
\State $m_2 = m_2 +  P_l x_{min}^2 + P_u x_{max}^2$
\State $m_3 = m_3 +  P_l x_{min}^3 + P_u x_{max}^3$
\State $m_4 = m_4 +  P_l x_{min}^4 + P_u x_{max}^4$
\State $m_1 = m_1 / m_0$
\Comment{Normalize probability measure to 1}
\State $m_2 = m_2 / m_0$
\State $m_3 = m_3 / m_0$
\State $m_4 = m_4 / m_0$
\State $variance = m_2 - m_1^2$
\State $skewness = (m_3 - 3 m_1 var - m_1^3) / var^{3 / 2}$
\State $kurtosis = (m_4 - 4 m_1 m_3 + 6 m_1^2 m_2 - 3 m_1^4) / var^2$
\end{algorithmic}
\end{algorithm}

\pagebreak

\section{2019-2023 Out-of-sample results for the U.S. }
\label{app:US_update}
\counterwithin{table}{section} 
\counterwithin{figure}{section}
\renewcommand{\thetable}{F\arabic{table}}

Empirical results presented in Section~\ref{sec:Results} cover
period from 1995 to 2018. Here, we provide
additional out-of-sample empirical evidence by extending
our analysis to cover period from 2019 to 2023 for the U.S stocks. 
We report evaluation of mean forecasts in Table~\ref{tab:Mean forecast evaluation OOS}, volatility forecasts in Table~\ref{tab:VolatilityForecast OOS}, quantile forecasts in Table~\ref{tab:avg_quantile_loss OOS} and profitability of mean and median portfolios formed on mean/median forecasts in Table~\ref{tab:MSEvsMean OOS}. \\

In terms of $R^2$, the two-stage quantile neural network continues to outperform the direct mean forecasts from the 2-hidden-layer neural network using mean squared error loss (2hNN MSE), particularly for the liquid sample, where the $R^2$ is $1.77$ compared to $0.13$ for the 2hNN MSE model. The $R^2$ for the full sample drops to $1$, with only a marginal advantage over competing methods. However, the difference in performance between the mean forecast from the two-stage quantile neural network and the median forecast from the two-stage quantile neural network is no longer statistically significant, which contrasts with the results from the 1995-2018 period, where outperformance was significant.

\begin{table}[H]
\caption{{\bf Mean forecast evaluation in the US 2019-2023.}
This table shows the $R^2$ of the mean and median forecasts of the next 22 days of stock returns for the full and liquid samples across all regions. The mean prediction based on the two-stage quantile model (Two-stage NN) is compared with the direct mean prediction from the two-hidden-layer neural network (2hNN MSE - Mean) and the median prediction from the two-stage model (Two-stage NN - Median). Results are obtained separately for full and liquid samples in the US covering the period from 2019 to 2023. Models are retrained annually using the expanding window. Note that losses are multiplied by 100, and values in parentheses are Diebold-Mariano test statistics for the difference between the model in a row and the two-stage NN mean. Standard errors are adjusted using the Newey-West standard deviation estimator with 12 lags.}
\centering
\resizebox{0.8\textwidth}{!}{
\begin{tabular}{ccclccc}
\toprule \hline \\ [-2ex]
\multicolumn{3}{c}{Full Sample} && \multicolumn{3}{c}{Liquid Sample}\\
\cmidrule{1-3} \cmidrule{5-7}
Two stage NN & Two stage NN & 2hNN MSE && Two stage NN & Two stage NN & 2hNN MSE \\
Mean & Median & Mean && Mean & Median & Mean \\
\hline \\ [-2ex]
1.00 & 0.20 & 0.30 &  & 1.77 & 1.72 & 0.13\\
& (-1.11) & (-1.49) &  &  & (-0.10) & (-1.43)\\
 \hline \bottomrule
\end{tabular}
}
\label{tab:Mean forecast evaluation OOS}
\end{table} 

The results for volatility forecasting, presented in Table~\ref{tab:VolatilityForecast OOS}, indicate that the two-stage quantile neural network remains effective in predicting the dispersion of returns in comparison with the GARCH benchmark. Across both the full and liquid samples, the model achieves statistically significant lower mean root mean squared error (RMSE) and absolute deviation (MAD) compared to the GARCH benchmark. For the full sample, the MAD decreases from $6.65$ under GARCH to $4.40$ under the two-stage quantile neural network, while the RMSE declines from $12.23$ to $7.39$. 
These results are consistent with the findings in Table~\ref{tab:VolatilityForecast}, where the two-stage quantile neural network significantly outperformed GARCH in volatility forecasting for the 1995-2018 period suggesting that while mean return predictability has weakened, the model's ability to predict volatility remains robust and statistically meaningful.

\begin{table}[H]
\caption{\textbf{Volatility forecast evaluation in the US 2019-2023.}
This table shows the mean absolute deviation (MAD), root mean squared error (RMSE), and Diebold Mariano test statistic (DM) comparing the corresponding losses of the two competing models for the 22-day volatility forecasts.
Volatility forecasting based on two-stage quantile forecasting (two-stage NN) is compared with GARCH model forecasting. Results are obtained separately for full and liquid samples in the US covering the period from 2019 to 2023. Neural network models are retrained annually using an expanding window. Diebold-Mariano test statistics are computed for the difference between the GARCH and two-stage NN model volatility forecasts. Note that losses are multiplied by 100 and the standard errors used in the Diebold-Mariano test statistics are adjusted using the Newey-West standard deviation estimator with 12 lags.}
\centering
\resizebox{0.6\textwidth}{!}{
\begin{tabular}{lcclcc}
\toprule \hline \\ [-2ex]
Measure & \multicolumn{2}{c}{Full Sample} && \multicolumn{2}{c}{Liquid Sample}\\
\cmidrule{2-3} \cmidrule{5-6} \\ [-2ex]
 & Two stage NN & GARCH && Two stage NN & GARCH \\
 \hline \\ [-2ex]
MAD & 4.40 & 6.65 &  & 3.24 & 3.95\\
RMSE & 7.39 & 12.23 &  & 5.47 & 7.33\\
DM &  & (7.72) &  &  & (4.13)\\
\hline \bottomrule
\end{tabular}
}
\label{tab:VolatilityForecast OOS}
\end{table}

The performance of the quantile forecasts in Table~\ref{tab:avg_quantile_loss OOS} suggests that the two-stage quantile neural network remains the best-performing model, although the margin of improvement over other neural network architectures has narrowed. In the full sample, the average quantile loss of the two-stage quantile neural network is $2.730$, only slightly lower than the $2hNN$ loss of $2.753$. A similar pattern holds for the liquid sample, where the loss is $1.871$ for the two-stage quantile neural network compared to $1.886$ for the 2hNN model. While the two-stage model still outperforms GARCH, LNN, and 1hNN, these differences are no longer statistically significant at the $5$\% level in the full sample. In the liquid sample, the two-stage model remains the best-performing specification, but the differences relative to the alternative models are also not statistically significant. This is in contrast to the 1995-2018 period in Table~\ref{tab:avg_quantile_loss}, where the two-stage quantile neural network significantly outperformed all benchmark models in quantile forecasting. Even though the improvements in quantile forecasts are not statistically significant, the two-stage quantile neural network still consistently achieves the lowest quantile loss, indicating that it retains some predictive advantages even in the more recent period. The statistical insignificance is also to a large degree due to the small length of the tested sample. Purely based on sample length, t-statistics should be 2.2 times smaller for the 5-years sample here versus 24-years sample originally.

\begin{table}[H]
\caption{{\bf Out-of-sample quantile forecasts evaluation in the US 2019-2023.}
This table shows the average out-of-sample quantile cross-sectional loss for quantile forecasts from the two-stage model and alternative models. The results are obtained separately for the full and liquid samples covering the out-of-sample period from 2019 to 2023. Neural network models are trained annually using the expanding window. The average loss is calculated as the time series average of the cross-sectional averages of the average quantile loss across all $\tau \in\{0. 00005,0.0001,0.001, 0.005$, $0.01,\dots,0.05$, $0.075$, $0.1, 0.15, 0.2,\dots,0.8,0.85, 0.9$,$0.925$, $0.95, \dots 0.99$,$ 0.995, 0.999, 0.9999, 0.99995\}$ as in (Eq. \ref{eq:avg_out_loss}) for the two-stage NN model, as well as benchmark neural network models (the linear model (LNN), the one-layer model (1hNN), the two-layer model (2hNN)) and the benchmark GARCH model. Note that losses are multiplied by 100, and the t-statistics in parentheses are computed for the average loss difference between the given model specification and the two-stage benchmark model using Newey-West standard errors with 12 lags.}
\centering
\resizebox{0.6\textwidth}{!}{
\begin{tabular}{llclc}
\toprule \hline \\ [-2ex]
Specification & Variable & Full Sample && Liquid Sample \\
\cmidrule{3-3} \cmidrule{5-5}
GARCH & $L_{avg}$ & 2.799 &  & 1.906\\
  & (t-stat) & (3.42) &  & (2.83)\\
\cmidrule{3-3} \cmidrule{5-5}
LNN & $L_{avg}$ & 2.783 &  & 1.889\\
  & (t-stat) & (4.39) &  & (2.95)\\
\cmidrule{3-3} \cmidrule{5-5}
1hNN & $L_{avg}$ & 2.752 &  & 1.887\\
  & (t-stat) & (3.02) &  & (1.59)\\
\cmidrule{3-3} \cmidrule{5-5}
2hNN & $L_{avg}$ & 2.753 &  & 1.886\\
  & (t-stat) & (1.67) &  & (1.08)\\
\cmidrule{3-3} \cmidrule{5-5}
Two stage NN & $L_{avg}$ & 2.730 &  & 1.871\\
  \hline\bottomrule
\end{tabular}
}
\label{tab:avg_quantile_loss OOS}
\end{table}

The out-of-sample performance of long-short portfolios formed using the predicted mean and median, reported in Table~\ref{tab:MSEvsMean OOS}, further supports the notion of declining return predictability. The full sample long-short portfolio generates an average monthly return of $4.67\%$, which is lower than the $6.44\%$ observed in the earlier period, see Table~\ref{tab:MSEvsMean}. Similarly, for the liquid sample, the average return declines from $2.35\%$ in 1995-2018 to $1.81\%$ in 2019-2023. A similar deterioration is observed in the Sharpe ratios, with the full sample ratio decreasing from $4.34$ to $2.61$. While the two-stage quantile neural network continues to deliver positive portfolio returns, the differences in performance compared to alternative models are no longer statistically significant at the $5\%$ level in either the full or liquid sample. This contrasts with the 1995-2018 period, where long-short portfolios formed using the two-stage quantile neural network mean forecasts significantly outperformed those based on alternative models. Even though statistical significance has diminished, the two-stage quantile neural network still delivers the highest average returns and Sharpe ratios, demonstrating that its predictive signals remain economically relevant, albeit to a lesser extent than in the 1995-2018 period. The decline in profitability aligns with prior literature, such as \cite{mclean2016does}, which documents a systematic drop in the out-of-sample predictive ability of stock return anomalies over time.

\begin{table}[H]
\caption{{\bf Mean and median portfolios.} This table shows average monthly returns and annualised Sharpe ratios for decile long-short equal-weighted portfolios based on the median and mean predictions from the two-stage quantile NN model, the mean prediction from the two-hidden-layer neural network (2hNN - Mean) and the median prediction from the two-stage NN model (2hNN - Median). The results are shown for the full and liquid samples, covering the period from 2019 to 2023.
The long portfolio is formed by taking the top 10\% of stocks with the highest forecast each month, and the short portfolio is formed by taking the bottom 10\% of stocks with the lowest forecast each month. Values in parentheses are t-statistics adjusted for Newey-West standard errors with 12 lags.
The significance of the difference in mean returns is tested using t-statistics with Newey-West standard errors with 12 lags.}
\centering
\resizebox{0.8\textwidth}{!}{
\begin{tabular}{lccclccc}
\toprule \hline \\ [-2ex]
&\multicolumn{3}{c}{Full Sample} && \multicolumn{3}{c}{Liquid Sample}\\
\cmidrule{2-4} \cmidrule{6-8}
&Two stage NN & Two stage NN & 2hNN MSE && Two stage NN & Two stage NN & 2hNN MSE \\
&Mean & Median & Mean && Mean & Median & Mean \\
\hline \\ [-2ex]
Avg Ret & 4.67 & 3.11 & 4.45 &  & 1.81 & 1.74 & 1.30\\
(t-stat) &  & (1.13) & (0.40) &  &  & (0.16) & (1.49)\\
SR & 2.61 & 1.13 & 3.02 &  & 1.18 & 0.93 & 0.89\\
 \hline \bottomrule
\end{tabular}
}
\label{tab:MSEvsMean OOS}
\end{table}


\pagebreak

\section{Additional distributional forecast evaluation metrics}
\label{app:US_update_quantile_breakdown}
\counterwithin{table}{section} 
\counterwithin{figure}{section}
\renewcommand{\thetable}{G\arabic{table}}

To statistically evaluate differences in forecast accuracy, we employ a modified Diebold–Mariano (DM) test following \cite{gu2020empirical}. At each forecast origin date $t$, we compare the cross-sectional average of squared forecast errors across models. Let $y_{i,t+1}$  denote the realized outcome in period $t+1$ for stock $i$, which can be either a return (Table~\ref{tab:Mean forecast evaluation} or a realized volatility (Table~\ref{tab:VolatilityForecast}). Let $\hat{y}^{(j)}_{i,t+1}$ be the corresponding forecast from model \( j \in \{1,2\} \), and define the prediction error as $e^{(j)}_{i,t+1} = y_{i,t+1} - \hat{y}^{(j)}_{i,t+1}$. The cross-sectional difference in squared errors is
\begin{equation}
d_t = \frac{1}{n_t} \sum_{i=1}^{n_t} \left[ \left( e^{(1)}_{i,t+1} \right)^2 - \left( e^{(2)}_{i,t+1} \right)^2 \right],
\end{equation}
where $n_t$ is the number of stocks with available forecasts at time $t$. The Diebold–Mariano test statistic is then computed as 
\begin{equation}
DM = \frac{\bar{d}}{\hat{\sigma}_{\bar{d}}},
\end{equation}
where $\bar{d}$ is the time-series mean of $d_t$ and $\hat{\sigma}_{\bar{d}}$ is the Newey–West standard error with 12 lags. This test is used throughout to compare the predictive performance of return and volatility forecasts. By aggregating forecast errors cross-sectionally at each time point, this formulation provides valid inference even in the presence of strong cross-sectional correlation. \\


Further, the evaluation of $\tau$-specific quantile forecast losses, reported in Table~\ref{tab:correlationsInt}, provides a detailed assessment of the model's ability to estimate the distribution of returns across different quantiles. The results indicate that the two-stage quantile neural network (2sNN) achieves the lowest quantile loss in 40 out of 43 (35 out of 43) quantiles for the full (liquid) sample,  confirming its general robustness in estimating the distribution of stock returns. However, its comparative advantage varies across different parts of the distribution, and it is notably weaker in the extreme tails.
Comparing across full and liquid samples, the relative ranking of the models remains largely unchanged across the two samples, with the two-stage quantile neural network generally yielding the lowest out-of-sample quantile loss.  


\begin{table}[H] 
\caption{{\bf $\tau$-specific out-of-sample quantile forecast losses.}
This table shows the average out-of-sample quantile cross-section loss for quantile forecasts from the two-stage model and alternative models. Quantile losses
for $\tau \in\{0. 00005,0.0001,0.001, 0.005$, $0.01,\dots,0.05$, $0.075$, $0.1, 0.15, 0.2,\dots,0.8,0.85, 0.9$,$0.925$, $0.95, \dots 0.99$,$ 0.995, 0.999, 0.9999, 0.99995\}$ are reported for the two-stage NN model, as well as benchmark neural network models (the linear model (LNN), the one-layer model (1hNN), the two-layer model (2hNN)) and the benchmark GARCH model.
The results are obtained separately for the full and liquid samples for US stocks covering the out-of-sample period from 1995 to 2018. Neural network models are trained annually using the expanding window on US data only, while out-of-sample forecasts are generated for both the US and international samples Note that losses are multiplied by 100.0.}

\resizebox{0.9\textwidth}{!}{
\begin{tabular}{lccccclccccc}

\toprule \hline \\ [-2ex]
& \multicolumn{5}{c}{Full Sample} && \multicolumn{5}{c}{Liquid Sample}\\
\cmidrule{2-6} \cmidrule{8-12}
& GARCH & LNN & 1hNN & 2hNN & 2sNN && GARCH & LNN & 1hNN & 2hNN & 2sNN \\
\midrule
0.00005 & 0.007 & 0.005 & 0.004 & 0.004 & 0.004 &  & 0.005 & 0.005 & 0.004 & 0.004 & 0.004\\
0.0001 & 0.012 & 0.010 & 0.008 & 0.008 & 0.008 &  & 0.009 & 0.010 & 0.007 & 0.007 & 0.007\\
0.001 & 0.074 & 0.069 & 0.061 & 0.060 & 0.059 &  & 0.053 & 0.054 & 0.048 & 0.048 & 0.047\\
0.005 & 0.272 & 0.251 & 0.238 & 0.237 & 0.232 &  & 0.197 & 0.191 & 0.186 & 0.186 & 0.181\\
0.01 & 0.469 & 0.433 & 0.419 & 0.419 & 0.409 &  & 0.342 & 0.331 & 0.328 & 0.329 & 0.319\\
0.02 & 0.800 & 0.747 & 0.729 & 0.728 & 0.712 &  & 0.588 & 0.576 & 0.571 & 0.575 & 0.558\\
0.03 & 1.085 & 1.019 & 0.999 & 0.998 & 0.978 &  & 0.803 & 0.790 & 0.785 & 0.789 & 0.768\\
0.04 & 1.343 & 1.267 & 1.244 & 1.243 & 1.219 &  & 0.998 & 0.984 & 0.979 & 0.984 & 0.960\\
0.05 & 1.580 & 1.495 & 1.471 & 1.470 & 1.443 &  & 1.177 & 1.163 & 1.158 & 1.164 & 1.137\\
0.075 & 2.106 & 2.006 & 1.977 & 1.975 & 1.946 &  & 1.579 & 1.566 & 1.559 & 1.566 & 1.534\\
0.1 & 2.563 & 2.450 & 2.420 & 2.416 & 2.387 &  & 1.929 & 1.916 & 1.909 & 1.915 & 1.882\\
0.15 & 3.325 & 3.198 & 3.166 & 3.161 & 3.134 &  & 2.517 & 2.503 & 2.497 & 2.504 & 2.469\\
0.2 & 3.935 & 3.803 & 3.772 & 3.768 & 3.745 &  & 2.990 & 2.975 & 2.971 & 2.978 & 2.944\\
0.25 & 4.425 & 4.300 & 4.270 & 4.264 & 4.246 &  & 3.370 & 3.353 & 3.352 & 3.358 & 3.327\\
0.3 & 4.813 & 4.699 & 4.673 & 4.668 & 4.653 &  & 3.669 & 3.651 & 3.653 & 3.660 & 3.631\\
0.35 & 5.112 & 5.016 & 4.993 & 4.988 & 4.975 &  & 3.896 & 3.878 & 3.883 & 3.889 & 3.863\\
0.4 & 5.331 & 5.256 & 5.236 & 5.231 & 5.220 &  & 4.055 & 4.038 & 4.045 & 4.051 & 4.028\\
0.45 & 5.477 & 5.425 & 5.407 & 5.400 & 5.392 &  & 4.150 & 4.135 & 4.144 & 4.149 & 4.130\\
0.5 & 5.558 & 5.522 & 5.506 & 5.500 & 5.493 &  & 4.184 & 4.172 & 4.182 & 4.186 & 4.170\\
0.55 & 5.578 & 5.551 & 5.537 & 5.530 & 5.522 &  & 4.160 & 4.150 & 4.157 & 4.162 & 4.149\\
0.6 & 5.539 & 5.511 & 5.495 & 5.486 & 5.478 &  & 4.076 & 4.066 & 4.071 & 4.075 & 4.064\\
0.65 & 5.428 & 5.395 & 5.375 & 5.365 & 5.355 &  & 3.930 & 3.919 & 3.921 & 3.923 & 3.914\\
0.7 & 5.233 & 5.194 & 5.170 & 5.159 & 5.145 &  & 3.717 & 3.705 & 3.701 & 3.702 & 3.696\\
0.75 & 4.938 & 4.894 & 4.866 & 4.853 & 4.837 &  & 3.432 & 3.416 & 3.409 & 3.408 & 3.403\\
0.8 & 4.525 & 4.478 & 4.447 & 4.433 & 4.414 &  & 3.067 & 3.046 & 3.036 & 3.033 & 3.029\\
0.85 & 3.966 & 3.919 & 3.884 & 3.870 & 3.849 &  & 2.605 & 2.581 & 2.567 & 2.562 & 2.560\\
0.9 & 3.208 & 3.164 & 3.130 & 3.116 & 3.093 &  & 2.025 & 1.998 & 1.983 & 1.975 & 1.974\\
0.925 & 2.726 & 2.683 & 2.651 & 2.639 & 2.616 &  & 1.676 & 1.647 & 1.633 & 1.625 & 1.623\\
0.95 & 2.141 & 2.101 & 2.072 & 2.062 & 2.040 &  & 1.269 & 1.242 & 1.230 & 1.221 & 1.220\\
0.96 & 1.867 & 1.828 & 1.801 & 1.792 & 1.772 &  & 1.085 & 1.059 & 1.048 & 1.039 & 1.038\\
0.97 & 1.559 & 1.522 & 1.498 & 1.490 & 1.472 &  & 0.884 & 0.859 & 0.849 & 0.841 & 0.840\\
0.98 & 1.203 & 1.170 & 1.149 & 1.144 & 1.128 &  & 0.659 & 0.636 & 0.628 & 0.620 & 0.620\\
0.99 & 0.768 & 0.743 & 0.725 & 0.722 & 0.710 &  & 0.395 & 0.378 & 0.371 & 0.365 & 0.366\\
0.995 & 0.489 & 0.474 & 0.455 & 0.454 & 0.446 &  & 0.235 & 0.223 & 0.217 & 0.213 & 0.214\\
0.999 & 0.182 & 0.181 & 0.157 & 0.156 & 0.153 &  & 0.072 & 0.073 & 0.062 & 0.060 & 0.061\\
0.9999 & 0.059 & 0.074 & 0.033 & 0.034 & 0.034 &  & 0.015 & 0.022 & 0.010 & 0.010 & 0.010\\
0.99995 & 0.042 & 0.057 & 0.021 & 0.023 & 0.022 &  & 0.010 & 0.016 & 0.006 & 0.006 & 0.006\\
\hline \bottomrule
\end{tabular}
}
\label{tab:tauSpecificQL}
\end{table} 

Further, to complement the analysis based on average quantile loss, the average out-of-sample continuous ranked probability score (CRPS) is also reported. A set of predicted quantiles $\{ \hat{Q}_{i,t}(\tau_k) \}$, defined over a potentially uneven grid of $\tau_k \in (0,1)$ as in the neural network specifications used in the main text, is first obtained. Cubic spline interpolation is then applied to map these quantile levels to a smooth cumulative distribution function (CDF) $F_{i,t}(x)$ over return space, which is evaluated over a dense and uniformly spaced grid. The CRPS is computed as

\begin{equation}
\text{CRPS}_{i,t} = \int_{-\infty}^{\infty} \left( F_{i,t}(x) - \mathbb{I}\{r_{i,t} \le x\} \right)^2 dx,
\end{equation}

To ensure numerical stability and enforce monotonicity, linear interpolation is used in cases where the estimated density becomes ill-behaved. The final CRPS is computed as the time series average of the cross-sectional mean across all predicted stocks.

The results in Table~\ref{tab:scoring_rule} confirm the findings based on average quantile loss. The two-stage neural network achieves the lowest CRPS in all regions and for both the full and liquid samples. For the full sample, improvements over GARCH are statistically significant at the 5\% level across all regions. The same holds for the comparison with 1hNN in the US, Europe and Japan, and with 2hNN in the US and Asia-Pacific. In the liquid sample, the two-stage model significantly outperforms GARCH in all regions, while differences relative to 1hNN and 2hNN are smaller and not statistically significant at conventional levels. Overall, the CRPS results are consistent with the earlier quantile loss evaluation and provide further support for the predictive accuracy of the two-stage model. This alignment is expected given the well-known relationship between the average quantile loss and the continuous ranked probability score.

\begin{table}[H]
\caption{{\bf Out-of-sample quantile forecasts evaluation with scoring rule.}
This table shows the average out-of-sample continuous ranked probability score (CRPS) for quantile forecasts from the two-stage model and alternative models. The results are obtained separately for the full and liquid samples for individual regions covering the out-of-sample period from 1995 to 2018. Neural network models are trained annually using the expanding window on US data only, while out-of-sample forecasts are generated for both the US and international samples. The average CRPS is calculated as the time series average of the cross-sectional averages of the CRPS for the two-stage NN model, as well as benchmark neural network models (the one-layer model (1hNN), the two-layer model (2hNN)) and the benchmark GARCH model. Note that CRPS is multiplied by 100, and the t-statistics in parentheses are computed for the average difference between the given model specification and the two-stage benchmark model using Newey-West standard errors with 12 lags.}
\resizebox{1\textwidth}{!}{
\begin{tabular}{llcccclcccc}
\toprule \hline \\ [-2ex]
Specification & Variable & \multicolumn{4}{c}{Full Sample} && \multicolumn{4}{c}{Liquid Sample}\\
\cmidrule{1-6} \cmidrule{8-11}
&& USA & Europe & Japan & Asia Pacific && USA & Europe & Japan & Asia Pacific\\
\cmidrule{3-6} \cmidrule{8-11} \\ [-2ex]
GARCH & Avg CRPS & 7.805 & 6.458 & 6.210 & 7.975 &  & 5.777 & 5.247 & 5.792 & 6.643\\
  & (t-stat) & (9.44) & (9.64) & (5.46) & (12.10) &  & (4.21) & (2.60) & (2.65) & (3.87)\\
\cmidrule{3-6} \cmidrule{8-11} \\ [-2ex]
1hNN & Avg CRPS & 7.646 & 6.309 & 6.092 & 7.752 &  & 5.741 & 5.217 & 5.759 & 6.589\\
  & (t-stat) & (3.91) & (2.18) & (2.85) & (1.98) &  & (2.48) & (1.05) & (1.09) & (1.39)\\
\cmidrule{3-6} \cmidrule{8-11} \\ [-2ex]
2hNN & Avg CRPS & 7.630 & 6.297 & 6.085 & 7.761 &  & 5.743 & 5.218 & 5.777 & 6.608\\
  & (t-stat) & (1.90) & (0.56) & (1.26) & (1.97) &  & (2.19) & (0.98) & (1.93) & (1.87)\\
\cmidrule{3-6} \cmidrule{8-11} \\ [-2ex]
Two Stage & Avg CRPS & 7.603 & 6.290 & 6.063 & 7.736 &  & 5.714 & 5.205 & 5.747 & 6.561\\
\hline\bottomrule
\end{tabular}
}
\label{tab:scoring_rule}
\end{table}

\pagebreak

\section{Simulation}
\label{app:US_update_simulations}
\counterwithin{table}{section} 
\counterwithin{figure}{section}
\renewcommand{\thetable}{H\arabic{table}}

We simulate 60 years of daily returns for 7500 stocks, with each month comprising 22 daily observations. The first 30 years are used as the training/estimation sample and the remaining 30 years as the test period. For each stock $i$ and day $t$, returns are generated as:

\begin{equation}
r_{i,t} = \beta_{i}r_{m,t} + \sigma_{i,t}\epsilon_{i,t} + J_{i,t},
\end{equation}


where $r_{m,t}$ is the simulated market return, $\beta_{i}$ is the market beta, $\sigma_{i,t}$ is the idiosyncratic volatility, $\epsilon_{i,t}$ follows a $t$-distribution, and $J_{i,t}$ represents occasional jump component. The idiosyncratic volatility $\sigma_{i,t}$ is simulated using a GJR-GARCH(1,1) model with a Student-$t$ distribution.

\begin{figure}[H]
\centering
\includegraphics[width=0.95\textwidth]{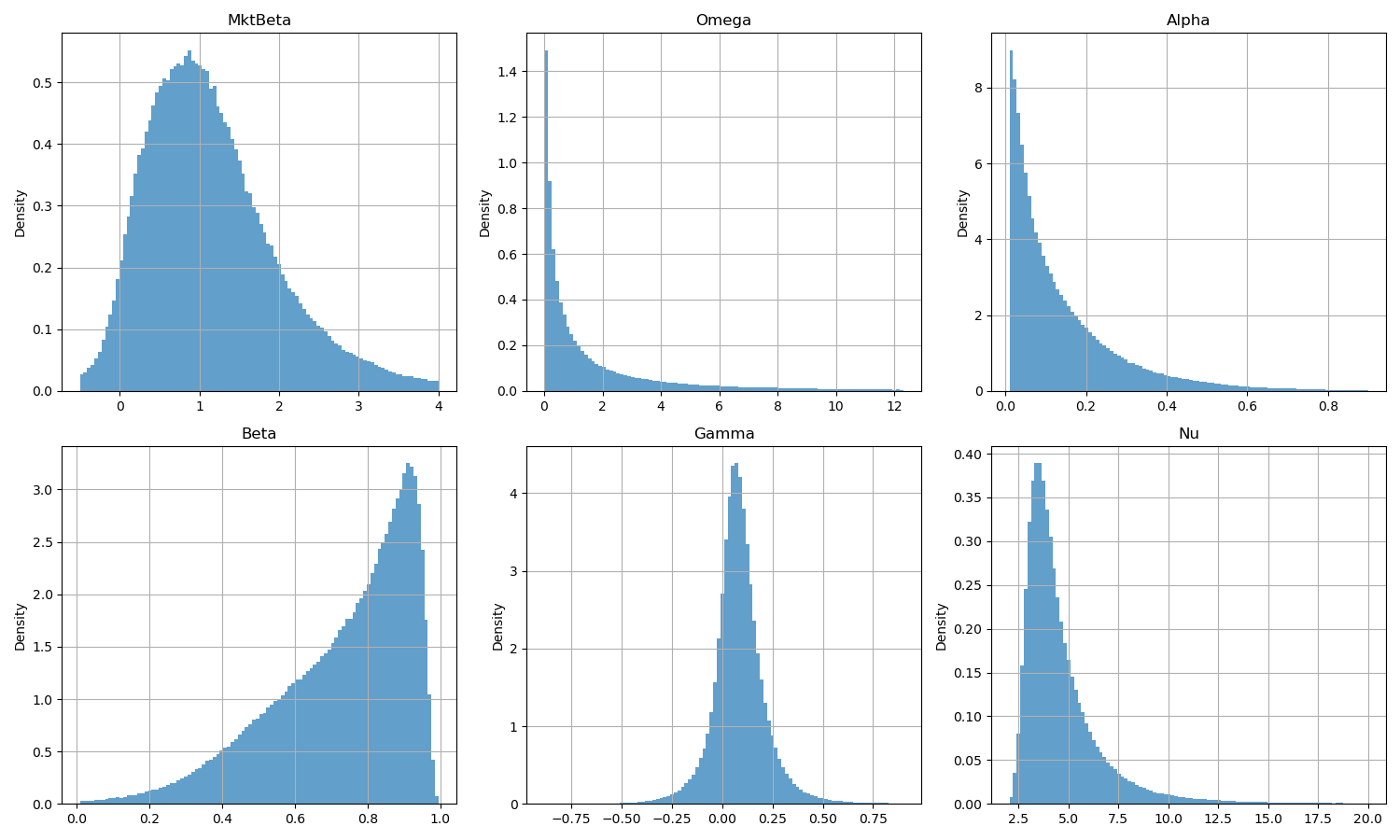}
\caption{\textbf{Empirical distributions of parameters used in the simulation.} 
This figure shows histograms of the key parameters used to simulate daily idiosyncratic volatility via GJR-GARCH with $t$-distributed innovations. The parameters: market beta, $\omega$, $\alpha$, $\beta$, asymmetry coefficient $\gamma$, and degrees of freedom $\nu$—are estimated from U.S. stock-level data over the period 1973–2023. }
\label{fig:figureSimVars}
\end{figure}

Parameters of the data-generating process are based on actual U.S. stock data. GJR-GARCH(1,1) model is estimated based on the past 3 years of data for all stocks with sufficient data each month over the period January 1973 to December 2023. Stock-specific $\beta_{i}$ values are also estimated from empirical data for each month. Individual stock-month observations are discarded if GARCH estimation does not converge or if estimated parameters fall outside reasonable bounds.\footnote{Specifically $\omega$ in the extreme 5\% tails, $\alpha > 0.01$, $\beta > 0.01$, $\alpha + \beta > 0.5$, degrees of freedom less than 20, or $\beta_{i,t} > 4$ or $\beta_{i,t} \leq -0.5$. Figure~\ref{fig:figureSimVars} displays the cross-sectional distributions of market beta (\textit{MktBeta}), the GARCH parameters $\omega$, $\alpha$, $\beta$, and the asymmetry term $\gamma$, as well as the degrees of freedom $\nu$ of the $t$-distribution.} Finally, a set of parameters is randomly sampled for each individual simulated stock from the filtered empirical stock-month parameters and kept constant throughout the simulation.

The market return $r_{m,t}$ is generated from a standard GARCH(1,1) process with parameters $\alpha = 0.06$, $\beta = 0.94$, $\omega = 0.0025$, and a $t$-distribution with $5$ degrees of freedom. The parameters were chosen so that the simulated returns mimic the historical volatility of the S\&P 500. The jump components $J_{i,t}$ occur with daily probability $0.01$ and are simulated as Student-$t$ random variables with $3$ degrees of freedom and standard deviation $0.25$. Daily returns are truncated to lie in $[-0.9, 10]$, and monthly returns, along with their forecasts, are further truncated to $[-1, 20]$ to avoid influence from extreme outliers.\\ 

We use the same model architecture, hyperparameters, and data transformations as in our empirical application. Features include exponentially weighted moving averages (EWMAs) of squared returns with decay factors $\{0.8, 0.9, 0.94, 0.96, 0.98, 0.99\}$,
EWMAs of squared negative returns with decay factors $\{0.8, 0.9, 0.94\}$, total volatilities over the past $3$, $6$, and $12$ months.
In addition, market-related characteristics that are feasible to estimate such as beta, idiosyncratic risk, maximum return, total volatility, and betting-against-beta are considered as are cross-sectional mean returns estimated via EWMA with decay rates $\{0.9, 0.94, 0.96, 0.99, 0.999\}$. 

To evaluate prediction accuracy, we first compute true quantiles of stock-month returns via bootstrap with 100,000 simulations based on the true underlying data-generating process. These monthly returns are calculated as the cumulative returns over 22 trading days.\footnote{Note that we discard stock-months where the true realized volatility exceeds $1$ from evaluation of predictions. These extreme cases can distort overall accuracy and introduce instability into loss metrics. These cases are more problematic for GARCH baseline rather than our neural network architecture.} This simulation setup lets us assess how well the two-stage quantile neural network forecasts conditional quantiles under a realistic, known data-generating process. It also helps us understand how the performance of the model scales with the sample size compared to a benchmark. We use the first five years of simulated data for signal construction (burn-in) and the next 25 years as a training window to estimate the model.

As a baseline, we estimate GARCH(1,1) models with $t$-distributed innovations using previous 3 years of daily returns for each stock-month and then compute monthly quantiles via simulation. Neural networks are estimated using varying cross-sectional sizes of the training sample: $n = 100$, $500$, $1000$, and $2500$ stocks. This corresponds to $12 \times 25 \times n$ observations, e.g.,  from 30,000 for $n=100$ up to 750,000 monthly stock-returns. The $n$ stocks are randomly sampled from the full population of 7500 stocks. Neural networks are estimated and their out-of-sample performance is saved. True quantiles metrics and baseline GARCH(1,1) prediction performance is saved for the same subsample of stocks. The sampling of $n$ stocks is repeated 20 times and results are reported based on averages over the 20 runs.\footnote{Note that this subsampling of precomputed values for 7500 stocks was introduced to make the exercise computationally feasible. A better solution without computational resource constraints would be to simulate completely new data for each of the runs.}\\

Table~\ref{tab:SimulationsEval} shows the root mean squared error (RMSE) of quantile forecasts across a wide range of $\tau$ values used in the empirical results, comparing the two-stage neural network (NN) model trained on varying sample sizes to a GARCH(1,1) benchmark. The evaluation is based on a data-generating process that embeds realistic features of stock returns, including time-varying idiosyncratic volatility (GJR-GARCH), market beta exposure, fat-tailed innovations, and jump components. 

Overall, the results indicate that the two-stage model delivers substantial improvements in quantile forecasting accuracy over the GARCH benchmark across almost all $\tau$-quantiles.
This improvement is most pronounced in the central quantiles and gradually narrows toward the extreme tails. For instance, at the median ($\tau = 0.5$), the RMSE of the GARCH model is $2.16$, whereas the two-stage quantile model trained on 2500 stocks achieves a much lower RMSE of $0.97$. Even with as few as 100 stocks in the training sample, the two-stage quantile model already outperforms GARCH, although with higher variability (RMSE = $1.59$, std = $0.16$). 

Similar patterns hold for both lower and upper quantiles. At $\tau = 0.1$, the GARCH RMSE is 4.29, compared to 3.96 for the NN with 2500 stocks. At $\tau = 0.9$, the two-stage quantile neural network again performs better (RMSE = $4.33$) than GARCH (RMSE = $5.05$), with performance improving steadily with sample size. 

However, in the right extreme tails—particularly at $\tau \in \{0.9999, 0.99995\}$, the advantage of the two-stage model diminishes. For example, at $\tau = 0.99995$, the GARCH model achieves an RMSE of $308.87$, while the best-performing two-stage quantile neural network (with 2500 stocks) yields an RMSE of $302.05$. Although the two-stage quantile neural network still performs slightly better, the relative gains are modest and the variability across simulations increases substantially, with standard errors above $10$ in these tail regions. This behaviour reflects the difficulty of accurately estimating extremely rare events, which occur roughly once every 20,000 months (for $\tau = 0.00005$) and are thus inherently hard to learn even in a controlled setting.

The results also confirm that the predictive accuracy of the two-stage neural network improves systematically with the cross-sectional size of the estimation sample. As the training set grows from 100 to 2500 stocks, corresponding to 30,000 up to 750,000 monthly observations, forecast precision increases across all quantiles.

Taken together, these results provide strong evidence that the two-stage neural network framework is effective at recovering the conditional distribution of returns even in complex, nonlinear environments. The model consistently outperforms GARCH across various parts of the distribution and performs competitively even in the far tails. The improvements are most pronounced when the network is trained on realistically sized panels of data (e.g., 1000 plus stocks over 25 years), as in our empirical application. \\

It is important to note that the data-generating process in our simulation is relatively favorable to the GARCH benchmark and imposes structural limitations on the potential benefits of cross-sectional learning. While all stocks are exposed to a common market factor through betas \( \beta_{i} \), there are no additional latent factors or shared idiosyncratic components beyond this market return \( r_{m,t} \). Each stock's idiosyncratic returns \( \sigma_{i,t}\epsilon_{i,t} \) are generated via a GJR-GARCH(1,1) process with independent parameters for each stock. As such, the simulation lacks any structured cross-sectional dependence in higher moments or residuals beyond the market beta channel.

This setup is particularly well aligned with the assumptions underlying univariate GARCH-type models, which are designed for conditionally heteroskedastic time series with stock-specific parameters. In contrast, our two-stage quantile neural network performs the best under the assumption of parameter sharing across the individual stocks where the parameters are a function of stock characteristics. Learning from a broad panel of instrument-level features—including historical volatilities, return patterns, and cross-sectional characteristics—is not possible because they are not explicitly informative in this setting. Despite this structural disadvantage, the neural network demonstrates substantial predictive gains over GARCH across a wide range of quantiles, especially in central and intermediate regions of the distribution. This highlights the flexibility and robustness of our two-stage quantile neural network architecture even when the data-generating process is not tailored to exploit its cross-sectional learning capacity.

\begin{table}[H]
\caption{\textbf{Simulation results - quantile forecasts evaluation} 
This table presents simulation results for monthly stock return quantile forecasts across quantiles $\tau \in \{0.00005, \ldots, 0.99995\}$, based on a realistic return-generating process with GJR-GARCH volatility and jump components calibrated to U.S. stock data. 
The average (Mean) and standard deviation (Std. Dev.) of simulated monthly stock return quantiles are derived from Monte Carlo simulations with 100,000 draws per stock-month. The GARCH forecast uses a GARCH(1,1) model with $t$-distributed errors, estimated on three years of daily data per stock and month. The two-stage quantile neural network forecasts are trained on datasets of increasing size (100, 500, 1,000, and 2,500 stocks), each with 25 years of monthly data. RMSE is calculated relative to the true quantiles, multiplied by 100, with standard deviations across 20 simulation runs in parentheses.}

\resizebox{1.\textwidth}{!}{
\begin{tabular}{lclclclcccc}
\toprule \hline \\ [-2ex]

& \multicolumn{3}{c}{Actuals} && GARCH && \multicolumn{4}{c}{Two-stage NN} \\
\cmidrule{2-4} \cmidrule{6-6} \cmidrule{8-11}
& Mean && Std.Dev. && RMSE && \multicolumn{4}{c}{RMSE} \\
\cmidrule{2-2} \cmidrule{4-4} \cmidrule{6-6} \cmidrule{8-11}
Quantile & N=2500 && N=2500 && N=2500 && N=100 & N=500 & N=1000 & N=2500 \\
\midrule
0.00005 & -95.54 &  & 3.32 (0.01) &  & 8.50 (0.18) &  & 4.37 (0.34) & 5.16 (0.33) & 5.10 (0.37) & 5.07 (0.33)\\
0.0001 & -94.67 &  & 3.39 (0.01) &  & 11.08 (0.19) &  & 4.83 (0.39) & 5.85 (0.34) & 5.76 (0.41) & 5.69 (0.37)\\
0.001 & -89.47 &  & 3.85 (0.03) &  & 24.18 (0.18) &  & 7.12 (1.03) & 5.90 (0.53) & 5.86 (0.55) & 5.85 (0.35)\\
0.005 & -62.39 &  & 13.16 (0.09) &  & 12.92 (0.10) &  & 9.22 (0.66) & 8.94 (0.53) & 8.82 (0.56) & 8.61 (0.23)\\
0.01 & -51.45 &  & 13.89 (0.10) &  & 9.58 (0.06) &  & 8.17 (0.58) & 7.80 (0.35) & 7.79 (0.42) & 7.59 (0.17)\\
0.02 & -41.58 &  & 13.59 (0.10) &  & 7.08 (0.03) &  & 6.91 (0.43) & 6.51 (0.24) & 6.50 (0.33) & 6.36 (0.12)\\
0.03 & -36.30 &  & 13.10 (0.10) &  & 6.01 (0.03) &  & 6.23 (0.34) & 5.81 (0.21) & 5.81 (0.27) & 5.69 (0.11)\\
0.04 & -32.76 &  & 12.62 (0.09) &  & 5.43 (0.03) &  & 5.77 (0.28) & 5.36 (0.20) & 5.36 (0.23) & 5.26 (0.09)\\
0.05 & -30.13 &  & 12.16 (0.09) &  & 5.08 (0.03) &  & 5.44 (0.25) & 5.01 (0.19) & 5.03 (0.21) & 4.94 (0.08)\\
0.075 & -25.54 &  & 11.11 (0.08) &  & 4.58 (0.03) &  & 4.82 (0.21) & 4.41 (0.17) & 4.44 (0.17) & 4.37 (0.07)\\
0.1 & -22.38 &  & 10.21 (0.08) &  & 4.29 (0.03) &  & 4.38 (0.19) & 3.99 (0.15) & 4.02 (0.14) & 3.96 (0.07)\\
0.15 & -17.91 &  & 8.71 (0.07) &  & 3.89 (0.03) &  & 3.74 (0.18) & 3.37 (0.12) & 3.41 (0.11) & 3.37 (0.06)\\
0.2 & -14.61 &  & 7.46 (0.06) &  & 3.58 (0.03) &  & 3.26 (0.16) & 2.91 (0.10) & 2.95 (0.08) & 2.92 (0.06)\\
0.25 & -11.88 &  & 6.36 (0.05) &  & 3.32 (0.04) &  & 2.86 (0.15) & 2.52 (0.09) & 2.56 (0.07) & 2.53 (0.06)\\
0.3 & -9.48 &  & 5.35 (0.05) &  & 3.08 (0.04) &  & 2.52 (0.14) & 2.17 (0.08) & 2.20 (0.06) & 2.19 (0.05)\\
0.35 & -7.28 &  & 4.38 (0.04) &  & 2.84 (0.04) &  & 2.22 (0.14) & 1.84 (0.08) & 1.87 (0.06) & 1.85 (0.04)\\
0.4 & -5.20 &  & 3.45 (0.04) &  & 2.61 (0.04) &  & 1.96 (0.14) & 1.53 (0.08) & 1.55 (0.05) & 1.54 (0.03)\\
0.45 & -3.18 &  & 2.54 (0.03) &  & 2.38 (0.04) &  & 1.74 (0.15) & 1.24 (0.08) & 1.25 (0.05) & 1.24 (0.03)\\
0.5 & -1.17 &  & 1.65 (0.02) &  & 2.16 (0.04) &  & 1.59 (0.16) & 0.97 (0.08) & 0.98 (0.05) & 0.97 (0.03)\\
0.55 & 0.86 &  & 0.90 (0.02) &  & 1.93 (0.04) &  & 1.53 (0.17) & 0.78 (0.09) & 0.78 (0.06) & 0.77 (0.03)\\
0.6 & 2.97 &  & 0.98 (0.01) &  & 1.73 (0.04) &  & 1.59 (0.17) & 0.75 (0.08) & 0.74 (0.05) & 0.73 (0.02)\\
0.65 & 5.20 &  & 1.90 (0.01) &  & 1.57 (0.04) &  & 1.78 (0.15) & 0.93 (0.05) & 0.93 (0.04) & 0.92 (0.02)\\
0.7 & 7.63 &  & 3.10 (0.02) &  & 1.52 (0.04) &  & 2.12 (0.14) & 1.28 (0.05) & 1.28 (0.05) & 1.27 (0.02)\\
0.75 & 10.36 &  & 4.50 (0.03) &  & 1.69 (0.03) &  & 2.60 (0.15) & 1.76 (0.06) & 1.76 (0.06) & 1.75 (0.03)\\
0.8 & 13.56 &  & 6.18 (0.04) &  & 2.19 (0.02) &  & 3.28 (0.19) & 2.36 (0.08) & 2.37 (0.07) & 2.35 (0.04)\\
0.85 & 17.58 &  & 8.32 (0.06) &  & 3.18 (0.02) &  & 4.25 (0.29) & 3.16 (0.12) & 3.20 (0.09) & 3.15 (0.05)\\
0.9 & 23.22 &  & 11.30 (0.09) &  & 5.05 (0.04) &  & 5.78 (0.56) & 4.36 (0.18) & 4.42 (0.15) & 4.33 (0.07)\\
0.925 & 27.33 &  & 13.42 (0.11) &  & 6.73 (0.07) &  & 6.97 (0.77) & 5.27 (0.25) & 5.38 (0.23) & 5.25 (0.12)\\
0.95 & 33.43 &  & 16.44 (0.15) &  & 9.68 (0.12) &  & 8.77 (1.10) & 6.72 (0.39) & 6.88 (0.34) & 6.70 (0.23)\\
0.96 & 37.01 &  & 18.14 (0.16) &  & 11.66 (0.16) &  & 9.86 (1.28) & 7.57 (0.47) & 7.76 (0.39) & 7.57 (0.28)\\
0.97 & 41.92 &  & 20.38 (0.19) &  & 14.64 (0.22) &  & 11.36 (1.56) & 8.81 (0.64) & 9.03 (0.51) & 8.81 (0.38)\\
0.98 & 49.46 &  & 23.76 (0.23) &  & 19.86 (0.33) &  & 13.74 (1.98) & 10.83 (0.99) & 11.13 (0.73) & 10.89 (0.51)\\
0.99 & 64.35 &  & 30.53 (0.30) &  & 32.28 (0.61) &  & 18.74 (2.71) & 15.37 (1.75) & 15.86 (1.22) & 15.53 (0.86)\\
0.995 & 82.47 &  & 39.44 (0.40) &  & 50.00 (0.97) &  & 25.60 (3.13) & 21.75 (3.00) & 22.56 (2.31) & 21.95 (1.35)\\
0.999 & 144.39 &  & 77.17 (0.93) &  & 105.69 (1.47) &  & 57.28 (5.20) & 50.56 (7.55) & 51.75 (3.99) & 50.66 (2.38)\\
0.9999 & 330.70 &  & 229.23 (3.26) &  & 242.02 (2.85) &  & 254.93 (21.25) & 199.92 (13.43) & 198.86 (9.06) & 198.60 (8.33)\\
0.99995 & 422.40 &  & 301.80 (4.06) &  & 308.87 (2.95) &  & 370.83 (27.29) & 304.33 (18.59) & 304.61 (12.35) & 302.05 (11.93)\\
 
\hline \bottomrule
\end{tabular}
}
\label{tab:SimulationsEval}
\end{table}

\pagebreak

\section{Comparison with tree-based machine learning benchmarks}
\label{app:US_update_tree_based}
\counterwithin{table}{section} 
\counterwithin{figure}{section}
\renewcommand{\thetable}{I\arabic{table}}

In this section, we extend our set of benchmark models to include tree-based machine learning methods previously used in the stock returns forecasting literature \citep{gu2020empirical,tobek2021does}, random forests (RF) and gradient-boosting regression trees (GBRT).

Table~\ref{tab:avg_quantile_loss_rf} reports the average out-of-sample quantile cross-sectional loss for quantile forecasts from the two-stage quantile neural network and alternative benchmark models, including RF, GBRT as well as linear model (LNN) and the benchmark GARCH model, across four regions and for both the full and liquid samples. The quantile loss is averaged across a representative set of quantiles $\tau \in \{0.01, 0.1, 0.25, 0.5, 0.75, 0.9, 0.99\}$. RF and GBRT models use 176 features, where the input features are the same as those for the standardised $\tau$-quantiles network in the two-stage model.

The results show that the two-stage quantile neural network outperforms both GBRT and RF across all regions and sample definitions. The differences in average quantile loss are statistically significantly different from $0$. For example, in the U.S. full sample, the two-stage model achieves a loss of $2.907$, compared to $2.984$ for RF and $2.994$ for GBRT, with corresponding t-statistics of $8.55$ and $4.32$, respectively. This pattern holds across Europe, Japan, and Asia-Pacific, where the two-stage NN model consistently delivers lower quantile losses than GBRT and RF. The same conclusion applies to the liquid sample, where the two-stage model achieves the lowest average loss in all regions.

\begin{table}[H]
\caption{{\bf Out-of-sample quantile forecasts evaluation.}
This table shows the average out-of-sample quantile cross-section loss for quantile forecasts from the two-stage model and alternative models. The results are obtained separately for the full and liquid samples for individual regions covering the out-of-sample period from 1995 to 2018. Neural network models are trained annually using the expanding window on US data only, while out-of-sample forecasts are generated for both the US and international samples. The average loss is calculated as the time series average of the cross-sectional averages of the average quantile loss across all $\tau \in\{0.01, 0.1, 0.25, 0.5, 0.75, 0.9, 0.99\}$ as in (Eq. \ref{eq:avg_out_loss}) for the two-stage NN model, as well as the linear model (LNN), the benchmark GARCH model, random forest (RF), and gradient boosting regression trees (GBRT). Note that losses are multiplied by 100.0, and the t-statistics in parentheses are computed for the average loss difference between the given model specification and the two-stage benchmark model using Newey-West standard errors with 12 lags.}
\resizebox{1\textwidth}{!}{
\begin{tabular}{llcccclcccc}
\toprule \hline \\ [-2ex]
Specification & Variable & \multicolumn{4}{c}{Full Sample} && \multicolumn{4}{c}{Liquid Sample}\\
\cmidrule{3-6} \cmidrule{8-11}
& & USA & Europe & Japan & Asia Pacific && USA & Europe & Japan & Asia Pacific\\
\cmidrule{3-6} \cmidrule{8-11} \\ [-2ex]
GBRT & $L_{avg}$ & 2.994 & 2.471 & 2.388 & 3.057 &  & 2.302 & 2.047 & 2.275 & 2.625\\
  & (t-stat) & (4.32) & (1.99) & (4.63) & (2.66) &  & (4.92) & (2.94) & (4.63) & (3.32)\\
 \cmidrule{3-6} \cmidrule{8-11} \\ [-2ex]
RF & $L_{avg}$ & 2.984 & 2.458 & 2.363 & 2.989 &  & 2.254 & 1.992 & 2.201 & 2.540\\
  & (t-stat) & (8.55) & (3.67) & (5.88) & (3.90) &  & (5.17) & (3.09) & (2.24) & (2.53)\\
 \cmidrule{3-6} \cmidrule{8-11} \\ [-2ex]
GARCH & $L_{avg}$ & 3.000 & 2.501 & 2.356 & 3.073 &  & 2.243 & 1.981 & 2.178 & 2.546\\
  & (t-stat) & (10.21) & (7.81) & (8.43) & (13.10) &  & (5.73) & (3.20) & (2.13) & (4.49)\\
 \cmidrule{3-6} \cmidrule{8-11} \\ [-2ex]
LNN & $L_{avg}$ & 2.948 & 2.436 & 2.333 & 2.958 &  & 2.225 & 1.968 & 2.165 & 2.507\\
  & (t-stat) & (7.63) & (1.16) & (6.09) & (1.64) &  & (3.21) & (1.21) & (0.24) & (0.79)\\
 \cmidrule{3-6} \cmidrule{8-11} \\ [-2ex]
Two stage NN & $L_{avg}$ & 2.907 & 2.426 & 2.291 & 2.947 &  & 2.213 & 1.962 & 2.161 & 2.501\\
\hline\bottomrule
\end{tabular}
}
\label{tab:avg_quantile_loss_rf}
\end{table}

Table~\ref{tab:MSEvsMean_rf} presents long-short portfolio performance based on mean and median forecasts derived from the two-stage quantile neural network, as well as the median predictions from RF and GBRT. Across all regions and for both full and liquid samples, the two-stage model outperforms both tree-based models by a wide margin in terms of average monthly returns and Sharpe ratios. In the U.S. full sample, the long-short portfolio based on the two-stage NN mean achieves an average monthly return of 6.44\% with a Sharpe ratio of 4.34, compared to 2.36\% (SR = 1.20) for GBRT and 0.57\% (SR = 0.33) for RF. The differences in mean returns between the two-stage NN and both tree-based models are statistically significant.
This pattern holds consistently in Europe, Japan, and Asia-Pacific, as well as in the liquid sample. 

\begin{table}[H]
\caption{{\bf Mean and median portfolios.} This table shows average monthly returns and annualised Sharpe ratios for decile long-short equal-weighted portfolios based on the mean predictions from the two-stage quantile NN model, the median prediction from the random forest model (RF - Median) and the median prediction from the gradient boosting regression trees model (GBR - Median). The results are shown for the full and liquid samples and for individual regions, covering the period from 1995 to 2018.
The long portfolio is formed by taking the top 10\% of stocks with the highest forecast each month, and the short portfolio is formed by taking the bottom 10\% of stocks with the lowest forecast each month. Values in parentheses are t-statistics adjusted for Newey-West standard errors with 12 lags.
The significance of the difference in mean returns is tested using t-statistics with Newey-West standard errors with 12 lags.}
\resizebox{1\textwidth}{!}{
\begin{tabular}{llccccclccccc}
\toprule \hline \\ [-2ex]
Specification & Variable & \multicolumn{5}{c}{Full Sample} && \multicolumn{5}{c}{Liquid Sample}\\
\cmidrule{3-7} \cmidrule{9-13}
&& USA & Europe & Japan & Asia Pacific & Global && USA & Europe & Japan & Asia Pacific & Global\\
\hline \\ [-2ex]
GBRT - Median & Avg Ret & 2.36 & 1.48 & 1.20 & 2.14 & 1.80 &  & 0.80 & 0.70 & 0.86 & 0.93 & 0.82\\
 & SR & 1.20 & 1.45 & 0.93 & 1.40 & 1.82 &  & 0.46 & 0.76 & 0.74 & 0.63 & 0.97\\
 & t-stat & 5.30 & 6.68 & 6.24 & 7.00 & 7.88 &  & 3.74 & 3.40 & 4.31 & 3.68 & 5.33\\
\hline \\ [-2ex]
RF - Median & Avg Ret & 0.57 & 0.10 & -0.22 & 0.46 & 0.23 &  & 0.33 & 0.15 & 0.25 & 0.48 & 0.30\\
 & SR & 0.33 & 0.12 & -0.22 & 0.39 & 0.28 &  & 0.28 & 0.20 & 0.29 & 0.53 & 0.58\\
 & t-stat & 6.43 & 8.51 & 8.19 & 9.77 & 9.28 &  & 3.84 & 3.80 & 6.09 & 5.11 & 5.60\\
 \hline \\ [-2ex]
Two stage NN - Mean & Avg Ret & 6.44 & 5.29 & 4.03 & 8.05 & 5.95 &  & 2.35 & 2.02 & 2.27 & 3.12 & 2.44\\
 & SR & 4.34 & 4.49 & 3.31 & 4.99 & 6.92 &  & 1.70 & 1.88 & 1.92 & 2.13 & 2.83\\
\hline\bottomrule
\end{tabular}
}
\label{tab:MSEvsMean_rf}
\end{table}

\pagebreak

\section{Motivation for two-stage architecture} \label{sec:AppendixTwoStageMotivation}
\counterwithin{table}{section} 
\counterwithin{figure}{section}
\renewcommand{\thetable}{C\arabic{table}}

Stock return prediction represents a challenging domain characterized by an extremely low signal-to-noise ratio, which poses significant difficulties for the effective application of machine learning techniques. Consequently, even modest improvements in the signal-to-noise ratio can yield substantial gains in predictive performance. The variation in stock returns can be decomposed into cross-sectional and time-series components. Applying cross-sectional standardisation to returns helps mitigate market-wide volatility fluctuations, thereby enhancing the signal-to-noise ratio. This approach is particularly beneficial during periods of financial crisis, when market-wide volatility tends to spike sharply, as it allows models to focus more effectively on idiosyncratic return signals.

A key factor contributing to the effectiveness of cross-sectional standardisation lies in the fact that all anomaly signals, used as predictive features, are themselves cross-sectionally standardised. This practice is well established in the literature, as anomaly signals typically capture relative performance across stocks within a given period and contain little to no time-series information. Consequently, standardising stock returns in a manner consistent with the anomaly signals ensures alignment with most of the predictive features.

Full cross-sectional standardisation of stock returns is formally defined as subtracting the cross-sectional mean of stock-level returns and scaling by the cross-sectional standard deviation at time $t$. Specifically,
\begin{equation} \label{eq:appendix standardisation}
    \widetilde{r}_{i,t+1} = \frac{r_{i,t+1} - \overline{r}_{t+1}}{\sigma_{t+1}},
\end{equation}
where $ \overline{r}_{t} = \frac{1}{N_t} \sum_{i=1}^{N_t} r_{i,t} $ denotes the cross-sectional average of stock returns at time $t$, $\sigma_t^2 = \frac{1}{N_t} \sum_{i=1}^{N_t} (r_{i,t} - \overline{r}_t)^2$ is the cross-sectional variance, and $N_t$ is the number of stocks in the sample at time $t$. The resulting standardised returns $\widetilde{r}_{i,t+1}$ have a cross-sectional mean of zero and unit volatility, effectively eliminating market-wide variations in level and scale. While this transformation improves the signal-to-noise ratio by removing common fluctuations over time, it introduces practical challenges: both the cross-sectional mean $\overline{r}_{t+1}$ and volatility $\sigma_{t+1}$ must be predicted to transform standardised predictions back into non-standardised returns. Moreover, the functional form in \autoref{eq:appendix standardisation} results in a complex gradient during backpropagation, which complicates the training of neural networks and often leads to suboptimal empirical performance.

To address these limitations, a simplified standardisation method is employed in the context of a two-stage network architecture:
\begin{equation} \label{eq:appendix standardisation2}
    \widetilde{r}_{i,t+1} = \frac{r_{i,t+1}}{\overline{\sigma}_{t}},
\end{equation}
where $\overline{\sigma}_t = \frac{1}{N_t} \sum_{i=1}^{N_t} \sigma_{i,t}$ represents the cross-sectional average of individual stock-level volatility estimates at time $t$. This average, $\overline{\sigma}_t$, serves as a practical proxy for the future cross-sectional volatility $\sigma_{t+1}$, leveraging only information available at time $t$, thus eliminating the need for its prediction. Empirically, using $\overline{\sigma}_t$ has been found to yield more robust and effective results than attempting to estimate $\sigma_{t+1}$ directly.

Notably, the cross-sectional mean return is not subtracted in this simplified approach. This omission is justified by the well-documented difficulty of accurately predicting aggregate market returns, particularly at short horizons. By focusing solely on volatility scaling, the method preserves interpretability and numerical stability while still mitigating the impact of time-varying market volatility, especially during periods of heightened turbulence, thereby improving the signal quality for predictive modeling.

\begin{figure}[H]
    \includegraphics[width = 1\textwidth]{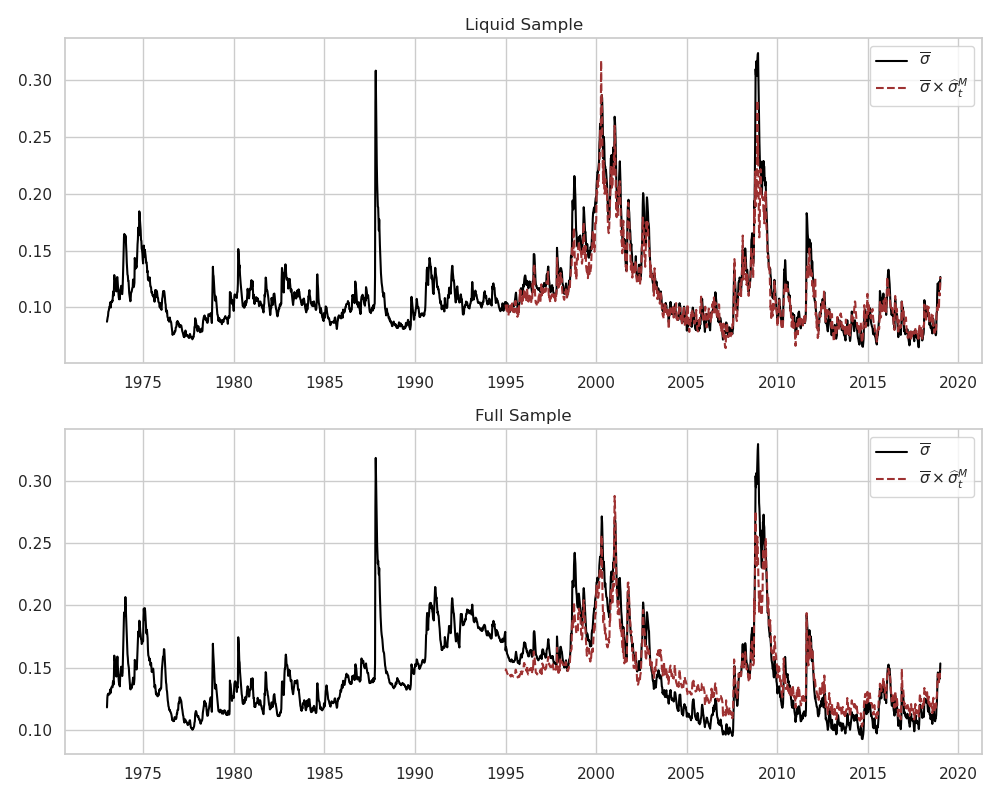}
    \caption{\textbf{The cross-sectional average of volatilities as a scaling factor.} 
    This figure shows the evolution over time of the cross-sectional average of stock-level volatilities $\overline{\sigma}$ for the full and liquid sample of U.S. stocks.
    Stock-level volatilities are calculated as exponential moving averages of squared daily returns with a smoothing factor of $0.94$.
    The scaling factor is used to standardise stock returns to be used as a label for the standardised $\tau$-quantile neural network or 
    stage 1 of the two-stage model defined in \autoref{subsubsec:TwoStage}. $\overline{\sigma} \times \widehat{\sigma}_t^M$ is a product of $\overline{\sigma}$ and output from the second stage of the two-stage model $\widehat{\sigma}_t^M$. $\overline{\sigma} \times \widehat{\sigma}_t^M$ is multiplied with the output from the first stage to get final predictions for return quantiles.
    }
    \label{fig:Scaler}
\end{figure}

In the two-stage model, Stage 2 introduces a multiplicative adjustment factor, denoted $\widehat{\sigma}_t^M$, which refines the market-wide volatility measure $\overline{\sigma}_t$ to enhance predictive performance. The final scaling factor applied to the outputs of Stage 1 is given by $\overline{\sigma}_t \times \widehat{\sigma}_t^M$. A visualization of both $\overline{\sigma}_t$ and the adjusted scalar $\overline{\sigma}_t \times \widehat{\sigma}_t^M$ over the period 1973–2018 in the U.S. market is provided in \autoref{fig:Scaler}. Note that $\widehat{\sigma}_t^M$ is displayed only from 1995 onward, as it is derived from out-of-sample model predictions.

Two key effects of $\widehat{\sigma}_t^M$ on the composite volatility scalar are evident over the full sample. First, it reduces the magnitude of the scalar during periods of market crisis, effectively moderating the impact of elevated historical volatility on the next month predictions. Second, it increases the scalar during unusually calm market regimes, thereby preventing excessive suppression of market-wide volatility predictions. Similarly, in the liquid stock sample, the peaks in market-wide volatility are noticeably dampened, contributing to more stable and robust predictions.

Overall, the contribution of Stage 2 output $\widehat{\sigma}_t^M$ to the final output is relatively modest, as intended. This design is inspired by the ResNet architecture \cite{he2016deep}, in which auxiliary pathways are expected to remain close to unity and serve primarily as fine-tuning mechanisms, which was found to improve the neural network learning process. In this context, $\widehat{\sigma}_t^M$ acts as a corrective refinement to $\overline{\sigma}_t$, making targeted adjustments only where necessary, rather than driving the overall scaling.

An alternative, yet unexplored modeling approach would involve employing two separate neural networks. The first network would correspond to Stage 1 of our current two-stage framework, generating predictions based on cross-sectionally standardised inputs. The role of Stage 2, currently fulfilled by the product $\overline{\sigma}_t \times \widehat{\sigma}_t^M$, could instead be replaced by an independent network trained to directly forecast the future cross-sectional volatility, $\sigma_{t+1}$. The output of this second network would then be used as the scaling factor in the return standardisation process, effectively substituting $\overline{\sigma}_t \times \widehat{\sigma}_t^M$ with a more flexible, end-to-end learned volatility estimate.

While the proposed two-stage architecture has demonstrated strong out-of-sample predictive performance, it comes with certain limitations regarding the assumptions it imposes on the distribution of predicted returns. Specifically, the rescaling mechanism inherent in the architecture constrains the modeled return distributions to the location-scale family. This implies that the model cannot capture features such as discrete probability masses outside zero return or other non-smooth distributional characteristics.

To address this limitation and provide a more flexible benchmark, we include results for a single-stage, two-hidden-layer neural network (2hNN) in all main tables, either in the main text or in the Appendix~\ref{sec:AppendixResults}. The 2hNN model does not employ any cross-sectional standardisation, allowing it to learn return distributions of arbitrary shape without structural constraints. While this approach may entail a modest reduction in out-of-sample predictive accuracy compared to the two-stage model, it offers greater distributional flexibility. As such, the 2hNN can be particularly useful in empirical applications where capturing the full complexity of return distributions is a primary objective, even at the cost of some predictive efficiency.

\end{document}